\documentclass[12pt]{article}

\usepackage{amsmath,amssymb,amsfonts,amsthm,bbm,cancel,wasysym}
\usepackage{epsfig,graphics,graphicx,epstopdf}
\usepackage{rotating}
\usepackage{colordvi,color}
\usepackage{array,booktabs,colortbl,colordvi,multirow}
\usepackage{axodraw}

\newcommand{\tinymath}[1]{{\tiny{\mbox{$#1$}}}}
\newcommand{\scriptmath}[1]{{\scriptsize{\mbox{$#1$}}}}
\newcommand{\smallmath}[1]{{\small{\mbox{$#1$}}}}

\newcommand{\bfmath}[1]{{\mbox{\boldmath{$#1$}}}}

\newcommand{\Bfmath}[1]{{\large{\mbox{\boldmath{$#1$}}}}}

\newcommand{\ctoprule}{\toprule[0.5mm]}
\newcommand{\cbottomrule}{\bottomrule[0.5mm]}
\newcommand{\cmrule}{\midrule[0.25mm]}

\newcommand{\crowcolor}{\rowcolor[rgb]{0.9,0.9,0.9}}
\newcommand{\whitecell}{\cellcolor[rgb]{1,1,1}}
\newcommand{\greycell}{\cellcolor[rgb]{0.9,0.9,0.9}}

\newcommand{\vtext}[1]{\begin{sideways}\small{#1}\end{sideways}}

\newcommand{\be}{\begin{equation}}
\newcommand{\ee}{\end{equation}}
\newcommand{\bea}{\begin{eqnarray}}  
\newcommand{\eea}{\end{eqnarray}}

\newcommand{\refeq}[1]{\mbox{Eq.~(\ref{#1})}}

\newcommand{\Dslash}{\cancel{D}~}

\renewcommand{\Re}[1]{\mathrm{Re}\left[#1\right]}
\renewcommand{\Im}[1]{\mathrm{Im}\left[#1\right]}

\newcommand{\hc}{\mathrm{h.c.}} 

\newcommand{\U}[1]{U(#1)}
\newcommand{\SU}[1]{SU(#1)}

\newcommand{\Adj}{\mathrm{Adj}}

\newcommand{\SM}{\mathrm{SM}}

\newcommand{\mt}[1]{\mathrm{#1}}

\newcommand{\WB}{W\!B}

\newcommand{\units}[1]{~\mathrm{#1}}

\newcommand{\Zprime}{{Z^\prime}}

\newcommand{\nn}{\nonumber \\} 

\newcommand{\trps}{\mathrm{T}}

\newcommand{\la}{\langle}
\newcommand{\ra}{\rangle}

\begin{document}


\begin{flushright}
UG-FT-272/10 \\
CAFPE-142/10 \\

\today
\end{flushright}
\vspace*{5mm}
\begin{center}

\renewcommand{\thefootnote}{\fnsymbol{footnote}}

{\Large {\bf  Electroweak Limits on General New Vector Bosons   
}} \\
\vspace*{1cm}
{\bf F.\ del Aguila}\footnote{E-mail: faguila@ugr.es},
{\bf J.\ de Blas}\footnote{E-mail: deblasm@ugr.es}
and
{\bf M.\ P\'erez-Victoria}\footnote{E-mail: mpv@ugr.es}

\vspace{0.5cm}

Departamento de F\'{\i}sica Te\'orica y del Cosmos and CAFPE,\\
Universidad de Granada, E-18071 Granada, Spain

\end{center}
\vspace{.5cm}

\begin{abstract}
 
\noindent 
We study extensions of the Standard Model with general new vector bosons. The full Standard Model gauge symmetry is used to classify the extra vectors and constrain their couplings. We derive the corresponding effective Lagrangian, valid at energies lower than the mass of the extra vectors, and use it to extract limits from electroweak precision observables, including LEP~2 data. We consider both universal and nonuniversal couplings to fermions. We study the interplay of several extra vectors, which can have the effect of opening new regions in parameter space. In particular, it allows to explain the anomaly in the bottom forward-backward asymmetry with perturbative couplings. Finally, we analyze quantitatively the implications for the Higgs mass.

\end{abstract}

\renewcommand{\thefootnote}{\arabic{footnote}}
\setcounter{footnote}{0}


\section{Introduction}
\label{Intro}

Renormalizable extensions of the Standard Model (SM) are associated with new particles of spin 0, 1/2 or 1. Therefore, in a weakly-coupled scenario, these are the new particles that can be expected to give the biggest signals in precision experiments and at large colliders. In this paper we study, from a model independent point of view, the case of new particles of spin 1, i.e. new massive vector bosons. We focus mainly on their impact on electroweak precision data (EWPD), from which we extract the corresponding limits on their masses and couplings. We include LEP 2 cross sections and asymmetries in our fits, as they give important restrictions beyond the ones of $Z$-pole and low-energy observables.

New vector bosons are a common occurrence in theories beyond the SM. They appear whenever the gauge group of the SM is extended, as the gauge bosons of the extra (broken) symmetries. This is the case of Grand Unified Theories (GUT) \cite{GUTs}, including string constructions, or Little Higgs models~\cite{LH}. They also occur in theories in extra dimensions~\cite{XDs}, when the gauge bosons propagate through the bulk~\cite{Antoniadis:1999bq}. 
Strongly-interacting theories~\cite{Hill:2002ap}, such as technicolor~\cite{Weinberg:1975gm,Susskind:1978ms,Farhi:1980xs} or composite Higgs models~\cite{Kaplan:1983sm}, often give rise to spin 1 resonances~\cite{Casalbuoni:1985kq, Casalbuoni:1986vq,Casalbuoni:1987ud,Peskin:1990zt,Chivukula:2003wj,Barbieri:2008cc,Barbieri:2009tx,Hernandez:2010iu}. This can be related to the previous possibilities via hidden gauge symmetry~\cite{Bando:1984ej} or holography~\cite{Maldacena:1997re,ArkaniHamed:2000ds}.

It is possible to classify vector bosons according to their electric charge: neutral vector bosons, called $Z^\prime$, charge $\pm 1$ vector bosons, called $W^\prime$ and vectors with other integer or fractional charges. On the other hand, the complete theory including the new vectors must be invariant under the full $\SU{3}_c \otimes \SU{2}_L \otimes \U{1}_Y$ gauge group. This imposes additional restrictions on the allowed couplings to the SM fields, and also implies that certain vectors must appear simultaneously and have similar masses. In order to make use of this information in a model-independent approach, we classify in this paper the new vectors into irreducible representations of the full SM gauge group, and we impose the corresponding gauge invariance on the Lagrangian. We further restrict the possible couplings by the phenomenological requirement that the effects of the new vectors should be visible at available energies.

As a straightforward example of the implications of the complete SM gauge invariance, as opposed to simple conservation of electric charge, consider the case of a sequential $Z^\prime$ boson, with couplings proportional to the ones of the SM $Z$ boson. Such a neutral vector boson is often included in electroweak fits and direct searches. In fact, this vector has different couplings to the two components of the $SU(2)_L$ doublets, and it cannot be a singlet under the SM group. Nevertheless, it can arise after electroweak symmetry breaking as a mixture of a singlet vector and the third component of a vector in the adjoint of $SU(2)_L$. This is the case of models with a replica of the SM gauge group, or in extra dimensions. Thus, gauge invariance implies that the sequential $Z^\prime$ boson necessarily comes together with two charged vectors and another neutral vector, the $\gamma^\prime$. All these new fields have similar masses, with splittings of the order of the Higgs vev. Similarly, the results in this paper imply that a new charged vector boson with sizable couplings to both leptons and quarks must be accompanied by at least one neutral vector, with mass of the same order.

The extra vector bosons that have been most extensively studied are neutral singlets, usually associated to an extra abelian gauge symmetry (see, for instance, the review~\cite{Langacker:2008yv}). In this paper we give general model-independent\footnote{We also give limits on some popular models to illustrate how the analyses of particular models fit in our general framework.} limits on these $Z^\prime$s. We go far beyond this particular case, and study all the representations that could in principle give observable effects. We study the case of universal couplings to all families of quarks and leptons, and also cases with nonuniversal couplings. Furthermore, we consider a few examples with more than one type of vectors, which is the actual situation in most explicit models. We will show that the cooperation of several extra vectors allows, in some cases, to extend the allowed regions in parameter space.

In order to analyze the implications of the new vectors at energies below their mass, we first integrate them out and obtain the corresponding effective theory, including only operators of dimension six. This is useful to isolate the important effects, and also to exhibit more clearly which combinations of parameters are constrained by the data. The accuracy of this approximation is high for masses around the TeV. The effective Lagrangian is also helpful to study the interplay of
the extra vector bosons with other new particles of different spin, but we will not perform that sort of analysis in this paper.

It is well known that $Z^\prime$ bosons contribute with a positive sign to the $\rho$ parameter (or, equivalently, to the Peskin-Takeuchi $T$ parameter), and can be used to improve considerably the SM fit when the Higgs boson is heavy. We will see that the same role can be played by a hypercharged triplet. We analyze this effect quantitatively and show that, in extensions with these two kinds of vector bosons, a Higgs heavier than $\sim 300$~GeV is allowed by EWPD. On the other hand, it turns out that charged singlets give a negative contribution to $\rho$. This opens the door to cancellations of the different contributions to this parameter. Due to this interplay between the effects of new vector bosons and the Higgs loops in EWPD, in our fits we leave the mass of the Higgs as a free parameter, and we include the available information from direct Higgs searches.

The paper is organized as follows. In Section \ref{GEVB}, we classify the different types of extra vector bosons, and write their interactions with SM fields. In Section \ref{EDNVB}, we integrate the new vectors out, and obtain the coefficients of the dimension-six gauge-invariant operators in the effective Lagrangian. In Section \ref{Limitsonevector} we perform fits to EWPD and find the limits on each type of vector boson. In Section \ref{section:several} we discuss the effect on EWPD of including several types of extra vectors simultaneously. We give some phenomenologically interesting examples with nonuniversal couplings, including vector-boson explanations of the observed forward-backward asymmetry of the $b$ quark. Section \ref{section:Higgs} is devoted to an analysis of the implications for the Higgs mass. We present our conclusions in Section \ref{Conclusions_NewVectors}. Finally, three appendices contain the explicit operators and coefficients in the effective Lagrangian, plus the observables and other details of our fits.


\section{General extra vector bosons}
\label{GEVB}
We want to study general vector bosons beyond the ones in the SM, with the only restrictions that they be heavier than LEP~2 energies, have perturbative couplings, and be potentially observable by their indirect effects on precision data or as resonances in colliders. 
The leading effects in EWPD arise from tree-level exchanges of just one heavy vector boson contributing to processes with four fermions in the external legs. This requires interactions that couple SM operators to the extra vector fields and are linear in the latter. Clearly, the interactions should be renormalizable to avoid further suppressions. From the point of view of the low-energy effective theory to be discussed below, these couplings produce dimension-six operators, while interactions with more than one new vector field in the same operator---and nonrenormalizable interactions---give rise to operators of higher scaling dimension. Moreover, vectors with linear interactions can be singly produced, and have the best chances of being observed at colliders.

The requirement of linear renormalizable couplings, together with Lorentz symmetry and invariance under the full SM gauge group, constrain the possible quantum numbers of the new vectors. In Table~\ref{table:newvectors}, we give the quantum numbers for the 15 irreducible representations of vector fields that can have linear and renormalizable interactions. This table also introduces our notation for each class of vector boson, which is partly inspired by the usual notation for SM fields.
Note that the representations with nonvanishing hypercharge are complex.
\begin{table}[t]
\begin{center}
{\small
\begin{tabular}{ l  c  c  c  c  c  c  c  c} 
\ctoprule
\crowcolor\!\!Vector\!\!\!\!&${\cal B}_\mu$\!\!\!\!&${\cal B}_\mu^1$\!\!\!\!&${\cal W}_\mu$\!\!\!\!&${\cal W}_\mu^1$\!\!\!\!&${\cal G}_\mu$\!\!\!\!&${\cal G}_\mu^1$\!\!\!\!&${\cal H}_\mu$\!\!\!\!&${\cal L}_\mu$\!\!\!\!\\
\cmrule
\!\!Irrep\!\!\!\!&$\left(1,1\right)_0$\!\!\!\!&$\left(1,1\right)_1$\!\!\!\!&$\left(1,\Adj\right)_0$\!\!\!\!&$\left(1,\Adj\right)_1$\!\!\!\!&$\left(\Adj,1\right)_0$\!\!\!\!&$\left(\Adj,1\right)_1$\!\!\!\!&$\left(\Adj,\Adj\right)_{0}$\!\!\!\!&$\left(1,2\right)_{-\frac 32}$\!\!\!\!\\
\cbottomrule
&&&&&&&&\\
\ctoprule
\crowcolor\!\!Vector\!\!\!\!&${\cal U}_\mu^2$\!\!\!\!&${\cal U}_\mu^5$\!\!\!\!&${\cal Q}_\mu^1$\!\!\!\!&${\cal Q}_\mu^5$\!\!\!\!&${\cal X}_\mu$\!\!\!\!&${\cal Y}_\mu^1$\!\!\!\!&${\cal Y}_\mu^5$\!\!\!\!&\\
\cmrule
\!\!Irrep\!\!\!\!&$\left(3,1\right)_{\frac 23}$\!\!\!\!&$\left(3,1\right)_{\frac 53}$\!\!\!\!&$\left(3,2\right)_{\frac 16}$\!\!\!\!&$\left(3,2\right)_{-\frac 56}$\!\!\!\!&$\left(3,\Adj\right)_{\frac 23}$\!\!\!\!&$\left(\bar 6,2\right)_{\frac 16}$\!\!\!\!&$\left(\bar 6,2\right)_{-\frac 56}$\!\!\!\!&\\
\cbottomrule
\end{tabular}
}
\caption{Vector bosons contributing to the dimension-six effective Lagrangian. The quantum numbers $(R_c,R_L)_Y$ denote the representation $R_c$ under $SU(3)_c$, the representation $R_L$ under $SU(2)_L$ and the hypercharge $Y$ (normalized such that the electric charge is $Q=Y+T^3$).}
\label{table:newvectors}
\end{center}
\end{table}

For our purposes, it is not important whether the new vector bosons are the gauge bosons of a broken extended gauge group or not. Nevertheless, it is interesting to note that all the types of vector bosons in Table~\ref{table:newvectors} can in principle be obtained as the gauge bosons of an extended gauge group broken down to the SM. We give explicit examples of the corresponding symmetry breaking patterns in Table~\ref{table:GUT}. Models with bigger gauge groups usually incorporate new fermions, which in particular are necessary to cancel anomalies.
Here, we will assume that these exotic fermions, if they exist at all, do not contribute to EWPD. At any rate, in our general low-energy formulation, we only impose the SM gauge invariance, and the absence of anomalies does not impose any restriction on the couplings of the new vectors to the SM fermions.
\begin{table}[!h]
\begin{center}
{\small
\begin{tabular}{ c  c } 
\ctoprule
Vector& Model \\
$ $&$ $\\[-0.3cm]
\cmrule
$ $&$ $\\[-0.3cm]
${\cal B}_\mu$&$U(1)^\prime,~\mbox{Extra Dimensions}$\\
$ $&$ $\\[-0.3cm]
${\cal B}_\mu^1$&$SU(2)_R\otimes U(1)_X\rightarrow U(1)_Y$\\
$ $&$ $\\[-0.3cm]
${\cal W}_\mu$&$SU(2)_1\otimes SU(2)_2\rightarrow SU(2)_D\equiv SU(2)_L,~\mbox{Extra Dimensions}$\\
$ $&$ $\\[-0.3cm]
${\cal W}_\mu^1$&$SU(4)\rightarrow U(1)\otimes\left(SU(3)\rightarrow SU(2)\right)$\\
$ $&$ $\\[-0.3cm]
${\cal G}_\mu$&$SU(3)_1\otimes SU(3)_2\rightarrow SU(3)_D\equiv SU(3)_c,~\mbox{Extra Dimensions}$\\
$ $&$ $\\[-0.3cm]
${\cal G}_\mu^1$&\!\!$SO(12)\rightarrow(SO(8)\rightarrow SU(3))\otimes(SU(2)\otimes SU(2)\rightarrow SU(2)_D\rightarrow U(1)_Y)$\\
$ $&$ $\\[-0.3cm]
${\cal H}_\mu$&$SU(6)\rightarrow SU(3)\otimes SU(2)$\\
$ $&$ $\\[-0.3cm]
${\cal L}_\mu$&$G_2\rightarrow SU(2)\otimes (SU(2)\rightarrow U(1)_Y)$\\
$ $&$ $\\[-0.3cm]
${\cal U}_\mu^2, ~{\cal U}_\mu^5$&$SU(4)\rightarrow SU(3)\otimes U(1)$\\
$ $&$ $\\[-0.3cm]
${\cal Q}_\mu^1,~{\cal Q}_\mu^5$&$SU(5)\rightarrow SU(3)_c\otimes SU(2)_L\otimes U(1)_Y$\\
$ $&$ $\\[-0.3cm]
${\cal X}_\mu$&$SU(6)\rightarrow U(1)\otimes SU(3)\otimes(SU(3)\rightarrow SU(2))$\\
$ $&$ $\\[-0.3cm]
${\cal Y}_\mu^1,~{\cal Y}_\mu^5$&$F_4\rightarrow SU(3)\otimes(SU(3)\rightarrow SU(2)\otimes U(1))$\\[-0.2cm]
$ $&$ $\\
\cbottomrule
\end{tabular}
}
\caption{Examples of symmetry breaking patterns giving rise to each type of vectors bosons in Table~\ref{table:newvectors}~\cite{Slansky:1981yr}. Generating the right Weinberg angle and accommodating the matter fields requires, in some cases, an extension of the gauge groups in this table and a more involved pattern of symmetry breaking.}
\label{table:GUT}
\end{center}
\end{table}
As indicated in Table~\ref{table:GUT}, some of these vector bosons---an infinite number of each type, actually---also appear in extra-dimensional theories when the gauge bosons of the corresponding SM group factor propagate in the bulk. In fact, the pattern of symmetry breaking in these cases is essentially a generalization of the one shown in the table, as can be best understood by dimensional deconstruction~\cite{ArkaniHamed:2001ca}. Other kinds of vectors can also appear in this context as well, when the higher-dimensional gauge group is bigger.  We should also point out that the representations $\mathcal{U}^{2,5}$, $\mathcal{Q}^{1,5}$ and $\mathcal{X}$ correspond to the vector leptoquarks classified by Buchm\"uller, R\"uckl and Wyler in \cite{Buchmuller:1986zs}.

Once the field content of the theory has been established, we proceed to construct the most general renormalizable theory invariant under $SU(3)_c\otimes SU(2)_L \otimes U(1)_Y$. The Lagrangian has the form
\be
\mathcal{L}=\mathcal{L}_\SM + \mathcal{L}_V + \mathcal{L}_{V-\SM} + ~\mbox{nonlinear,} \label{Lagrangian}
\ee
where $\mathcal{L}_\SM$ is the SM Lagrangian, $\mathcal{L}_V$ contains the quadratic terms for the heavy vector bosons (with kinetic terms covariantized with respect to the SM group) and $\mathcal{L}_{V-\SM}$ contains the possible interaction or kinetic terms formed as products of SM fields and a single vector field. Mass mixing terms of SM and new vectors are forbidden by gauge invariance\footnote{There are, nevertheless, interactions with the Higgs doublet that give rise to mass mixing of the $Z$ and $W$ bosons with the new vectors when the electroweak symmetry is broken. This effect is in fact crucial for $Z$-pole observables.}.
``Nonlinear'' in Eq.~(\ref{Lagrangian}) refers to interaction terms that are nonlinear in the heavy vector fields. As we have argued above, those terms can be safely neglected. 

Before writing the different pieces in \refeq{Lagrangian}, let us introduce some notation. The gauge bosons of the SM are generically called $A$, i.e $A=B,W,G$. The new vectors are represented by the specific symbols in Table~\ref{table:newvectors} or generically by $V$. The calligraphic letter $\mathcal{A}$ denotes any of the three extra vectors in the same representation as the SM gauge bosons, namely $\mathcal{B}$, $\mathcal{W}$ and $\mathcal{G}$.
Covariant derivatives are always covariant with respect to the SM gauge group, and are defined as
\be
D_\mu X = \left(\partial_\mu+ig_sG_\mu^A\bold{T}_A+i g W_\mu^aT_a+ig'Y B_\mu\right) X ,
\ee
with $\mathbf{T}_A$ and $T_a$, respectively, the $SU(3)_c$ and $SU(2)_L$ generators in the representation in which the field or operator $X$ lives, and $Y=Q-T_3$ its hypercharge. 
We use matrix notation to write the singlet product of two objects in a given representation and its complex conjugate: in the product $A^\dagger B$, $A^\dagger$ and $B$ are row and column vectors, respectively, made out of the components of $A^\dagger$ and $B$ in some orthonormal basis of the vector space for their representation. For real representations the invariants can also be written as $A^\trps B$.
This notation is standard for the fundamental representations in the SM, but can be used for any representation. For the adjoints, we can use the usual basis given by the generators and write, for instance, $G^\trps_{\mu\nu} G^{\mu\nu} \equiv G^A_{\mu\nu}G^{A\,\mu\nu}$. Finally, $[.]_R$ denotes a projection into the irreducible representation $R$.
In the detailed formulas in the appendix, we shall be more explicit and use color and isospin indices.

With these definitions, the SM part of the Lagrangian reads
\be
\begin{split}
\mathcal{L}_{\SM}=&-\frac 14 G^\trps_{\mu\nu}G^{\mu\nu}-\frac 14 W^\trps_{\mu\nu}W^{\mu\nu}-\frac 14 B_{\mu\nu}B^{\mu\nu}+ \\
+&\overline{l_L^i} i \Dslash  l_L^i+\overline{q_L^i} i\Dslash  q_L^i+\overline{e_R^i} i\Dslash e_R^i+\overline{u_R^i} i\Dslash u_R^i+\overline{d_R^i} i\Dslash d_R^i+\\
+&\left(D_\mu \phi\right)^\dagger D^\mu \phi-V\left(\phi\right)-\\
-&\left(y^e_{(i)}\overline{l_L^i}\phi e^i_R+y^d_{(i)}\overline{q_L^i}\phi d^i_R+V^\dagger_{ij}y^u_{(j)} \overline{q_L^i} \tilde{\phi} u^j_R+\hc\right),
\label{SMLag}
\end{split}
\ee
where a sum is understood on repeated indices. As usual, we have defined $\tilde{\phi}=i\sigma_2 \phi^*$. The Higgs potential, $V(\phi)=-\mu_\phi^2\left|\phi\right|^2+\lambda_\phi \left|\phi\right|^2$, gives a vev to the Higgs field: $\la \phi \ra^\trps = 1/\sqrt{2}\,(0~v)$, $v\approx 246$~GeV. We have assumed a minimal Higgs sector, as we are not considering extra scalars in this paper\footnote{
Nevertheless, only the vev of the scalar is relevant for the new couplings that enter precision tests. 
Therefore, our equations are valid for a more general symmetry breaking sector with any number of scalar doublets and singlets. In this case, the ``Higgs'' field $\phi$ refers to the linear combination of doublets that acquires the vev. Similarly, our analysis applies to the usual nonlinear realization of electroweak breaking. However, we have assumed a single elementary scalar doublet in the SM loop corrections that enter our fits.}. The indices $i$ and $j$ in the fermion fields label the different families.
Without loss of generality, we have chosen a fermion basis such that $y^e$ and $y^d$ are diagonal,
with the diagonal elements specified by the corresponding index within parentheses. Then, the up Yukawa couplings write $y^u_{ij}=V^\dagger_{ij} y^u_{(j)}$, where $V$ is the CKM matrix.
  
The quadratic terms for the new vector bosons are given by\footnote{Note that the most general kinetic term is $D_\mu V_{\nu}^\dagger D^\mu V^{\nu}+\beta D_\mu V_{\nu}^\dagger D^\nu V^{\mu}$, with $\beta$ an arbitrary parameter. However, in this paper we restrict ourselves to spin-1 degrees of freedom. Then we must take $\beta=-1$, for otherwise $\partial_\mu V^\mu$ would propagate as an independent scalar field.\label{footn:betafn}}
\be
{\cal L}_V= - \sum_V \eta_V \left(\frac{1}{2} D_\mu V_{\nu}^\dagger D^\mu V^{\nu}-\frac{1}{2}D_\mu V_{\nu}^\dagger D^\nu V^{\mu}+\frac{1}{2}M_{V}^2 V_\mu^\dagger V^\mu \right),
\label{VLag}
\ee
The sum is over all new vectors $V$, which can be classified into the different irreducible representations of Table~\ref{table:newvectors}. We set $\eta_V=1 \;(2)$ when $V$ is in a real (complex) representation, in order to use the usual normalization. Even though the kinetic terms of the extra vectors incorporate SM covariant derivatives to keep manifest gauge invariance, the corresponding interactions among SM gauge bosons and two new vectors could be moved to the ``nonlinear'' terms of \refeq{Lagrangian}. On the other hand, we have written explicit mass terms for the new vectors. The masses can arise, in particular, from vacuum expectation values of extra scalar fields, but this is not necessary. In writing \refeq{VLag}, we have chosen a basis with diagonal, canonically normalized kinetic terms and diagonal masses. Mass terms often appear in nondiagonal form in explicit models. In these cases, it is necessary to diagonalize them before using our formulas. 
Finally, the couplings of the new vectors to the SM are given by
\be
\mathcal{L}_{V-\SM}= - \sum_V \frac{\eta_V}{2}\left({V^\mu}^\dagger J^V_\mu +\hc\right).
\ee
The vector currents $J^V_\mu$ have the form
\be
J^V_\mu = \sum_k g_V^k j^{Vk}_\mu,
\ee
where $g_V^k$is a coupling constant and $j^{Vk}_\mu$ is a vector operator of scaling dimension 3 in the same representation as $V$. Actually, the different currents that can be built with the SM fields determine the possible representations of the extra vectors. We can distinguish three kinds of SM currents:
\begin{itemize}
\item {\em With two fermions\/}. Schematically, $j^{V\psi_1\psi_2}_\mu = [\overline{\psi_1} \otimes \gamma_\mu \psi_2]_{R_V}$, with $\psi_1$, $\psi_2$ (different in principle) fermion multiplets, $R_V$ the representation of $V$ and $\otimes$ a product of representations.
\item {\em With two scalars and a covariant derivative\/}: $j^{V\phi}_\mu = [\Phi^\dagger \otimes D_\mu \phi]_{R_V}$, where $\Phi$ denotes either $\phi$ or $\tilde{\phi}$.
\item {\em With a gauge boson and two covariant derivatives\/}: $j^{\mathcal{A}}_\mu = D^\nu A_{\nu\mu}$.
\end{itemize}
The couplings to currents of the third type induce a kinetic mixing of the SM gauge bosons $A$ with the heavy vectors $\mathcal{A}$~\cite{delAguila:1988jz}. It turns out that the corresponding terms in $\mathcal{L}_{V-\SM}$ are redundant. In the case of only one extra vector multiplet, they can be eliminated by the field redefinition
\be
\begin{split}
A_\mu &\rightarrow A_\mu + g_\mathcal{A}^A \mathcal{A}_\mu,\\
{\cal A}_\mu&\rightarrow \left(1+ g_{\cal A}^{A~2} \right)^{-\frac 12}{\cal A}_\mu.
\end{split}
\ee
This redefines the mass $M_{\cal A}$ and currents $J^{\cal A}$ in the following way:
\be
\begin{split}
M_{\cal A}&\rightarrow \left(1+ g_{\cal A}^{A~2} \right)^{-\frac 12}M_{\cal A},\\
J^{\cal A}_\mu&\rightarrow \left(1+ g_{\cal A}^{A~2} \right)^{-\frac 12}\left(J^{\cal A}_\mu + g_\mathcal{A}^A J^A_\mu\right).
\end{split}
\label{redefinition}
\ee
In addition, new ``nonlinear'' terms are generated. In the following we shall work in the basis without kinetic $A$-$\mathcal{A}$ mixing (hence, with redefined couplings). This is a consistent choice that simplifies a lot many of the expressions below. At any rate, when kinetic mixing with just one extra vector is found in any particular model, it is possible to use our formulas as they stand, and simply perform the substitution in \refeq{redefinition} (or \refeq{App_replacement} in Appendix~\ref{app: NV_OpCoeff}) at the end.

We write explicitly all the possible currents in the tables of Appendix~\ref{app: NV_OpCoeff}.


\section{Effective description of new vector bosons}
\label{EDNVB}
Effective theories are a very convenient tool to analyze the indirect effects of new physics above the electroweak scale~\cite{Burgess:1993vc,Han:2004az}. In this section, we obtain at leading order an effective Lagrangian that describes a completely general extension of the SM with new vector bosons. This Lagrangian is written in terms of SM fields alone, and can be used at energies much smaller than the masses of the extra vectors. 

The effective Lagrangian is computed by integrating the heavy degrees of freedom out. The theory \refeq{Lagrangian} belongs to a decoupling scenario, in which the conditions for the Appelquist-Carazone theorem are satisfied~\cite{Appelquist:1974tg}. In this case the resulting nonlocal Lagrangian can be Taylor expanded as
\be
{\cal L}_{\mathrm{eff}}={\cal L}_4+\frac{1}{\Lambda}{\cal L}_5+\frac{1}{\Lambda^2}{\cal L}_6+\dots ,
\ee
where $\mathcal{L}_d$ contains gauge-invariant operators of dimension $d$ and $\Lambda$ is of the order of the smallest scale (other than the Higgs vev) in the complete theory. In our case $\Lambda$ corresponds to the mass of the heavy vector bosons. The operators in each ${\cal L}_d$ give observable contributions of order $\left(E/\Lambda\right)^n$, with $n\ge d-4$ and $E$ the typical energy of the process, which in some cases is $v$. In this manner, the effective Lagrangian provides a neat classification of the size of the different interactions. Moreover, the restrictions from the SM gauge invariance are manifest.

The list of operators up to dimension six, with the requirement of independent conservation of baryon ($B$) and lepton ($L$) number, was given long ago by Buchmuller and Wyler in~\cite{Buchmuller:1985jz,Arzt:1994gp}. We will use the notation in that reference. The original list contained one operator of dimension five and 81 of dimension six, but later some of them have been found to be redundant \cite{Grzadkowski:2003tf}, so the list is slightly shorter. The operators can be further classified depending on whether they arise at tree or at the loop level when a heavy particle is integrated out \cite{Arzt:1994gp}. For our purposes it is sufficient to treat the new physics at tree level, so our Lagrangian will contain only operators of the tree-level kind. On the other hand, we relax the assumption of $B$ and $L$ number conservation. Finally, we remark that flavor effects can be easily taken into account by treating the operators as matrices in flavor space~\cite{delAguila:2000aa}. We will write explicit flavor indices in our general formulas, which are thus completely general in this sense.  All the independent dimension-six operators that can be induced by extra vector bosons are collected in Appendix~\ref{app: Operators}.

Our task in this section is to perform the integration explicitly, starting with the theory~\refeq{Lagrangian}, write the result in the operator basis of Appendix~\ref{app: Operators}, and obtain the coefficients of each operator in the case that the new particles are vector bosons. These coefficients will depend on the masses $M_V$ and the couplings $g_V^k$. The integration of general extra leptons and general extra quarks has been performed in~\cite{delAguila:2008pw} and~\cite{delAguila:2000rc}, respectively.

\begin{figure}[thb]
\begin{center}
\begin{picture}(150,100)(-5,-50)
\Text(-43,35)[!]{$\psi_2$}
\ArrowLine(-35,35)(0,1)
\ArrowLine(0,-1)(-35,-35)
\Text(-43,-35)[!]{$\overline{\psi_1}$}
\Vertex(0,0){2}
\Text(25,15)[!]{$V^\mu$}
\Photon(50,1)(0,1){5}{4}
\Photon(50,-1)(0,-1){5}{4}
\Vertex(50,0){2}
\Text(93,35)[!]{$\psi_4$}
\ArrowLine(50,1)(85,35)
\ArrowLine(85,-35)(50,-1)
\Text(93,-35)[!]{$\overline{\psi_3}$}
\Text(25,-60)[!]{$(a)$}
\end{picture}
%
%
\begin{picture}(150,100)(-85,-50)
\Text(-25,15)[!]{$W_\mu^a,B_\mu$}
\Photon(-50,0)(0,0){5}{4}
\Text(9,35)[!]{$\Phi^\dagger$}
\DashLine(0,35)(0,1){3}
\DashLine(0,-35)(0,-1){3}
\Vertex(0,0){2}
\Text(7,-35)[!]{$\phi$}
\Text(25,15)[!]{$V^\mu$}
\Photon(0,1)(50,1){5}{4}
\Photon(0,-1)(50,-1){5}{4}
\Vertex(50,0){2}
\Text(59,35)[!]{$\Phi^\dagger$}
\DashLine(50,35)(50,1){3}
\DashLine(50,-1)(50,-35){3}
\Text(57,-35)[!]{$\phi$}
\Photon(50,0)(100,0){5}{4}
\Text(75,15)[!]{$W_\mu^a,B_\mu$}
\Text(25,-60)[!]{$(b)$}
\end{picture}
\begin{picture}(150,110)(-50,-45)
\Text(-43,35)[!]{$\psi_2$}
\ArrowLine(-35,35)(0,1)
\ArrowLine(0,-1)(-35,-35)
\Text(-43,-35)[!]{$\overline{\psi_1}$}
\Vertex(0,0){2}
\Text(25,15)[!]{$V^\mu$}
\Photon(0,1)(50,1){5}{4}
\Photon(0,-1)(50,-1){5}{4}
\Vertex(50,0){2}
\Text(59,35)[!]{$\Phi^\dagger$}
\DashLine(50,35)(50,1){3}
\DashLine(50,-1)(50,-35){3}
\Text(57,-35)[!]{$\phi$}
\Photon(50,0)(100,0){5}{4}
\Text(75,15)[!]{$W_\mu^a,B_\mu$}
\Text(25,-60)[!]{$(c)$}
\end{picture}
\end{center}
\caption{Feynman diagrams relevant for the dimension-six effective Lagrangian. \label{f_Feynmandiagrams}}
\end{figure}
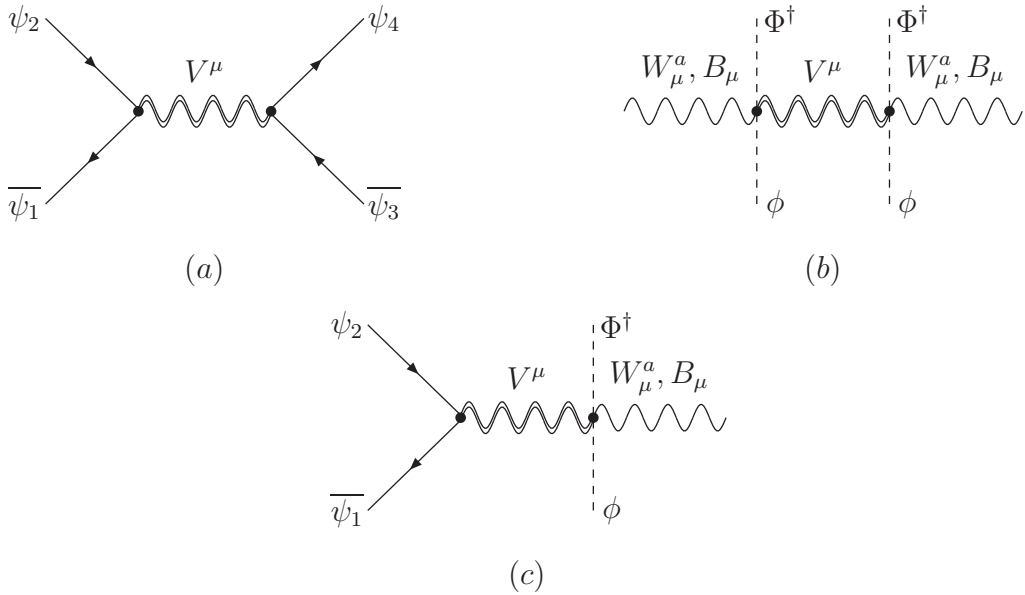

At the classical level, the integration can be carried out by computing the tree-level Feynman diagrams in Fig.~\ref{f_Feynmandiagrams} and matching to the corresponding amplitudes in the effective theory. 
Equivalently, we can solve the classical equations of motion for the heavy vectors and substitute the solutions into the Lagrangian. Proceeding in this algebraic manner, we readily find from $\mathcal{L}_V+\mathcal{L}_{V-\SM}$ the on-shell vector fields
\be
V_\mu =  \frac{1}{p^2-M_V^2} \left[\frac{p_\mu p_\nu}{M_V^2}-\eta_{\mu\nu} \right]\left( J^{V\,\nu} + O(V^\nu)\right).\label{ExactSol}
\ee
The $O(V_\mu)$ terms arise from the ``nonlinear'' terms in $\mathcal{L}_{V-\SM}$. The next step is to expand this equation in powers of $p^2/M_V^2$ and solve for $V_\mu$. At the leading order we simply have
\be
V_\mu = \frac{1}{M_V^2} J^V_\mu +O\left(\frac{1}{M_V^4}\right)  \label{onshellvector},
\ee
where the order $1/M_V^4$ terms follow from both the $O(V_\mu)$ terms in (\ref{ExactSol})  and the higher-order terms in the inverse propagator expansion. Then, we substitute \refeq{onshellvector} into the Lagrangian $\mathcal{L}$ and find
\be
\mathcal{L}_\mathrm{eff}= \mathcal{L}_\mathrm{SM} - \frac{\eta_V}{2M_V^2} \,
(J_\mu^V)^\dagger J^{V\,\mu} + O\left(\frac{1}{M_V^4}\right) 
\label{oureffective}
\ee
The terms of order $1/M_V^4$ contribute to operators of dimension eight and higher, and will be neglected in the following. In particular, we see that, as promised, the ``nonlinear'' terms in $\mathcal{L}_{V-\SM}$ do not contribute to the effective Lagrangian up to dimension six, and can be ignored. The result \refeq{oureffective} includes a few operators that are not in the list of Appendix~\ref{app: Operators}. In order to compare with previous work, it is convenient to express the result in our basis, performing some Fierz reorderings and field redefinitions. The final result can then be written as
\be
\mathcal{L}_6^V=  - \frac{\eta_V}{2M_V^2} \,
(J_\mu^V)^\dagger J^{V\,\mu} = \sum_i \frac{\alpha_i}{M_V^2} \mathcal{O}_i ,
\ee
where $\mathcal{O}_i$ are the operators collected in Appendix~\ref{app: Operators}, and $\alpha_i$ are dimensionless numerical coefficients.
It is clear from the general expression \refeq{oureffective}, and also from the Feynman diagrams in Fig.~\ref{f_Feynmandiagrams}, that the terms in the effective Lagrangian can be of three basic forms:
\begin{itemize}
\item[$a$)] {\em Four fermions\/}:~~ $\frac{g_V^{\psi_1\psi_2} g_V^{\psi_3\psi_4}}{M_V^2} [\overline{\psi_1}\otimes \gamma_\mu \psi_2]_{R_V} [\overline{\psi_3}\otimes \gamma^\mu \psi_4]_{R_V}$.
\item[$b$)] {\em Oblique\/}:~~ $\frac{g_V^\phi g_V^\phi}{M_V^2} [\Phi^{\dagger} \otimes D_\mu \phi]_{R_V} [ D^\mu \phi^\dagger \otimes \Phi]_{R_V}$.
\item[$c$)] {\em Scalars, vectors and fermions\/}:~~ $\frac{ g_V^\phi g_V^{\psi_1\psi_2}}{M_V^2}  [\Phi^{\dagger} \otimes D^\mu \phi]_{R_V}[\overline{\psi_1}\otimes \gamma_\mu \psi_2]_{R_V}$.
\end{itemize}
In addition there are operators that arise from the field redefinitions, which just redefine the fermion masses and Yukawa interactions, and the Higgs potential. The four-fermion terms are relevant for LEP~2 and low-energy observables. Upon electroweak symmetry breaking, the oblique terms modify the gauge boson propagators\footnote{The only observable oblique operator in our approximation is ${\cal O}^{(3)}_\phi$. Its coefficient is proportional to the $\rho$ parameter. The effect of the operator ${\cal O}^{(1)}_\phi$ is absorbed into the input parameter $M_Z$. On the other hand, removing the kinetic mixing between SM and new vectors, as discussed in the previous section, prevents the operator ${\cal O}_{\WB}$, related to the Peskin-Takeuchi $S$ parameter~\cite{Peskin:1991sw}, from being generated at tree level. The same formulas would follow had we left the kinetic mixing terms and performed a perturbative field redefinition in the effective Lagrangian to eliminate ${\cal O}_{\WB}$. Let us also observe that the field basis we are using is different from the ``oblique'' basis introduced by Barbieri et al.\ in~\cite{Barbieri:2004qk} for oblique new physics. Thus, in particular, we cannot identify directly ${\cal O}^{(3)}_\phi$ with the $\hat{T}$ parameter of~\cite{Barbieri:2004qk}. The relation between these two operator bases, for arbitrary new physics, has been given in~\cite{Cacciapaglia:2006pk}. The predictions for physical observables are, of course, basis independent.} and those of the third type change the fermion-gauge trilinear couplings. Hence, the last two kinds of operators contribute mainly to observables at the $Z$ pole (and the $W$ mass, for the oblique operators). On the other hand, note that the coefficients of all the operators are given by the sum of the contributions of the different vector bosons, and the contribution of each vector is the product of two of its couplings divided by its mass squared.

The explicit result of the integration of the new vector bosons is given in Appendix~\ref{app: NV_OpCoeff}. In Tables \ref{Table: B0Table} to \ref{Table: Y5Table} of that appendix, we collect the contributions of each kind of extra vector boson to the coefficients  $\alpha_i$ of the dimension-six operators ${\cal O}_i$.

\section{Limits on new vector bosons}
\label{Limitsonevector}

We can now use the effective Lagrangian to perform fits to EWPD, and extract limits on the new vector bosons in Table~\ref{table:newvectors}. To start with, we assume that only one new vector gives sizable contributions. We will see in Section~\ref{section:several} that such ``one-at-a-time'' analysis is often justified (but not always).
\begin{table}[t]
\begin{center}
{\small
\begin{tabular}{ c | c  c  c  c  c  c  c  c }
\ctoprule
$\mbox{Vector}$&$Z$ pole&$M_W$&CKM&$\nu N$&$\nu e$&$\mbox{APV}$&M\o ller&$\mbox{LEP 2}$\\
\cmrule
\crowcolor$ $&$ $&$ $&$ $&$ $&$ $&$ $&$ $&$ $\\[-0.3cm]
\crowcolor${\cal B}_\mu $&$\checkmark $&$\checkmark $&$\checkmark$&$\checkmark $&$\checkmark $&$\checkmark $&$\checkmark $&$\checkmark $\\
$ $&$ $&$ $&$ $&$ $&$ $&$ $&$ $&$ $\\[-0.3cm]
${\cal W}_\mu $&$\checkmark  $&$\checkmark $&$\checkmark$&$\checkmark  $&$\checkmark  $&$\checkmark  $&$\checkmark  $&$\checkmark  $\\
\crowcolor$ $&$ $&$ $&$ $&$ $&$ $&$ $&$ $&$ $\\[-0.3cm]
\crowcolor${\cal G}_\mu $&$ $&$ $&$ $&$ $&$ $&$ $&$ $&$ $\\
$ $&$ $&$ $&$ $&$ $&$ $&$ $&$ $&$ $\\[-0.3cm]
${\cal H}_\mu $&$ $&$ $&$ $&$ $&$ $&$ $&$ $&$ $\\
\crowcolor$ $&$ $&$ $&$ $&$ $&$ $&$ $&$ $&$ $\\[-0.3cm]
\crowcolor${\cal B}_\mu^1 $&$\checkmark $&$\checkmark $&$\checkmark$&$\checkmark  $&$\checkmark $&$\checkmark $&$\checkmark $&$\checkmark $\\
$ $&$ $&$ $&$ $&$ $&$ $&$ $&$ $&$ $\\[-0.3cm]
${\cal W}_\mu^1 $&$\checkmark $&$\checkmark  $&$ $&$\checkmark $&$\checkmark $&$\checkmark $&$\checkmark $&$\checkmark$\\
\crowcolor$ $&$ $&$ $&$ $&$ $&$ $&$ $&$ $&$ $\\[-0.3cm]
\crowcolor${\cal G}_\mu^1 $&$ $&$ $&$ $&$ $&$ $&$ $&$ $&$ $\\
$ $&$ $&$ $&$ $&$ $&$ $&$ $&$ $&$ $\\[-0.3cm]
${\cal L}_\mu $&$ $&$ $&$ $&$  $&$\checkmark $&$  $&$  $&$\checkmark $\\
\crowcolor$ $&$ $&$ $&$ $&$ $&$ $&$ $&$ $&$ $\\[-0.3cm]
\crowcolor${\cal U}_\mu^2 $&$ $&$ $&$\checkmark  $&$\checkmark  $&$  $&$\checkmark $&$  $&$\checkmark $\\
$ $&$ $&$ $&$ $&$ $&$ $&$ $&$ $&$ $\\[-0.3cm]
${\cal U}_\mu^5 $&$ $&$ $&$ $&$ $&$ $&$\checkmark  $&$ $&$\checkmark  $\\
\crowcolor$ $&$ $&$ $&$ $&$ $&$ $&$ $&$ $&$ $\\[-0.3cm]
\crowcolor${\cal Q}_\mu^1 $&$ $&$ $&$ $&$\checkmark $&$ $&$\checkmark  $&$ $&$\checkmark  $\\
$ $&$ $&$ $&$ $&$ $&$ $&$ $&$ $&$ $\\[-0.3cm]
${\cal Q}_\mu^5 $&$ $&$ $&$ $&$\checkmark $&$ $&$\checkmark  $&$ $&$\checkmark  $\\
\crowcolor$ $&$ $&$ $&$ $&$ $&$ $&$ $&$ $&$ $\\[-0.3cm]
\crowcolor${\cal X}_\mu $&$ $&$ $&$\checkmark $&$\checkmark  $&$ $&$\checkmark  $&$ $&$\checkmark  $\\
$ $&$ $&$ $&$ $&$ $&$ $&$ $&$ $&$ $\\[-0.3cm]
${\cal Y}_\mu^1 $&$ $&$ $&$ $&$ $&$ $&$ $&$ $&$ $\\
\crowcolor$ $&$ $&$ $&$ $&$ $&$ $&$ $&$ $&$ $\\[-0.3cm]
\crowcolor${\cal Y}_\mu^5 $&$ $&$ $&$ $&$ $&$ $&$ $&$ $&$ $\\
\cbottomrule
\end{tabular}
}
\caption{Experimental data constraining (directly or indirectly) the couplings of the vector bosons.}
\label{table: ExpVConstr}
\end{center}
\end{table}

The observables and free parameters that enter the fits and the details of our fitting procedure are described in Appendix~\ref{app: fits}. In particular, LEP~2 cross sections and asymmetries are important to lift some flat directions that would remain in the fits to the other observables. In Table~\ref{table: ExpVConstr}, we summarize which sets of data can constrain each kind of vector boson. We see that five types of vectors, ${\cal G}$, ${\cal H}$, ${\cal G}^1$, ${\cal Y}^1$ and ${\cal Y}^5$, are invisible to all the precision observables, as they couple to quarks only. These vectors could in principle be produced at hadron colliders, and the non-observation of the corresponding resonances at Tevatron places limits on their masses. In the following we focus on EWPD, so we restrict our attention to the cases that can modify these data. 

All parameters are assumed to be real, as is the case in known models. On the other hand, for general coupling matrices, flavor-changing neutral currents (FCNC) are induced (except for $\mathcal{W}^1$, which does not couple to fermions). These give, generally, bounds much stronger than the ones derived from EWPD. Avoiding these bounds requires fine tuning, or some mechanism that imposes a certain structure on the coupling matrices. 

The hypothesis of diagonal and universal couplings is sufficient to avoid all FCNC for the vector fields that connect each fermion multiplet with its adjoint, i.e. $\mathcal{B}$ and $\mathcal{W}$. For the other types of vectors with couplings to fermions, universality does not guarantee the absence of FCNC. Another 
possibility is that the new vectors couple to just one family of fermions, in the fermion basis with maximally diagonal Yukawas (before electroweak breaking). 
In this case, there are still FCNC if the vector leptoquarks couple to left-handed (LH) quarks
(in the up sector, for our choice of $q_L$ basis), but they are suppressed by CKM off-diagonal entries. In particular, the FCNC are under control if the vectors couple to the third family of quarks only, as in the examples of Section~\ref{section_non_universal}. This particular structure of couplings is fine-tuned, since it breaks the $U(3)^5$ flavor symmetry of the SM, which allowed us to choose freely the fermion basis. Nevertheless, it can be explained by some mechanism in the complete theory. For instance, warped extra dimensions with bulk fermions incorporate a GIM-like mechanism in a natural manner~\cite{Agashe:2004cp}.

In the fits of this section, we assume diagonal, universal couplings for $\mathcal{B}$, $\mathcal{W}$ and $\mathcal{B}^1$~\footnote{Note that the addition of ${\cal B}$ and ${\cal B}^1$ can result in the ${\cal W}$ case in the appropriate limit. In particular, for vanishing fermionic couplings and related ${\cal B}$ and ${\cal B}^1$ masses and couplings there is a flat direction along the Higgs coupling with no EWPD constraints at this order (see below). In this case, the main constraints will come about from the non-observation at the Large Hadron Collider~\cite{Barbieri:2009tx,Hernandez:2010iu}.}. For the representations in which this flavor structure would generate dangerous FCNC, we assume couplings to just one family, as explained above, and explore all possibilities. The basic results are given in Table~\ref{table_OneVLimits}. The fits depend on the quadratic products of the different ratios $G^k_V\equiv g_V^k /M_V$. Therefore, relative signs among the different couplings are relevant, but the results are invariant under a global change of sign. 
The limits\footnote{Our one-dimensional (two-dimensional) 95\% confidence limits (regions) are defined by requiring a change of 3.84 (5.99) in $\chi^2$ with respect to the minimal value.} on the parameters in Table~\ref{table_OneVLimits}, unlike the best values, do not depend on the signs of these ratios, as the other parameters are integrated. In the rest of this section, we discuss these results and give additional details. 
Nonuniversal couplings for $\mathcal{B}$, $\mathcal{B}^1$ and $\mathcal{W}$ are studied in Section \ref{section_non_universal}.
\begin{table}[p]
\begin{center}
{\footnotesize
\begin{tabular}{c c c c c c}
\ctoprule
Vector&~$-\Delta \chi^2_\mathrm{min}$~&~~Parameter~~&~~Best Fit~~&~~Bounds~~&\!~C.L.~\\
$V_\mu$&$(\chi^2_\mt{min}/\mt{d.o.f.})$&$G_V^k\equiv g_V^k/M_V$& [TeV$^{-1}$]&[TeV$^{-1}$]&\\
\cmrule
\crowcolor&&&&&\\[-0.4cm]
\crowcolor${\cal B}_\mu $&$7.35$&$G^\phi_{{\cal B}_{\phantom{G}}}$&$-0.045$&$\left[-0.098,\phantom{+}0.098\right]$&95$\%$\\
\crowcolor                           &(0.77)&$G^l_{{\cal B}_{\phantom{G}}}$&$\phantom{+}0.021$&$\left[-0.210,\phantom{+}0.210\right]$&95$\%$\\
\crowcolor                           &&$G^q_{{\cal B}_{\phantom{G}}}$&$-0.89\phantom{0}$&-&-\\
\crowcolor                           &&$G^e_{{\cal B}_{\phantom{G}}}$&$\phantom{+}0.048$&$\left[-0.300,\phantom{+}0.300\right]$&95$\%$\\
\crowcolor                           &&$G^u_{{\cal B}_{\phantom{G}}}$&$-2.6\phantom{00}$&-&-\\
\crowcolor                           &&$G^d_{{\cal B}_{\phantom{G}}}$&$-6.0\phantom{00}$&-&-\\
&&&&&\\[-0.4cm]
${\cal W}_\mu $&$1.51$&$~G_{{\cal W}_{\phantom{G}}}^\phi$&$\phantom{+}0.002$&$\left[-0.12,\phantom{+}0.12\right]$&$1~\sigma$\\
                            &(0.79)&$~G_{{\cal W}_{\phantom{G}}}^l$&$\phantom{+}0.004$&$\left[-0.26,\phantom{+}0.26\right]$&$95\%$\\
                            &&$~G_{{\cal W}_{\phantom{G}}}^q$&$-9.6\phantom{00}$&-&-\\
&&&&&\\[-0.4cm]
\crowcolor&&&&&\\[-0.4cm]
\crowcolor${\cal B}^{1}_\mu $&$0.16$&$G_{{\cal B}^1_{\phantom{G}}}^\phi$&$\phantom{~+}6\cdot\! 10^{-4}$&$\left[-0.11,\phantom{+}0.11\right]$&$95\%$\\
\crowcolor                                   &(0.79)&$G_{{\cal B}^1_{\phantom{G}}}^{du}$&\phantom{+}6.6\phantom{00}&-&-\\
&&&&&\\[-0.4cm]
${\cal W}^{1}_\mu $&$0.65$&$|G_{{\cal W}^1}^\phi|$&$\phantom{+}0.18$&$<0.50$&$95\%$\\
&(0.78)&&&&\\
&&&&&\\[-0.4cm]
\crowcolor&&&&&\\[-0.4cm]
\crowcolor${\cal L}_\mu $&$\begin{array}{c}\phantom{0}\\0\\(0.79)\end{array}$&$|G_{\cal L}^{el}|$&$\phantom{+}0\phantom{.00}$&$<\left(\begin{array}{c c c}0.29&0.33&0.39\\0.34&\mbox{-}&\mbox{-}\\0.39&\mbox{-}&\mbox{-}\end{array}\right)$&$95\%$\\
&&&&&\\[-0.4cm]
${\cal U}^{2}_\mu $&$\begin{array}{c}\phantom{0}\\0\\(0.79)\end{array}$&$|G_{{\cal U}^2}^{ed}|$&$\phantom{+}0\phantom{.00}$&$<\left(\begin{array}{c c c}0.21&0.49 &0.49\\\mbox{-}&\mbox{-}&\mbox{-}\\\mbox{-}&\mbox{-}&\mbox{-}\end{array}\right)$&$95\%$\\
$ $&$ $&$|G_{{\cal U}^2}^{lq}|$&$\phantom{+}0\phantom{.00}$&$<\left(\begin{array}{c c c}0.12&0.29 &0.29\\ 0.56&0.65&\mbox{-}\\\mbox{-}&\mbox{-}&\mbox{-}\end{array}\right)$&$95\%$\\
\crowcolor&&&&&\\[-0.4cm]
\crowcolor${\cal U}^{5}_\mu $&$\begin{array}{c}\phantom{0}\\\le2.77\\(0.77)\end{array}$&$|G_{{\cal U}^5}^{eu}|$&$\begin{array}{c} \\ \phantom{+}0.43\\\phantom{+}\left[1,2\right]\end{array}$&$<\left(\begin{array}{c c c}0.25&0.62&\mbox{-}\\\mbox{-}&\mbox{-}&\mbox{-}\\\mbox{-}&\mbox{-}&\mbox{-}\end{array}\right)$&$95\%$\\
&&&&&\\[-0.4cm]
${\cal Q}^{1}_\mu $&$\begin{array}{c}\phantom{0}\\\le0.45\\(0.79)\end{array}$&$|G_{{\cal Q}^1}^{ul}|$&$\begin{array}{c} \\ \phantom{+}0.27\\\phantom{+}\left[1,2\right]\end{array}$&$<\left(\begin{array}{c c c}0.22&0.54&\mbox{-}\\0.57 &\mbox{-}&\mbox{-}\\\mbox{-}&\mbox{-}&\mbox{-}\end{array}\right)$&$95\%$\\
\crowcolor&&&&&\\[-0.4cm]
\crowcolor${\cal Q}^{5}_\mu $&$\begin{array}{c}\phantom{0}\\\le3.36\\(0.78)\end{array}$&$|G_{{\cal Q}^5}^{dl}|$&$\begin{array}{c} \\ \phantom{+}0.87\\\phantom{+}\left[1,1\right]\end{array}$&$<\left(\begin{array}{c c c}1.06&0.58 &\mbox{-} \\1.07 &\mbox{-} &\mbox{-} \\1.07 &\mbox{-} &\mbox{-} \end{array}\right)$&$95\%$\\
\crowcolor                                   &&$|G_{{\cal Q}^5}^{eq}|$&$\begin{array}{c} \\ \phantom{+}0.64\\\phantom{+}\left[1,1\right]\end{array}$&$<\left(\begin{array}{c c c}0.78&1.0\phantom{0}&1.2\phantom{0} \\\mbox{-}&\mbox{-}&\mbox{-}\\\mbox{-}&\mbox{-}&\mbox{-}\end{array}\right)$&$95\%$\\
&&&&&\\[-0.4cm]
${\cal X}_\mu $&$\begin{array}{c}\phantom{0}\\\le2.86\\(0.77)\end{array}$&$|G_{\cal X}^{lq}|$&$\begin{array}{c} \\ \phantom{+}0.65\\\phantom{+}\left[1,2\right]\end{array}$&$<\left(\begin{array}{c c c}0.27&0.93 &0.57 \\1.04 &1.40&\mbox{-}\\\mbox{-}&\mbox{-}&\mbox{-}\end{array}\right)$&$95\%$\\
\cbottomrule
\end{tabular}
\caption{ Results of the fit to EWPD for the extra vector bosons. We give $\Delta \chi^2_\mathrm{min}=\chi^2_\mathrm{min}-\chi^2_\SM$ values, together with the best fit values and bounds on the interactions of the new vectors. The results for the last six representations are obtained from a fit to each of the entries of the coupling matrices at a time. $[i,j]$ refers to the entries in the family matrices that give the best fit. See text for more details. 
\label{table_OneVLimits}}
}
\end{center}
\end{table}


\subsection{Neutral singlet $\mathcal{B}$}

Extra neutral vector bosons, known as $Z^\prime$ bosons, have been extensively studied in the past (see~\cite{Langacker:2008yv} for a review). In our gauge-invariant formalism, the only way neutral vectors can arise alone, without charged partners, is from the SM vector singlets $\mathcal{B}$. Electroweak symmetry breaking can then mix these fields with the SM $Z$ boson, with a mixing angle proportional to the coupling of the $\mathcal{B}$ to the Higgs doublet. We reserve the name $Z^\prime$ to denote the corresponding heavy mass eigenstates. The physical $Z'$ mass and the $Z$-$Z^\prime$ mixing, $\sin{\theta_{ZZ^\prime}}$, are related to the mass parameter $M_\mathcal{B}$ and the Higgs coupling $g_{\cal B}^\phi$ by
\be
M_{Z^\prime}^2 \approx M_\mathcal{B}^2 \left[1 + (g_{\cal B}^\phi)^2\frac{v^2}{M_{\cal B}^2}\right]~~~\mbox{and}~~~\sin{\theta_{ZZ^\prime}}\approx \frac{g_{\cal B}^\phi\sqrt{g^2+g'^2}}{2}\frac{v^2}{M_{\cal B}^2},
\ee
where we are assuming $M_\mathcal{B}\gg gv, g_\mathcal{B}^\phi v$.
These singlets appear in many extensions of the SM. They are usually associated with an extra abelian factor in the gauge group, which is broken down at a scale higher than electroweak (but hopefully small enough to allow for their eventual observation). This is the case of GUT/string and Little Higgs models, when the rank of the gauge group is higher than 4, and of theories with gauge fields in extra dimensions. 

In our model-independent analysis, with the assumption of universality, the ${\cal B}$ scenario has six new free parameters: the couplings to each matter multiplet divided by the mass of the $\mathcal{B}$. The result of the fit, displayed in Table~\ref{table_OneVLimits}, can be understood as follows. First, the $W$ mass and the $Z$-pole data constrain the (effective) Higgs coupling to be small. The direct Tevatron Higgs limits also contribute in the same direction, as discussed in Section~\ref{section:Higgs}. 

\begin{figure}[th]
\begingroup
  \makeatletter
  \providecommand\color[2][]{%
    \GenericError{(gnuplot) \space\space\space\@spaces}{%
      Package color not loaded in conjunction with
      terminal option `colourtext'%
    }{See the gnuplot documentation for explanation.%
    }{Either use 'blacktext' in gnuplot or load the package
      color.sty in LaTeX.}%
    \renewcommand\color[2][]{}%
  }%
  \providecommand\includegraphics[2][]{%
    \GenericError{(gnuplot) \space\space\space\@spaces}{%
      Package graphicx or graphics not loaded%
    }{See the gnuplot documentation for explanation.%
    }{The gnuplot epslatex terminal needs graphicx.sty or graphics.sty.}%
    \renewcommand\includegraphics[2][]{}%
  }%
  \providecommand\rotatebox[2]{#2}%
  \@ifundefined{ifGPcolor}{%
    \newif\ifGPcolor
    \GPcolortrue
  }{}%
  \@ifundefined{ifGPblacktext}{%
    \newif\ifGPblacktext
    \GPblacktexttrue
  }{}%
  \let\gplgaddtomacro\g@addto@macro
  \gdef\gplbacktexta{}%
  \gdef\gplfronttexta{}%
  \gdef\gplbacktextb{}%
  \gdef\gplfronttextb{}%
  \makeatother
  \ifGPblacktext
    \def\colorrgb#1{}%
    \def\colorgray#1{}%
  \else
    \ifGPcolor
      \def\colorrgb#1{\color[rgb]{#1}}%
      \def\colorgray#1{\color[gray]{#1}}%
      \expandafter\def\csname LTw\endcsname{\color{white}}%
      \expandafter\def\csname LTb\endcsname{\color{black}}%
      \expandafter\def\csname LTa\endcsname{\color{black}}%
      \expandafter\def\csname LT0\endcsname{\color[rgb]{1,0,0}}%
      \expandafter\def\csname LT1\endcsname{\color[rgb]{0,1,0}}%
      \expandafter\def\csname LT2\endcsname{\color[rgb]{0,0,1}}%
      \expandafter\def\csname LT3\endcsname{\color[rgb]{1,0,1}}%
      \expandafter\def\csname LT4\endcsname{\color[rgb]{0,1,1}}%
      \expandafter\def\csname LT5\endcsname{\color[rgb]{1,1,0}}%
      \expandafter\def\csname LT6\endcsname{\color[rgb]{0,0,0}}%
      \expandafter\def\csname LT7\endcsname{\color[rgb]{1,0.3,0}}%
      \expandafter\def\csname LT8\endcsname{\color[rgb]{0.5,0.5,0.5}}%
    \else
      \def\colorrgb#1{\color{black}}%
      \def\colorgray#1{\color[gray]{#1}}%
      \expandafter\def\csname LTw\endcsname{\color{white}}%
      \expandafter\def\csname LTb\endcsname{\color{black}}%
      \expandafter\def\csname LTa\endcsname{\color{black}}%
      \expandafter\def\csname LT0\endcsname{\color{black}}%
      \expandafter\def\csname LT1\endcsname{\color{black}}%
      \expandafter\def\csname LT2\endcsname{\color{black}}%
      \expandafter\def\csname LT3\endcsname{\color{black}}%
      \expandafter\def\csname LT4\endcsname{\color{black}}%
      \expandafter\def\csname LT5\endcsname{\color{black}}%
      \expandafter\def\csname LT6\endcsname{\color{black}}%
      \expandafter\def\csname LT7\endcsname{\color{black}}%
      \expandafter\def\csname LT8\endcsname{\color{black}}%
    \fi
  \fi
  \begin{tabular}{c c}
  \setlength{\unitlength}{0.03875bp}%
  \begin{picture}(7200.00,5040.00)(600,0)%
    \gplgaddtomacro\gplbacktexta{%
    }%
    \gplgaddtomacro\gplfronttexta{%
      \csname LTb\endcsname%
      \put(1656,800){\makebox(0,0){\footnotesize \strut{}-1}}%
      \put(2628,800){\makebox(0,0){\footnotesize \strut{}-0.5}}%
      \put(3600,800){\makebox(0,0){\footnotesize \strut{} 0}}%
      \put(4572,800){\makebox(0,0){\footnotesize \strut{} 0.5}}%
      \put(5544,800){\makebox(0,0){\footnotesize \strut{} 1}}%
      \put(3600,470){\makebox(0,0){\strut{}$\smallmath{G_{\cal B}^l}$~{\small [TeV$\smallmath{^{-1}}$]}}}%
      \put(998,1395){\makebox(0,0)[r]{\footnotesize \strut{}-1}}%
      \put(998,2013){\makebox(0,0)[r]{\footnotesize \strut{}-0.5}}%
      \put(998,2630){\makebox(0,0)[r]{\footnotesize \strut{} 0}}%
      \put(998,3247){\makebox(0,0)[r]{\footnotesize \strut{} 0.5}}%
      \put(998,3865){\makebox(0,0)[r]{\footnotesize \strut{} 1}}%
      \put(404,2630){\rotatebox{90}{\makebox(0,0){\strut{}$\smallmath{G_{\cal B}^e}$~{\small [TeV$\smallmath{^{-1}}$]}}}}%
    }%
    \gplbacktexta
    \put(0,0){\includegraphics[scale=0.775]{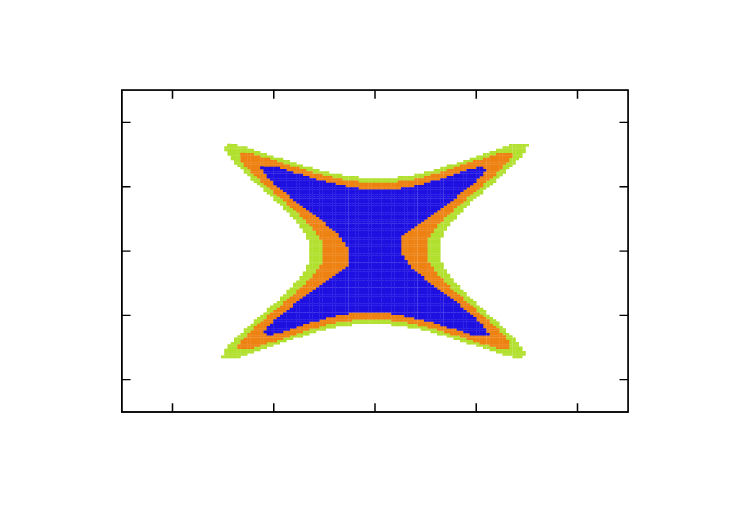}}%
    \gplfronttexta
  \end{picture}%
  &
  \setlength{\unitlength}{0.03875bp}%
  \begin{picture}(7200.00,5040.00)(2200,0)%
    \gplgaddtomacro\gplbacktextb{%
    }%
    \gplgaddtomacro\gplfronttextb{%
      \csname LTb\endcsname%
      \put(1656,800){\makebox(0,0){\footnotesize \strut{}-1}}%
      \put(2628,800){\makebox(0,0){\footnotesize \strut{}-0.5}}%
      \put(3600,800){\makebox(0,0){\footnotesize \strut{} 0}}%
      \put(4572,800){\makebox(0,0){\footnotesize \strut{} 0.5}}%
      \put(5544,800){\makebox(0,0){\footnotesize \strut{} 1}}%
      \put(3600,470){\makebox(0,0){\strut{}$\smallmath{G_{\cal B}^l}$~{\small [TeV$\smallmath{^{-1}}$]}}}%
      \put(998,1395){\makebox(0,0)[r]{\footnotesize \strut{}-1}}%
      \put(998,2013){\makebox(0,0)[r]{\footnotesize \strut{}-0.5}}%
      \put(998,2630){\makebox(0,0)[r]{\footnotesize \strut{} 0}}%
      \put(998,3247){\makebox(0,0)[r]{\footnotesize \strut{} 0.5}}%
      \put(998,3865){\makebox(0,0)[r]{\footnotesize \strut{} 1}}%
      \put(404,2630){\rotatebox{90}{\makebox(0,0){\strut{}$\smallmath{G_{\cal B}^e}$~{\small [TeV$\smallmath{^{-1}}$]}}}}%
    }%
    \gplbacktextb
    \put(0,0){\includegraphics[scale=0.775]{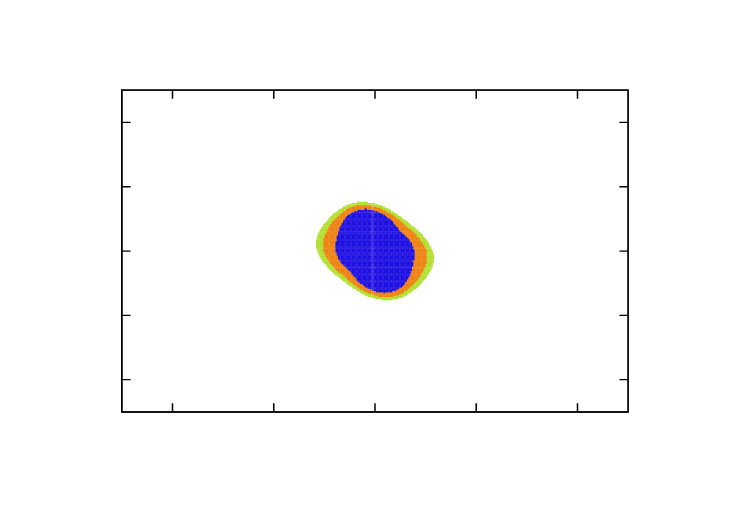}}%
    \gplfronttextb
  \end{picture}%
  \end{tabular}
\endgroup

\caption{From darker to lighter, confidence regions with $\Delta \chi^2\leq$ $2$ (blue), $4$ (orange) and $6$ ($95\%$ C.L.) (green), respectively, for the ${\cal B}$ couplings to leptons assuming no couplings to quarks. The region in the left plot results from the fit to EWPD without LEP 2 data. This is further constrained into the smaller region in the right plot by adding the LEP 2 cross sections and asymmetries to the fit.}
\label{fig_B0_le}
\end{figure}
\begin{figure}[th]
\begingroup
  \makeatletter
  \providecommand\color[2][]{%
    \GenericError{(gnuplot) \space\space\space\@spaces}{%
      Package color not loaded in conjunction with
      terminal option `colourtext'%
    }{See the gnuplot documentation for explanation.%
    }{Either use 'blacktext' in gnuplot or load the package
      color.sty in LaTeX.}%
    \renewcommand\color[2][]{}%
  }%
  \providecommand\includegraphics[2][]{%
    \GenericError{(gnuplot) \space\space\space\@spaces}{%
      Package graphicx or graphics not loaded%
    }{See the gnuplot documentation for explanation.%
    }{The gnuplot epslatex terminal needs graphicx.sty or graphics.sty.}%
    \renewcommand\includegraphics[2][]{}%
  }%
  \providecommand\rotatebox[2]{#2}%
  \@ifundefined{ifGPcolor}{%
    \newif\ifGPcolor
    \GPcolortrue
  }{}%
  \@ifundefined{ifGPblacktext}{%
    \newif\ifGPblacktext
    \GPblacktexttrue
  }{}%
  \let\gplgaddtomacro\g@addto@macro
  \gdef\gplbacktexta{}%
  \gdef\gplfronttexta{}%
  \gdef\gplbacktextb{}%
  \gdef\gplfronttextb{}%
  \makeatother
  \ifGPblacktext
    \def\colorrgb#1{}%
    \def\colorgray#1{}%
  \else
    \ifGPcolor
      \def\colorrgb#1{\color[rgb]{#1}}%
      \def\colorgray#1{\color[gray]{#1}}%
      \expandafter\def\csname LTw\endcsname{\color{white}}%
      \expandafter\def\csname LTb\endcsname{\color{black}}%
      \expandafter\def\csname LTa\endcsname{\color{black}}%
      \expandafter\def\csname LT0\endcsname{\color[rgb]{1,0,0}}%
      \expandafter\def\csname LT1\endcsname{\color[rgb]{0,1,0}}%
      \expandafter\def\csname LT2\endcsname{\color[rgb]{0,0,1}}%
      \expandafter\def\csname LT3\endcsname{\color[rgb]{1,0,1}}%
      \expandafter\def\csname LT4\endcsname{\color[rgb]{0,1,1}}%
      \expandafter\def\csname LT5\endcsname{\color[rgb]{1,1,0}}%
      \expandafter\def\csname LT6\endcsname{\color[rgb]{0,0,0}}%
      \expandafter\def\csname LT7\endcsname{\color[rgb]{1,0.3,0}}%
      \expandafter\def\csname LT8\endcsname{\color[rgb]{0.5,0.5,0.5}}%
    \else
      \def\colorrgb#1{\color{black}}%
      \def\colorgray#1{\color[gray]{#1}}%
      \expandafter\def\csname LTw\endcsname{\color{white}}%
      \expandafter\def\csname LTb\endcsname{\color{black}}%
      \expandafter\def\csname LTa\endcsname{\color{black}}%
      \expandafter\def\csname LT0\endcsname{\color{black}}%
      \expandafter\def\csname LT1\endcsname{\color{black}}%
      \expandafter\def\csname LT2\endcsname{\color{black}}%
      \expandafter\def\csname LT3\endcsname{\color{black}}%
      \expandafter\def\csname LT4\endcsname{\color{black}}%
      \expandafter\def\csname LT5\endcsname{\color{black}}%
      \expandafter\def\csname LT6\endcsname{\color{black}}%
      \expandafter\def\csname LT7\endcsname{\color{black}}%
      \expandafter\def\csname LT8\endcsname{\color{black}}%
    \fi
  \fi
  \begin{tabular}{ c c}
  \setlength{\unitlength}{0.03875bp}%
  \begin{picture}(7200.00,5040.00)(600,0)%
    \gplgaddtomacro\gplbacktexta{%
    }%
    \gplgaddtomacro\gplfronttexta{%
      \csname LTb\endcsname%
      \put(1656,800){\makebox(0,0){\strut{}-0.4}}%
      \put(2628,800){\makebox(0,0){\strut{}-0.2}}%
      \put(3600,800){\makebox(0,0){\strut{} 0}}%
      \put(4572,800){\makebox(0,0){\strut{} 0.2}}%
      \put(5544,800){\makebox(0,0){\strut{} 0.4}}%
      \put(3600,470){\makebox(0,0){\strut{}$\smallmath{G_{\cal B}^\phi}$~{\small [TeV$^{-1}$]}}}%
      \put(998,1086){\makebox(0,0)[r]{\strut{}-1}}%
      \put(998,1858){\makebox(0,0)[r]{\strut{}-0.5}}%
      \put(998,2630){\makebox(0,0)[r]{\strut{} 0}}%
      \put(998,3402){\makebox(0,0)[r]{\strut{} 0.5}}%
      \put(998,4174){\makebox(0,0)[r]{\strut{} 1}}%
      \put(404,2630){\rotatebox{90}{\makebox(0,0){\strut{}$\smallmath{G_{\cal B}^l}$~{\small [TeV$^{-1}$]}}}}%
    }%
    \gplbacktexta
    \put(0,0){\includegraphics[scale=0.775]{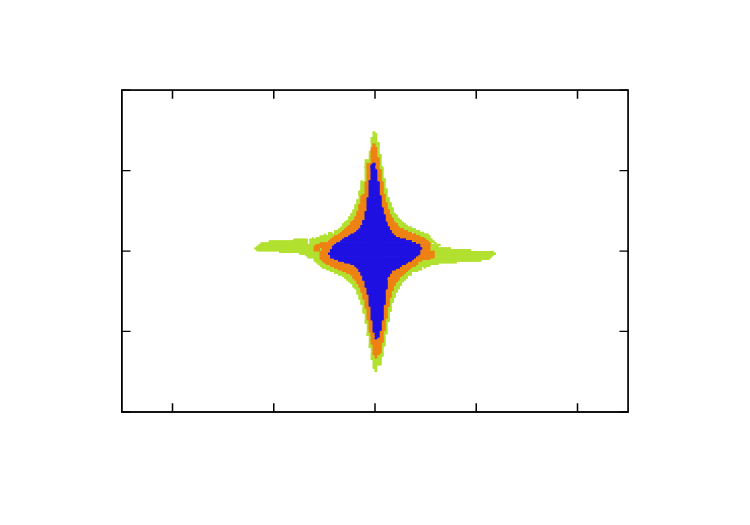}}%
    \gplfronttexta
  \end{picture}%
  &
  \setlength{\unitlength}{0.03875bp}%
  \begin{picture}(7200.00,5040.00)(2200,0)%
    \gplgaddtomacro\gplbacktextb{%
    }%
    \gplgaddtomacro\gplfronttextb{%
      \csname LTb\endcsname%
      \put(1656,800){\makebox(0,0){\strut{}-0.4}}%
      \put(2628,800){\makebox(0,0){\strut{}-0.2}}%
      \put(3600,800){\makebox(0,0){\strut{} 0}}%
      \put(4572,800){\makebox(0,0){\strut{} 0.2}}%
      \put(5544,800){\makebox(0,0){\strut{} 0.4}}%
      \put(3600,470){\makebox(0,0){\strut{}$\smallmath{G_{\cal B}^\phi}$~{\small [TeV$^{-1}$]}}}%
      \put(998,1086){\makebox(0,0)[r]{\strut{}-1}}%
      \put(998,1858){\makebox(0,0)[r]{\strut{}-0.5}}%
      \put(998,2630){\makebox(0,0)[r]{\strut{} 0}}%
      \put(998,3402){\makebox(0,0)[r]{\strut{} 0.5}}%
      \put(998,4174){\makebox(0,0)[r]{\strut{} 1}}%
      \put(404,2630){\rotatebox{90}{\makebox(0,0){\strut{}$\smallmath{G_{\cal B}^l}$~{\small [TeV$^{-1}$]}}}}%
    }%
    \gplbacktextb
    \put(0,0){\includegraphics[scale=0.775]{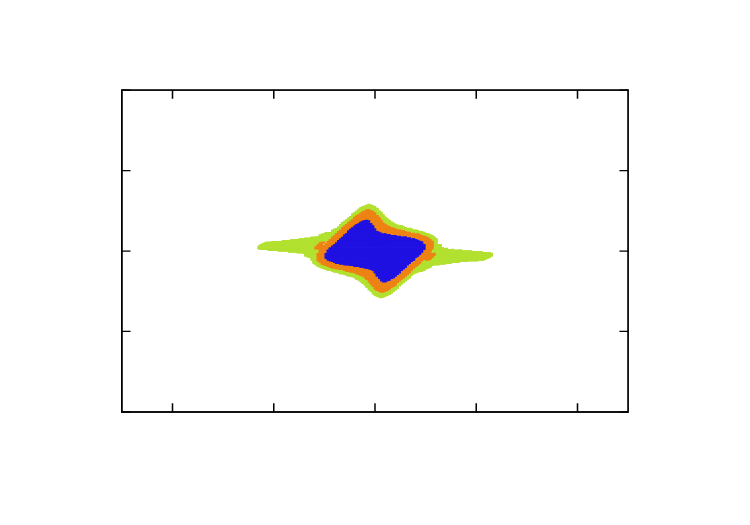}}%
    \gplfronttextb
  \end{picture}%
  \end{tabular}
\endgroup

\caption{From darker to lighter, confidence regions with $\Delta \chi^2\leq$ $2$(blue), $4$ (orange) and $6$ ($95\%$ C.L.) (green), respectively, for the ${\cal B}$ couplings to the Higgs and LH leptons assuming no couplings to quarks. The regions in the plot on the left are obtained from a fit to EWPD without LEP 2 data. They are reduced to smaller regions when the LEP 2 cross sections and asymmetries are added to the fit, as shown in the plot on the right.}
\label{fig_B0_Hl}
\end{figure}

Second, the low-energy data and the measurements of cross sections and asymmetries at LEP 2 impose significant constraints on the leptonic couplings, mostly independent of the Higgs coupling. This effect is apparent in Figs.~\ref{fig_B0_le} and~\ref{fig_B0_Hl}, where for simplity we consider hadrophobic vectors. We display several confidence regions for fits with and without LEP~2 data, in planes parametrized by different couplings. In the left-hand plot of Fig.~\ref{fig_B0_le}, we see that the regions with relatively large couplings along the diagonals, with equal absolute value of LH and right-handed (RH) couplings, are allowed by EWPD without LEP 2 data. In particular, this nonchiral combination avoids the constraints from parity violation in M{\o}ller scattering. However, the right-hand plot in this figure shows that these regions get excluded when the LEP 2 data are taken into account. In Fig.~\ref{fig_B0_Hl}, we see how the LEP 2 data help in constraining the lepton couplings, but not the couplings to the Higgs. As a matter of fact, a small nonvanishing Higgs coupling is favored by LEP 2 data, as a modification of trilinear couplings of the $Z$ boson can soften a bit the effect of four-fermion operators. Anyway, this effect is erased by $Z$-pole data.

Finally, for very small Higgs and lepton couplings, the fit is approximately flat along the directions of the quark couplings, due to the fact that no electroweak observable depends on their square. 
This implies the absence of limits on quark couplings in Table~\ref{table_OneVLimits}.

It is also worth noting that, at the minimum, the RH quark couplings are pretty large. The reason is that these couplings raise the prediction for the hadronic cross section measured at LEP 2, which in many energy bins is around $1~\sigma$ above the SM prediction. This results in a global $1.7~\sigma$ discrepancy when correlations are taken into account.  The LH counterparts, on the other hand, are more tightly constrained by the $Z$-pole observables, and stay smaller. This preference for large RH couplings is stronger in the case of $d$ quarks, due to the SM discrepancy in the bottom forward-backward asymmetry. Indeed, at the minimum we find $A_{FB}^b=0.1016$, and the pull in this observable is reduced to $1.5~\sigma$. We should stress, nevertheless, that such large couplings could drive the theory into a nonperturbative regime. We shall come back to this point in Section~\ref{section:several}.

There is a lot of work on the electroweak limits for particular $Z^\prime$ models (see for instance~\cite{Durkin:1985ev}), so it may be useful to discuss at this point the relation between model-dependent and model-independent fits. Each particular model imposes correlations among the couplings, and corresponds to some lower-dimensional manifold in the complete model-independent fit. Therefore, the latter contains all the necessary information. However, for obvious reasons, in the paper we are just showing partial information, in terms of one or two coupling-to-mass ratios at a time. In general, these will not be the free parameters in a particular model, so the translation of our results is not direct. Nevertheless, the one-dimensional limits and two-dimensional plots we have shown above provide basic guidelines to understand the constraints on explicit models. For instance, the allowed regions in the "minimal" class of $Z^\prime$ models considered in Ref.~\cite{Salvioni:2009mt,Salvioni:2009jp}, when given in terms of coupling-to-mass ratios, agree with those of Fig.~\ref{fig_B0_Hl} above.

\begin{table}[t]
\begin{center}
{\small
\begin{tabular}{c c c c c c c c c}
\ctoprule
& &\multicolumn{7}{c}{\underline{$95\%$ C.L. Electroweak Limits on}}\\
$ $&$ $&$ $&$ $&$ $&$ $&$ $&$ $&$ $\\[-0.3cm]
&&\multicolumn{3}{c}{\underline{$\sin{\theta_{Z\Zprime}}\left[\times 10^{-4}\right]$}}&&\multicolumn{3}{c}{\underline{$M_\Zprime$ [TeV]}}\\
$ $&$ $&$ $&$ $&$ $&$ $&$ $&$ $&$ $\\[-0.3cm]
Model&&EWPD        &LEP 2&All Data& &EWPD       &LEP 2&All Data\\
           &&(no LEP 2) &           &              & &(no LEP 2)&           &               \\
\cmidrule{1-1}\cmidrule{3-5}\cmidrule{7-9}
\crowcolor$ $&\whitecell$ $&$ $&$ $&$ $&\whitecell$ $&$ $&$ $&$ $\\[-0.3cm]
\crowcolor$Z^\prime_\chi $&\whitecell$ $&$\left[-10,\phantom{0}7\right] $&$\left[-80,118\right] $&$\left[-11,\phantom{0}7\right]  $&\whitecell$ $&$1.123 $&$0.772$&$1.022 $\\
$ $&$ $&$ $&$ $&$ $&$ $&$ $&$ $&$ $\\[-0.3cm]
$Z^\prime_\psi $&$ $&$\left[-19,\phantom{0}7\right]  $&$\left[-196,262\right]  $&$\left[-19,\phantom{0}7\right]  $&$ $&$0.151 $&$0.455$&$0.476 $\\
\crowcolor$ $&\whitecell$ $&$ $&$ $&$ $&\whitecell$ $&$ $&$ $&$ $\\[-0.3cm]
\crowcolor$Z^\prime_\eta $&\whitecell$ $&$\left[-22,25\right]  $&$\left[-150,164\right]  $&$\left[-23,27\right]  $&\whitecell$ $&$0.422 $&$0.460$&$0.488 $\\
$ $&$ $&$ $&$ $&$ $&$ $&$ $&$ $&$ $\\[-0.3cm]
$Z^\prime_I $&$ $&$\left[-\phantom{0}5,\phantom{0}9\right]  $&$\left[-144,96\right]  $&$\left[-\phantom{0}5,10\right]  $&$ $&$1.207 $&$0.652 $&$1.105 $\\
\crowcolor$ $&\whitecell$ $&$ $&$ $&$ $&\whitecell$ $&$ $&$ $&$ $\\[-0.3cm]
\crowcolor$Z^\prime_N $&\whitecell$ $&$\left[-14,\phantom{0}6\right]  $&$\left[-165,223\right]  $&$\left[-14,\phantom{0}6\right]  $&\whitecell$ $&$0.635 $&$0.421$&$0.699 $\\
$ $&$ $&$ $&$ $&$ $&$ $&$ $&$ $&$ $\\[-0.3cm]
$Z^\prime_S $&$ $&$\left[-\phantom{0}9,\phantom{0}5\right]  $&$\left[-85,129\right]  $&$\left[-10,\phantom{0}5\right]  $&$ $&$1.249 $&$0.728 $&$1.130 $\\
\crowcolor$ $&\whitecell$ $&$ $&$ $&$ $&\whitecell$ $&$ $&$ $&$ $\\[-0.3cm]
\crowcolor$Z^\prime_R $&\whitecell$ $&$\left[-17,\phantom{0}7\right]  $&$\left[-166,177\right]  $&$\left[-15,\phantom{0}5\right]  $&\whitecell$ $&$0.439 $&$0.724$&$1.130 $\\
$ $&$ $&$ $&$ $&$ $&$ $&$ $&$ $&$ $\\[-0.3cm]
$Z^\prime_{LR} $&$ $&$\left[-13,\phantom{0}5\right]  $&$\left[-147,189\right]  $&$\left[-12,\phantom{0}4\right]  $&$ $&$0.999 $&$0.667$&$1.162 $\\
\cbottomrule
\end{tabular}
\caption{ Comparison of 95$\%$ C.L. limits on $\sin{\theta_{Z\Zprime}}$ and $M_\Zprime$ obtained for several popular $Z^\prime$ models from a fit to standard EWPD without LEP 2, to LEP 2 cross sections and asymmetries, and to all data. The gauge coupling constants are taken equal to the GUT-inspired value, $\sqrt{5/3}~g^\prime \approx 0.46$.
\label{table: ZpLimits}}
}
\end{center}
\end{table}
%

To be more explicit, we show in Table~\ref{table: ZpLimits}, as an example, the limits on $Z^\prime$ masses and mixings that we find for some popular models usually considered in the literature~\cite{Langacker:2008yv,DelAguila:1995fa}~\footnote{Leptophobic neutral gauge bosons derived, for example, from $E_6$~\cite{del Aguila:1986ez} are not constrained by EWPD, except for their possible mixing with the $Z$ boson. If their coupling to the Higgs is nonvanishing, the $Z$-pole data can provide lower limits on $M_{Z^\prime_{\not L}}$ around 1~TeV~\cite{Erler:2009jh}.}. The mixing is fixed in some models~\cite{delAguila:1986ad} but can vary continuously in others. We leave it as a free parameter.
We give limits from three data sets:  i) EWPD excluding LEP 2 cross sections and asymmetries; ii) LEP 2 cross sections and asymmetries\footnote{Unlike \cite{Alcaraz:2006mx}, where the $Z$-$Z^\prime$  mixing is fixed to zero, we let $\sin{\theta_{ZZ^\prime}}$ vary in the fit to LEP 2 data.}; iii) all data. Our results for the first data set agree with the ones in the recent update~\cite{Erler:2009jh}, except for some differences in the limits for the mixing, which arise from our inclusion of Tevatron Higgs searches in the fit.

\begin{figure}[thb]
\input{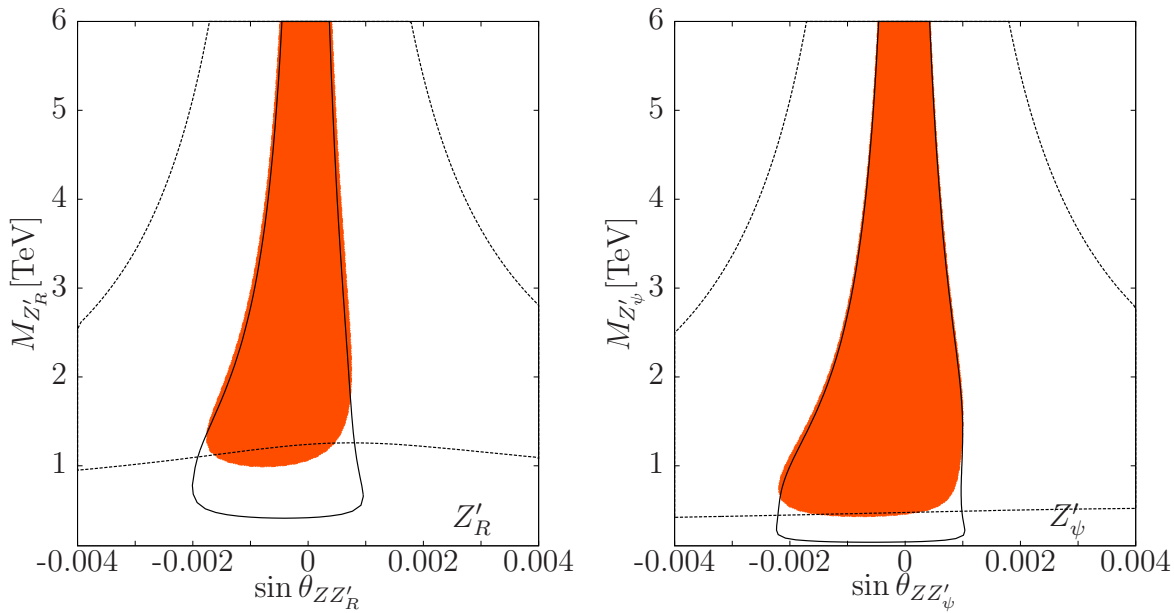}
\caption{$95\%$ C.L. contour in the $M_\Zprime$ - $\sin{\theta_{Z\Zprime}}$ plane for the $Z^\prime_{R}$ model (left) and $Z^\prime_\psi$ (right). The different contours correspond to the fit to EWPD without LEP 2 cross sections and asymmetries (solid line), to LEP 2 cross sections and asymmetries (dashed line), and to all data (solid region).
\label{f_E6chi}}  
\end{figure}

For most models, EWPD without LEP~2 are sufficient to constrain significantly both the mixing and the mass of the new vectors. The two exceptions are the $\psi$ and $R$ models, for which LEP~2 data are decisive to raise the limit on the mass. The correlations between the mixing and mass in these two models are illustrated in Fig.~\ref{f_E6chi}. For the $\psi$ model, this behaviour can be inferred from Fig.~\ref{fig_B0_le}, observing that the leptonic couplings are axial and lie along one diagonal of the plots in that figure.

\subsection{Left-handed triplet: ${\cal W}$}

\begin{figure}[!]
\input{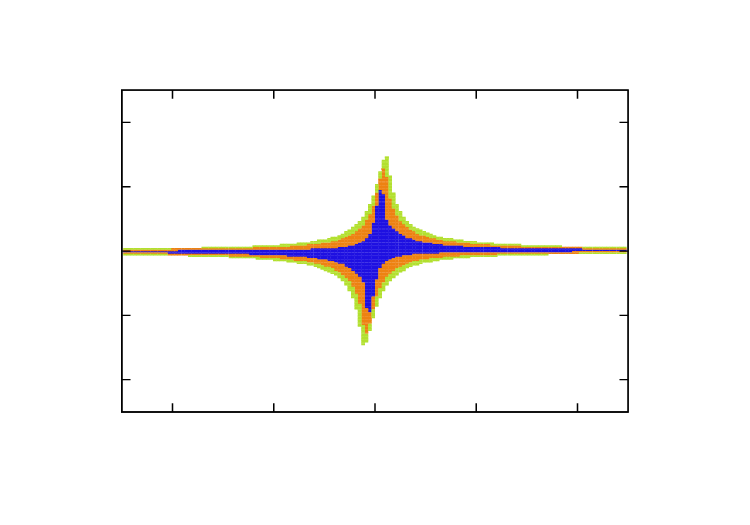}
\caption{From darker to lighter, confidence regions with $\Delta \chi^2\leq$ $2$ (blue), $4$ (orange) and $6$ ($95\%$ C.L.) (green), respectively, for the ${\cal W}$ couplings to LH leptons and to the Higgs boson. Notice the flat direction along the Higgs coupling axis when the lepton charge vanishes.}
\label{fig:W0lH}
\end{figure}

This $SU(2)_L$ triplet decomposes after electroweak breaking into a neutral vector boson and a charge $\pm 1$ complex vector boson.
This representation appears, for instance, in theories with extra dimensions and in some Little Higgs models or composite vector models. The most general case has three new parameters: the couplings to the Higgs and to the lepton and quark doublets. As in the previous case, the parameter space has a flat direction along the quark coupling direction when the interactions with the Higgs and leptons vanish. On the other hand, the coupling of this vector to the Higgs does not appear quadratically, since the $\mathcal{W}$ field preserves custodial symmetry, which forbids the operator ${\cal O}_{\phi}^{(3)}$. Therefore, there is an extra flat direction in the Higgs coupling for vanishing couplings to the fermions. This is illustrated in Figure~\ref{fig:W0lH}, where we plot several confidence regions in the plane spanned by the lepton and Higgs couplings. 

Note also that the $\chi^2$ at the minimum, which is placed over both flat directions, is less than 2 units smaller than for the SM. Thus, any value of the Higgs and quark couplings is allowed by EWPD at that confidence level. For $G_{\cal W}^\phi$, the $1~\sigma$ interval is finite, as reported in Table~\ref{table_OneVLimits}, whereas there are no limits on $G_{\cal W}^q$. As in the case of the singlet $\mathcal{B}$, there is a preference for large values of the quark coupling. However, for the ${\cal W}$ boson there are additional constraints on these couplings, for instance from $K^0$-$\overline{K^0}$ mixing, which are complementary to those presented here~\cite{Blanke:2008zb}.

\subsection{Charged singlet: ${\cal B}^1$}
This complex isosinglet vector has electric charge $\pm 1$. It appears, for instance, in left-right models. After electroweak symmetry breaking, it mixes with the SM charged bosons, with the mixing proportional to the Higgs coupling. With our assumption that the new vector is heavier than the $W$ boson, this mixing decreases $M_W$, and gives a negative contribution to the $\rho$ parameter. In the effective formalism, this effect is clear from the positive sign of the contribution of this vector to the operator ${\cal O}_\phi^{(3)}$.
Therefore, the presence of this vector with a nonvanishing scalar coupling favors a value for the Higgs mass yet lower than in the SM, in contrast with the case of singlets of zero hypercharge, $\mathcal{B}$. The LEP 2 lower bound on the Higgs mass then forces the Higgs coupling to be very small. The other parameter in this scenario is the coupling of the $\mathcal{B}^1$ to the RH quarks. This coupling induces RH charged currents, via the operator ${\cal O}_{\phi u d}$. 
Taking into account the preference for small Higgs coupling, the electroweak data cannot constrain the RH quark couplings. However, there are further constraints from $K^0$-$\overline{K^0}$ mixing on these couplings. They are typically more stringent than the ones for ${\cal W}$, since they induce operators of mixed chirality, but depend strongly on the flavor structure of the RH quarks~\cite{Langacker:1989xa}. (Our fits neither incorporate the analysis of RH quark couplings from kaon physics described in Ref.~\cite{Bernard:2006gy}.) On the other hand, we have assumed that there are no light RH neutrinos. If there were, there would be extra constraints on the additional coupling to RH leptons~\cite{Langacker:1989xa}.

\subsection{Fermiophobic triplet: ${\cal W}^1$}
The triplet with hypercharge 1 contains two real neutral vectors, which mix with the $Z$ boson upon electroweak symmetry breaking, a complex vector of charge $\pm 1$, which mixes with the $W$, and a complex vector of charge $\pm 2$, which gives no observable effect. The characteristic feature of this representation is that it cannot couple to any SM fermions. Hence, its only visible effects are oblique. Moreover, the net contribution to the $\rho$ parameter is positive, which makes EWPD consistent with a heavy Higgs. Therefore, the fit prefers a nonzero value of the coupling, in order to compensate the effect on EWPD of the direct LEP lower bound on the Higgs mass. The interplay with the Higgs mass is further discussed in Section~\ref{section:Higgs}.

\subsection{Leptophilic vector: ${\cal L}$}
This representation contains complex vectors of charges $\pm 1$ and $\pm 2$. Since it does not couple to the Higgs, the charge $\pm 1$ components do not mix with the $W$ boson at tree level. The vector field is coupled to a $\Delta L=2$ current mixing the LH and RH lepton multiplets. Despite this, no trace of lepton number violation remains in the effective Lagrangian, thanks to the absence of any other couplings. This fact allows to recover this symmetry by assigning lepton number $L=2$ to the field ${\cal L}$.  There can be, however, lepton {\em flavor} violation, even for diagonal couplings, as these create (destroy) two same-flavor anti-leptons (leptons), allowing for processes like $e^-e^- \to \mu^-\mu^-$. The only operator in the effective Lagrangian for this vector is the four-lepton interaction ${\cal O}_{le}$. This can contribute to $\nu^\mu e$ scattering as well as to $e^+e^-\rightarrow \ell^+\ell^-$ data at LEP 2. There are no restrictions from parity violating observables measured in M{\o}ller scattering, since ${\cal O}_{le}$ does not contribute to V-A couplings. In the case of couplings to only one flavor per SM multiplet, the weakest constraint shown in Table \ref{table_OneVLimits} occurs for couplings between electrons and taus. Similar bounds apply to couplings to electrons and muons.

\subsection{Singlet vector leptoquarks: ${\cal U}^2$ and ${\cal U}^5$}

The two colored $SU(2)_L$ singlets ${\cal U}^2$ and ${\cal U}^5$ decompose into complex vectors of fractional charges $\pm 2/3$ and $\pm 5/3$, respectively. The associated currents carry nonvanishing $L$, $B$ and $B-L$ numbers. But again, these global symmetries are preserved in the effective Lagrangian. For ${\cal U}^2$, this is so because the two terms in the current have the same $B$ and $L$ charges. The integration of ${\cal U}^2$ generates the operators ${\cal O}_{ed}$, ${\cal O}_{lq}^{(1,3)}$ and ${\cal O}_{qde}$, while for ${\cal U}^5$, only ${\cal O}_{eu}$ is generated. With the exception of ${\cal O}_{qde}$, which does not interfere with any of the SM amplitudes, these operators contribute to atomic parity violation (APV) in atoms and to the inclusive hadronic cross section at LEP 2. ${\cal O}_{lq}^{(1,3)}$ can also contribute to neutrino-nucleon scattering if ${\cal U}^2$ couples to muons and first family quarks. Finally, ${\cal O}_{lq}^{(3)}$ can modify the unitarity relation of the CKM matrix. In particular, this is the only constraint when ${\cal U}^2$ couples to the second family. The precise determination of the weak charge for Cesium and its good agreement with the SM prediction is the strongest constraint when these operators are coupled to the first family (together with the CKM unitarity for $G_{{\cal U}^2}^{lq}$). 
It is worth noting that the negative contribution to ${\cal O}_{eu}$ is favored by LEP 2 data, as it increases the total hadronic cross section above the $Z$ pole. For this reason, the fit with ${\cal U}^{5}$ gives some improvement in $\chi^2$, with just one extra free parameter.

\subsection{Doublet vector leptoquarks: ${\cal Q}^1$ and ${\cal Q}^5$}
The $SU(2)_L$ doublet ${\cal Q}^1$ contains two complex vectors, of charges $\pm 1/3$ and $\pm 2/3$, whereas the doublet ${\cal Q}^5$ is made of complex vectors of charges $\pm 1/3$ and $\pm 4/3$. Again, the corresponding currents carry nontrivial $B$, $L$ and $B-L$ numbers. Of these, $B-L$ is actually conserved, since all the terms in the current have the same charge, $\Delta (B-L)=-2/3$. On the other hand, there are dangerous contributions to baryon and lepton number violating operators.
In order to avoid proton decay while allowing for contributions to EWPD, we consider here the case without the $B$ violating couplings $g_{{\cal Q}^1}^{dq}$ and $g_{{\cal Q}^5}^{uq}$. Then, the vector ${\cal Q}^1$ generates only the operator ${\cal O}_{lu}$, while ${\cal Q}^5$ induces three operators: ${\cal O}_{ld}$, ${\cal O}_{qe}$ and ${\cal O}_{qde}$.
 
For ${\cal Q}^{1}$, the dominant constraint comes again from APV. Weaker bounds are obtained when we couple the new vector to electrons and $c$ quarks, or to muons and $u$ quarks, so that they affect LEP 2 data and the low-energy effective coupling $g_R^2$ in deep-inelastic neutrino-nucleon scattering, respectively. In the later case there is a small improvement in $\chi^2$.

In the case of the vector ${\cal Q}^{5}$, the bounds from APV are mild for couplings $g_{{\cal Q}^5}^{dl}$ and $g_{{\cal Q}^5}^{eq}$ to the first family. This is so because the independent contributions to the atomic weak charges from ${\cal O}_{ld}$ and ${\cal O}_{qe}$ can be adjusted to approximately cancel.
The strongest bound on $G_{{\cal Q}^5}^{dl}$ comes again from $g_R^2$, when $\mathcal{Q}^{5}$ couples to muons and down quarks. When it couples to electrons and $s$ or $b$ quarks, only the LEP 2 constraints apply. These are weaker, as a sizable value for these couplings is again favored. For $G_{{\cal Q}^5}^{eq}$, LEP 2 data give stronger constraints, which can be relaxed by the interplay with the other coupling, when the latter is not tied by other data.

The best minimum occurs for couplings to the first family. Besides the better agreement with LEP 2 hadronic data, there is an improvement in the combinations of the parity-violating $eq$ effective parameters $C_{1u}$ and $C_{1d}$ that appear in Table~\ref{Table: Exp-SM}. As a result, the $\chi^2$ of the global fit is decreased by 3.4, with two extra parameters.

\subsection{Triplet vector leptoquark: ${\cal X}$}
Finally, this $SU(2)_L$ triplet decomposes into complex vectors of charges $\pm 1/3$, $\pm 2/3$ and $\pm 5/3$. It connects LH quarks with LH leptons. Even though the current carries lepton and baryon numbers, the $B$ and $L$ symmetries are preserved in the effective Lagrangian. ${\cal X}$ only generates the operators ${\cal O}_{lq}^{(1)}$ and ${\cal O}_{lq}^{(3)}$, whose effects are described in the ${\cal U}^2$ subsection. The coefficients are different, however, and so are the constraints.

The strongest bound in Table~\ref{table_OneVLimits} is also provided by APV, in the case with couplings to electrons and first family quarks. The weakest nontrivial bound corresponds to the assumption that this vector only couples to the second family and, as in the case of ${\cal U}^2$, it comes from the CKM constraints. On the other hand, a $\mathcal{X}$ coupling muons to the LH $u$ and $d$ quarks allows to reduce to $1~\sigma$ the SM $2~\sigma$ discrepancy in the $g_L^2$ coupling extracted from deep-inelastic neutrino-nucleon scattering. A further reduction is prevented again by the precise measurement of unitarity in the first row of the CKM matrix. The global decrease in the $\chi^2$ is, however, marginal: $\Delta \chi^2_\mt{min}\approx -1.4$. In fact, a better improvement is found by choosing couplings between electrons and the second quark family, even if this does not modify $g_L^2$. It may seem surprising that, for this vector, the limits on couplings of electrons to the third family of quarks are significantly stronger than the corresponding ones for the second family. The explanation is that, for this representation, the contributions to the hadronic cross section at LEP 2 from the up (down) quarks are favored (disfavored), and in the case of the third family only the $b$ quark contributes.


\section{Several extra vectors}
\label{section:several}
In this section we discuss scenarios with several new vector bosons, both in the same and in different SM representations.

It seems quite natural that an extra vector boson around the TeV scale will come accompanied by other new particles, in particular additional new vectors. This occurs in many explicit models beyond the SM.
At the dimension-six order, the coefficients of the operators are just given by the sum of the contributions of each new vector, as we have shown explicitly before. Moreover, because the leading new effects come from the interference of SM amplitudes and diagrams with insertions of dimension-six operators, the interference between the contributions of different vectors to observables is negligible. On the other hand, opposite signs may occur in the sums. Hence, some (partial) cancellations are possible, both between contributions of different new vectors to a given operator, and between the contributions of different gauge-invariant operators to an observable.

One consequence of having more than one vector simultaneously is that some restrictions on the operator coefficients, which hold necessarily for just one vector, are removed. For example, for just one extra singlet ${\cal B}$ with arbitrary couplings, the following relations are always satisfied:
\bea
\left(\alpha_{\phi \psi}^{(1)}\right)^2\!\!\!\!&\sim&\!\!\!\!\frac 12 \alpha_\phi^{(3)}  \alpha_{\psi\psi},\\
\left(\alpha_{\psi \psi^\prime}\right)^2\!\!\!\!&\sim&\!\!\!\!\alpha_{\psi \psi} \alpha_{\psi^\prime \psi^\prime}, 
\eea
where $\psi$ and $\psi^\prime$ stand for any SM fermion multiplet. These relations do not  hold any longer if there are two singlets $\mathcal{B}$. For instance, two vectors with the same couplings to the lepton doublet and opposite Higgs couplings, 
\bea
g_{V_1}^l\!\!\!\! &=&\!\!\!\! \phantom{+}g_{V_2}^l , \nonumber  \\
g_{V_1}^\phi\!\!\!\! &=&\!\!\!\! -g_{V_2}^\phi , \label{mirror}
\eea
will have a vanishing coefficient $\alpha_{\phi l}^{(1)}$, but nonvanishing $\alpha_\phi^{(3)}$ and $\alpha_{ll}^{(1)}$. This would be impossible with only one extra vector, and it is an example of a cancellation of the effects of several vectors. The point is that models with more than one extra vector may have observable effects that cannot be reproduced by any model with just one.

An interesting possibility is that cancellations of this sort give rise to weaker bounds from EWPD on each vector. For the leptonic couplings, it turns out that there is little room for this effect, at least in the universal case. The reason is that the coefficients of the four-lepton operators $({\cal O}_{ll}^{(1)})_{iiii}$ and $({\cal O}_{ee})_{iiii}$ induced by any extra vector are negative definite, since they are given by minus sums of squares. Hence, no cancellations are possible here. For i=1 (first family), these coefficients are constrained to be very small by the differential cross sections in Bhabha scattering measured at LEP 2\footnote{Cancellations between the contributions of a $\mathcal{B}$ and a $\mathcal{L}$ are possible in the four-lepton operator with mixed chiralities, ${\cal O}_{le}$, but the angular distributions allow to isolate the effects of each individual operator on the cross sections.}. Furthermore, the operators modifying the trilinear couplings, which could counteract the action of the four-fermion operators, are independently constrained to be small by the $Z$-pole data.

Therefore, in universal scenarios with combinations of several vector bosons, the limits on the ratios of leptonic couplings to masses of each new vector are at least as stringent as the corresponding ``one-at-a-time'' limits in Table~\ref{table_OneVLimits}~\cite{delAguila:1986iw}. In other words, new vector bosons must be, to a certain degree, leptophobic (or more precisely, electrophobic~\cite{Salvioni:2009jp})\footnote{Remember that we are always working with the assumption that no other kind of new physics modifies EWPD.}.

\begin{figure}[!]
\input{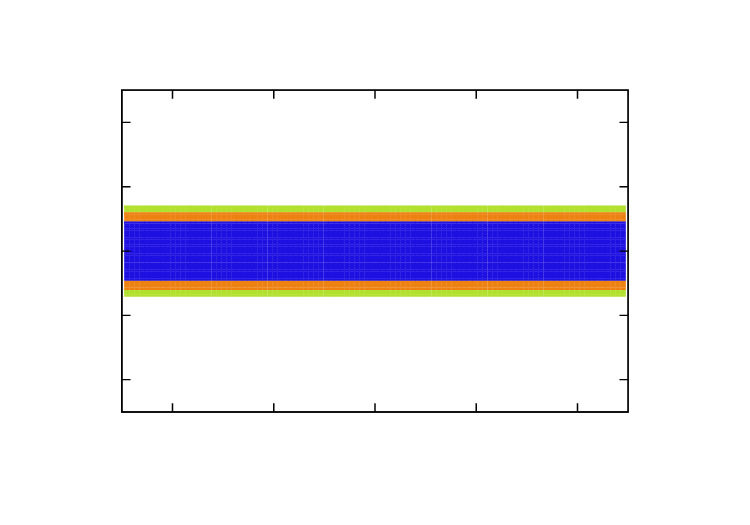}
\caption{From darker to lighter, confidence regions with $\Delta \chi^2\leq$ $2$ (blue), $4$ (orange) and $6$ ($95\%$ C.L.) (green), respectively, in the $G_{\mathcal{W}_1}^l$ - $G_{\mathcal{W}_1}^\phi$ plane of an extension with two mirror left-handed triplets, $\mathcal{W}_1$ and $\mathcal{W}_2$.}
\label{fig:W0lHMirr}
\end{figure}
Despite these limitations, the cooperation of several extra vectors can open new regions in the parameter space of couplings and masses. A simple example is the case of two left-handed triplets $\mathcal{W}$, with universal couplings as in \refeq{mirror}. We show in Fig.~\ref{fig:W0lHMirr} several confidence regions in the $G_{\mathcal{W}_1}^l$ - $G_{\mathcal{W}_1}^\phi$ plane. We see that the combination of these two ``mirror'' vectors make the EWPD blind to the coupling $g_\mathcal{W}^\phi$.  This figure is to be compared with the corresponding plot for just one $\mathcal{W}$ in Fig.~\ref{fig:W0lH}. A similar outcome is found in a model with ``mirror'' neutral singlets $\mathcal{B}$, when we also add a $\mathcal{B}^1$ vector boson to cancel the effect of the $\mathcal{B}$ bosons on the $\rho$ parameter. In the next subsection, we give some examples of cancellations between the contributions of different types of vector bosons.

\subsection{Nonuniversal couplings and the bottom forward-backward asymmetry}
\label{section_non_universal}

In some classes of models, the extra vector bosons couple in a nonuniversal way to the different families. Large couplings to the third family are expected, for instance, in models of dynamical electroweak symmetry breaking. In extra-dimensional theories, nonuniversal couplings appear when the fermions are separated in the extra dimension, with heavier fermions having naturally bigger couplings to the new vectors. For the most common types of vector bosons, $\mathcal{B}$, $\mathcal{W}$ and $\mathcal{B}^1$, we have assumed so far family universal couplings. In this section, we explore the impact of dropping this assumption, both for a unique singlet vector and for a few interesting combinations.  As we will see, the extra freedom allows to better reproduce the experimental data and even improve the SM fits.
For simplicity, we assume in the following that all the new couplings are small, except the ones to the third family of quarks and to the Higgs doublet. This has the advantage of making FCNC innocuous, due to CKM suppression\footnote{The limits from $\mu$ and $K$ decays can generally be satisfied if the couplings to the first two families are universal (not necessarily zero). This has the implication that possible anomalies in $B_s$ mixing and charmless $B$ decays could be accounted for by $b$ - $d$ and $b$ - $s$ vector boson couplings~\cite{Barger:2009eq}.}. Tuning the couplings of an extra singlet $\mathcal{B}$ to the RH bottom, it is possible to correct the deviation in the bottom forward-backward asymmetry at the $Z$ pole~\cite{He:2002ha,He:2003qv} (no other vector boson can play this role). We have actually seen some improvement in the prediction for this observable with universal singlets, but the nonuniversal scenario works better and allows to completely remove the discrepancy without modifying the observables that agree with the SM. This produces a significant decrease in the $\chi^2$ of the global fits, as shown in the first column of Table~\ref{table: Nonuniversal Comb}~\footnote{Even if we are focussing on the couplings to $b$ quarks, we do not include LEP 2 $b$ data in the fits we present here, because the reported values carry the hypothesis of no new physics beyond the SM~\cite{Alcaraz:2006mx}. At any rate, we have checked that these data have very little impact in the results of the fits.}.

This "solution" to the $A_{FB}^b$ anomaly puzzle suffers, however, from an important deficiency. 
In order to shift the asymmetry without modifying the $Z\rightarrow \overline{b}b$ partial decay width, we need a large correction to the $Z\overline{b}_Rb_R$ vertex and a small nonvanishing correction to the $Z\overline{b}_Lb_L$ vertex. These corrections are produced by the 3-3 entries of the operators ${\cal O}_{\phi d}^{(1)}$ and ${\cal O}_{\phi q}^{(1)}$, with coefficients proportional to the couplings of the extra vector to the Higgs, $g^\phi_\mathcal{B}$, and to the RH bottom and the LH top-bottom doublet, respectively. As we can see in Table~\ref{table: Nonuniversal Comb}, the ratio $G^b_\mathcal{B}=g^b_\mathcal{B}/M_\mathcal{B}$ is rather big at the minimum. Unless $M_\mathcal{B} \lesssim 1$~TeV, this can spoil perturbation theory in the complete theory, rendering the whole calculation meaningless\footnote{Remember also that we cannot trust our approximations for very light vectors, which in addition are subject to Tevatron bounds.}. In the second column of Table~\ref{table: Nonuniversal Comb}, we have forced $G^b_\mathcal{B}$ to be smaller than 1, and we see that the anomaly is then recovered.

The reason for the large coupling of the singlet to $b_R$, apart from the requisite of a big effect, is that it needs to compensate the smallness of the coupling to the Higgs. This is enforced by the data just as in the universal case. So, it is clear that we can alleviate this problem if we allow the Higgs coupling to become larger. This can be achieved in two ways. First, as we discuss in the next section in more detail, the coupling to the Higgs prefers to be larger when the Higgs mass increases. Therefore, if the Higgs were found to be heavy, smaller $b_R$ couplings would be required. We show this in the third and fourth columns of Table~\ref{table: Nonuniversal Comb}. We see that the coupling-to-mass ratio $G^b_\mathcal{B}$ is still greater than 1 for $M_H\leq 500$~GeV.

\begin{table}[h]
\begin{center}
{\small
\begin{tabular}{c | c c c c | c c }
\ctoprule
\crowcolor&\multicolumn{4}{c}{\greycell${\cal B}$}& \multicolumn{2}{c}{\greycell${\cal B}+{\cal B}^1$}\\
\crowcolor&Free&$G_{\cal B}^b\equiv1$&$M_H\!=\!200\!\units{GeV}$&$M_H\!=\!500\!\units{GeV}$&Free&$G_{\cal B}^b\equiv1$\\
\cmrule
$-\Delta \chi^2_\mt{min}$&$\phantom{-}8.2\phantom{00}$&$\phantom{-}2.7\phantom{00}$&$~\!14.1\phantom{0}$&$~\!47.7\phantom{0}$&$\phantom{-}8.2\phantom{0}$&$\phantom{-}8.2\phantom{0}$\\
Pull$[A_{FB}^b]$&$-0.5\phantom{00}$&$-2.5\phantom{00}$&$-0.4\phantom{0}$&$-0.4\phantom{0}$&$-0.5$\phantom{0}&$-0.5$\phantom{0}\\
\cmrule
$G_{\cal B}^b$~[TeV$^{-1}$]&$\phantom{-}6.4\phantom{00}$ &$\phantom{-}1\phantom{.000}$&$\phantom{-}3.8\phantom{0}$&$\phantom{-}2.4\phantom{0}$&$\phantom{-}3.2\phantom{0}$&$\phantom{-}1\phantom{.00}$\\
$G_{\cal B}^\phi$~[TeV$^{-1}$]&$\phantom{-}0.082$&$\phantom{-}0.078$&$\phantom{-}0.13$&$\phantom{-}0.19$&$\phantom{-}0.16$&$\phantom{-}0.53$\\
$G_{{\cal B}^1}^\phi$~\!\![TeV$^{-1}$]&-&-&-&-&$\phantom{-}0.20$&$\phantom{-}0.73$\\
\cbottomrule
\end{tabular}
\caption{ Effect of the SM singlets on the forward-backward asymmetry for the $b$ quark from the nonuniversal fit. The improvement in the $\chi^2$ for the cases of $M_H=200,500\units{GeV}$ is given with respect to the SM with the same values of the Higgs mass. The last two columns correspond to different points along a flat direction.
\label{table: Nonuniversal Comb}}
}
\end{center}
\end{table}

A more efficient way of reproducing the experimental asymmetry without too large couplings is to combine different new vectors. The only operator where the Higgs coupling enters quadratically is ${\cal O}_\phi^{(3)}$. While the coefficient of this operator is always positive when induced by a singlet neutral vector $\mathcal{B}$, the hypercharged vector $\mathcal{B}^1$ gives a negative contribution to it. Hence, if the theory contains a $\mathcal{B}$ and a $\mathcal{B}^1$, both contributions may cancel out\footnote{They cancel out automatically if the extension of the SM preserves custodial symmetry.}. Furthermore, the extra vector $\mathcal{B}^1$ gives no other observable effect in the fits if coupled only to the third family. In the last two columns of Table~\ref{table: Nonuniversal Comb}, we display the result of a global fit to a scenario with a neutral singlet $\mathcal{B}$ and a charged singlet $\mathcal{B}^1$, both coupled to the Higgs and to the third family of quarks. We see that in this case the coupling to $b_R$ can be made smaller than 1 at no cost in $\chi^2$. In fact, there is an almost flat direction for fixed values of the product $G^b_\mathcal{B} G^\phi_\mathcal{B}$ (and the correct $G^\phi_{\mathcal{B}^1}$ to counteract the effect of $G^\phi_\mathcal{B}$). 
\begin{figure}[!]
\begin{center}
\input{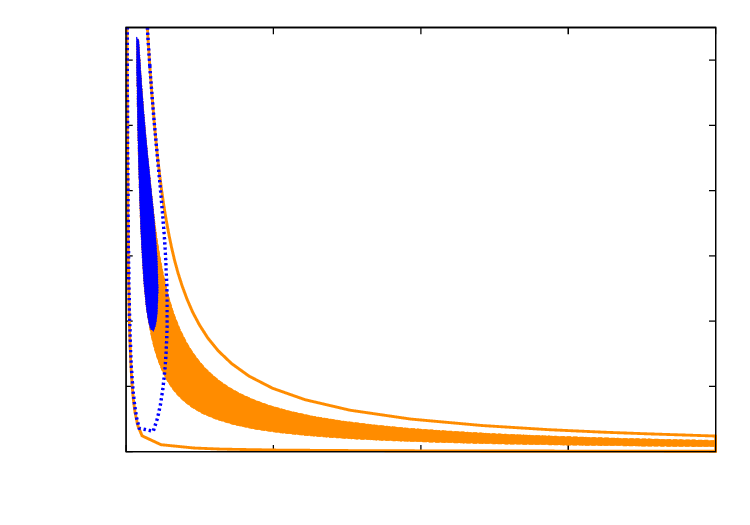}
\caption{Allowed regions of the ${\cal B}$ couplings to the RH bottom and to the Higgs at $1~\sigma$ (solid regions) and at $95\%$ C.L. (regions between lines) from the nonuniversal ${\cal B}$ fit (blue, dotted) and the ${\cal B}+{\cal B}^1$ fit (orange, solid).}
\label{fig: Minimal}
\end{center}
\end{figure}
The impact of including the second vector boson is also manifest in Figure~\ref{fig: Minimal}, where we plot the allowed values for the Higgs and $b_R$ couplings to the ${\cal B}$ singlet, with and without an additional ${\cal B}^1$ boson. We observe that introducing the charged singlet opens a new favored region in which $G^b_\mathcal{B}$ is smaller and $G^\phi_\mathcal{B}$ is larger. This very same mechanism is at work in the explicit extra-dimensional model in \cite{Djouadi:2006rk}.

\begin{figure}[!]
\begin{center}
\input{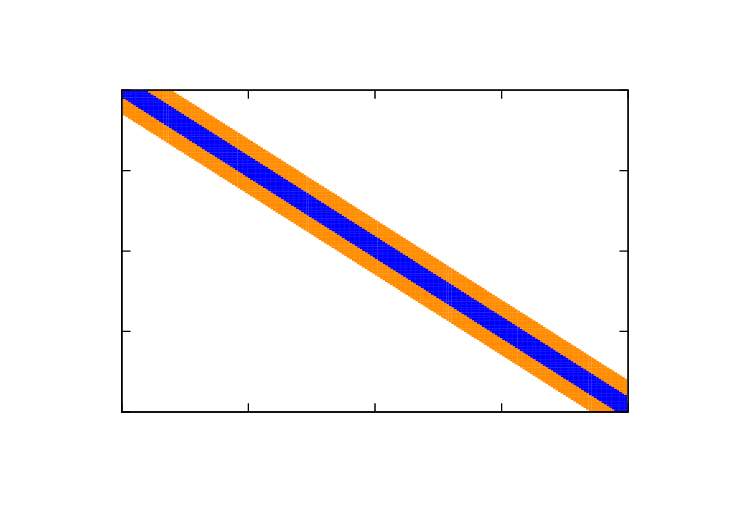}
\caption{Confidence regions at $1~\sigma$ (blue, dark)  and $95\%$ C.L. (orange, light) on the plane determined by the corrections to the $Z{\bar b}_Lb_L$ coming from a singlet ${\cal B}$ and a triplet ${\cal W}$.}
\label{fig:W0B0lH}
\end{center}
\end{figure}
In definite models, having small couplings of the new vectors to $b_L$ can be troublesome from a model building perspective, especially if the couplings to $b_R$ are large. This issue can be addressed by adding another type of extra vector that balances the effect of the singlet. One possibility is a triplet $\mathcal{W}$. This vector boson generates the operator ${\cal O}_{\phi q}^{(3)}$. Because the correction to the $Z \overline{b}_L b_L$ vertex is proportional to
$\alpha_{\phi q}^{(1)} + \alpha_{\phi q}^{(3)}$ and these coefficients do not have a definite sign, a cancellation is again possible. This mechanism is shown in Fig.~\ref{fig:W0B0lH}. It can be made natural if the $\mathcal{B}$ and $\mathcal{W}$ couplings are related by some symmetry~\cite{delAguila:2000aa}, as in
the custodial protection proposed in \cite{Agashe:2006at}. Note that this protection requires new fermions, which might modify the electroweak fits~\cite{Carena:2006bn}. We also point out that the correction to the $Z \overline{t}_L t_L$ vertex is proportional to $\alpha_{\phi q}^{(1)} - \alpha_{\phi q}^{(3)}$, so it can never be cancelled at the same time~\cite{delAguila:2000aa}. This has consequences for top physics at LHC~\cite{Berger:2009hi}.


\section{New vector bosons and the Higgs mass}\label{section:Higgs}

All the fits in the paper have been performed leaving the mass of the Higgs boson as a free parameter (unless otherwise indicated), and imposing the direct constraints from Higgs searches. In this section we focus on the implications of new vector bosons on the Higgs mass. 

LEP 2  and Tevatron  have put limits on the possible values of $M_H$. The LEP 2 experiments discard a light Higgs in a quite robust manner, with $M_H>114.4$ GeV at 95\% C.L.~\cite{Barate:2003sz}. A slight preference for values at around 116 GeV was also observed, due to the few Higgs-like events observed at the end of the LEP operation. The searches at Tevatron have been used to exclude the small window 163~GeV~$\le M_H\le 166$~GeV at 95 \% C.L.~\cite{Collaboration:2009je}, which is expected to be extended to $159$ GeV $\le M_H\le 168$ GeV~\footnote{It has been noted recently, however, that these exclusion limits should be reconsidered in light of the large theoretical uncertainties in the production cross sections of the Higgs at Tevatron~\cite{Baglio:2010um}.}. On the other hand, the global electroweak fit of the SM shows a preference for a light Higgs, $M_H= 101^{+32}_{-26}$ GeV. It is important to recall that these indirect limits result from averaging over partly inconsistent data. Related to this, if the $\sim 3~\sigma$ deviation in $A_{FB}^b$ were due to a systematic error, it should be removed from the fit, and the minimum would have a Higgs mass lower than the LEP lower bound, $M_H=73^{+28}_{-22}$~GeV. In this sense, there is a mild tension between indirect and direct limits~\cite{Chanowitz:2003hx}.

As we have seen before, $\mathcal{B}$ vector bosons generate the oblique operator ${\cal O}_\phi^{(3)}$ with a negative coefficient, and since
\begin{equation}
\Delta \rho = -\frac{\alpha_\phi^{(3)}}{2}\frac{v^2}{\Lambda^2},
\end{equation}
they give a positive contribution to the $\rho$ parameter. This has the right sign to neutralize the effect of increasing $M_H$ on $\rho$. In fact, $Z^\prime$ bosons have been used in the past to render a heavy Higgs consistent with EWPD~\cite{Peskin:2001rw}, and to release the tension with the LEP lower bound~\cite{Chanowitz:2008ix}. The other extra vectors that contribute to ${\cal O}_\phi^{(3)}$ are $\mathcal{B}^1$ and $\mathcal{W}^1$. The first one gives a contribution of opposite sign, so it favours smaller values of $M_H$. The contribution of the hypercharged triplet $\mathcal{W}^1$ has the same sign as for $\mathcal{B}$, and can be used to raise the allowed values $M_H$. It also has the virtue of not generating any other observable operator, which could worsen the quality of the fit. However, the appearance of this representation seems to require a rather contrived model building.

\begin{figure}[!]
\begingroup
  \makeatletter
  \providecommand\color[2][]{%
    \GenericError{(gnuplot) \space\space\space\@spaces}{%
      Package color not loaded in conjunction with
      terminal option `colourtext'%
    }{See the gnuplot documentation for explanation.%
    }{Either use 'blacktext' in gnuplot or load the package
      color.sty in LaTeX.}%
    \renewcommand\color[2][]{}%
  }%
  \providecommand\includegraphics[2][]{%
    \GenericError{(gnuplot) \space\space\space\@spaces}{%
      Package graphicx or graphics not loaded%
    }{See the gnuplot documentation for explanation.%
    }{The gnuplot epslatex terminal needs graphicx.sty or graphics.sty.}%
    \renewcommand\includegraphics[2][]{}%
  }%
  \providecommand\rotatebox[2]{#2}%
  \@ifundefined{ifGPcolor}{%
    \newif\ifGPcolor
    \GPcolortrue
  }{}%
  \@ifundefined{ifGPblacktext}{%
    \newif\ifGPblacktext
    \GPblacktexttrue
  }{}%
  \let\gplgaddtomacro\g@addto@macro
  \gdef\gplbacktexta{}%
  \gdef\gplfronttexta{}%
  \gdef\gplbacktextb{}%
  \gdef\gplfronttextb{}%
  \makeatother
  \ifGPblacktext
    \def\colorrgb#1{}%
    \def\colorgray#1{}%
  \else
    \ifGPcolor
      \def\colorrgb#1{\color[rgb]{#1}}%
      \def\colorgray#1{\color[gray]{#1}}%
      \expandafter\def\csname LTw\endcsname{\color{white}}%
      \expandafter\def\csname LTb\endcsname{\color{black}}%
      \expandafter\def\csname LTa\endcsname{\color{black}}%
      \expandafter\def\csname LT0\endcsname{\color[rgb]{1,0,0}}%
      \expandafter\def\csname LT1\endcsname{\color[rgb]{0,1,0}}%
      \expandafter\def\csname LT2\endcsname{\color[rgb]{0,0,1}}%
      \expandafter\def\csname LT3\endcsname{\color[rgb]{1,0,1}}%
      \expandafter\def\csname LT4\endcsname{\color[rgb]{0,1,1}}%
      \expandafter\def\csname LT5\endcsname{\color[rgb]{1,1,0}}%
      \expandafter\def\csname LT6\endcsname{\color[rgb]{0,0,0}}%
      \expandafter\def\csname LT7\endcsname{\color[rgb]{1,0.3,0}}%
      \expandafter\def\csname LT8\endcsname{\color[rgb]{0.5,0.5,0.5}}%
    \else
      \def\colorrgb#1{\color{black}}%
      \def\colorgray#1{\color[gray]{#1}}%
      \expandafter\def\csname LTw\endcsname{\color{white}}%
      \expandafter\def\csname LTb\endcsname{\color{black}}%
      \expandafter\def\csname LTa\endcsname{\color{black}}%
      \expandafter\def\csname LT0\endcsname{\color{black}}%
      \expandafter\def\csname LT1\endcsname{\color{black}}%
      \expandafter\def\csname LT2\endcsname{\color{black}}%
      \expandafter\def\csname LT3\endcsname{\color{black}}%
      \expandafter\def\csname LT4\endcsname{\color{black}}%
      \expandafter\def\csname LT5\endcsname{\color{black}}%
      \expandafter\def\csname LT6\endcsname{\color{black}}%
      \expandafter\def\csname LT7\endcsname{\color{black}}%
      \expandafter\def\csname LT8\endcsname{\color{black}}%
    \fi
  \fi
  \begin{tabular}{c c}
  \setlength{\unitlength}{0.0300bp}%
  \begin{picture}(7200.00,5040.00)(450,0)%
    \gplgaddtomacro\gplbacktexta{%
      \csname LTb\endcsname%
      \put(1210,1383){\makebox(0,0)[r]{\strut{} 155}}%
      \put(1210,2231){\makebox(0,0)[r]{\strut{} 160}}%
      \put(1210,3079){\makebox(0,0)[r]{\strut{} 165}}%
      \put(1210,3928){\makebox(0,0)[r]{\strut{} 170}}%
      \put(1210,4776){\makebox(0,0)[r]{\strut{} 175}}%
      \put(1592,484){\makebox(0,0){\strut{}116}}%
      \put(2899,484){\makebox(0,0){\strut{}200}}%
      \put(5102,484){\makebox(0,0){\strut{}500}}%
      \put(6769,484){\makebox(0,0){\strut{}1000}}%
      \put(440,2740){\rotatebox{90}{\makebox(0,0){\strut{}$\chi^2$}}}%
      \put(4106,154){\makebox(0,0){\strut{}$M_H$ [GeV]}}%
    }%
    \gplgaddtomacro\gplfronttexta{%
      \csname LTb\endcsname%
      \put(5883,4603){\makebox(0,0)[r]{\strut{}{\tiny SM}}}%
      \csname LTb\endcsname%
      \put(5883,4383){\makebox(0,0)[r]{\strut{}$\tinymath{{\cal W}^1}\!$}}%
      \csname LTb\endcsname%
      \put(5883,4163){\makebox(0,0)[r]{\strut{}$\tinymath{{\cal B}}~$}}%
    }%
    \gplbacktexta
    \put(0,0){\includegraphics[scale=0.6]{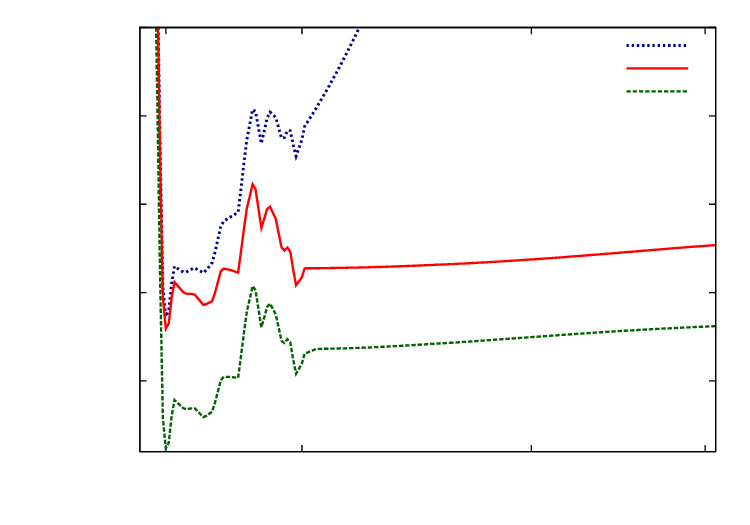}}%
    \gplfronttexta
  \end{picture}%
  &
  \setlength{\unitlength}{0.03875bp}%
  \begin{picture}(7200.00,5040.00)(770,520)%
    \gplgaddtomacro\gplbacktextb{%
    }%
    \gplgaddtomacro\gplfronttextb{%
      \csname LTb\endcsname%
      \put(1389,800){\makebox(0,0){\strut{}    116}}%
      \put(2539,800){\makebox(0,0){\strut{}200}}%
      \put(4476,800){\makebox(0,0){\strut{}500}}%
      \put(5941,800){\makebox(0,0){\strut{}1000}}%
      \put(3600,470){\makebox(0,0){\strut{}$M_H$ [GeV]}}%
      \put(998,1086){\makebox(0,0)[r]{\strut{} 0}}%
      \put(998,1527){\makebox(0,0)[r]{\strut{} 0.1}}%
      \put(998,1969){\makebox(0,0)[r]{\strut{} 0.2}}%
      \put(998,2410){\makebox(0,0)[r]{\strut{} 0.3}}%
      \put(998,2850){\makebox(0,0)[r]{\strut{} 0.4}}%
      \put(998,3291){\makebox(0,0)[r]{\strut{} 0.5}}%
      \put(998,3733){\makebox(0,0)[r]{\strut{} 0.6}}%
      \put(998,4174){\makebox(0,0)[r]{\strut{} 0.7}}%
      \put(404,2630){\rotatebox{90}{\makebox(0,0){\strut{}$G_{{\cal W}^1}^\phi$[TeV$^{-1}$]}}}%
    }%
    \gplbacktextb
    \put(0,0){\includegraphics[scale=0.775]{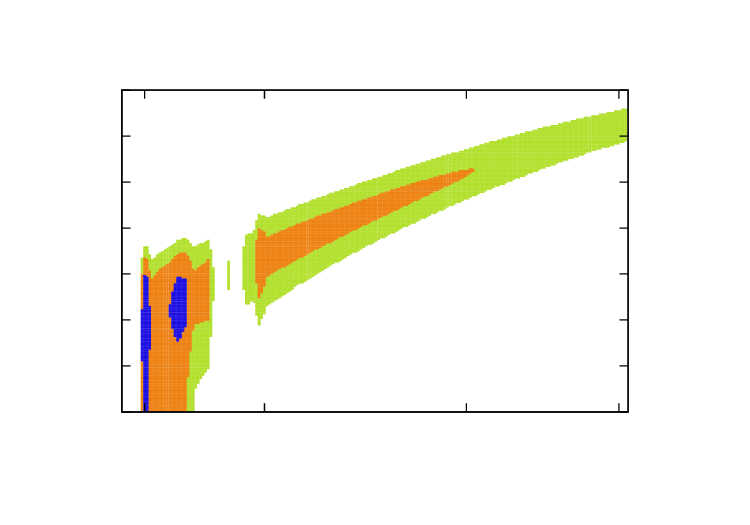}}%
    \gplfronttextb
  \end{picture}%
  \end{tabular}
\endgroup

\caption{{\em Left:\/} Minimum of the $\chi^2$ as a function of the Higgs mass for the SM fit, the ${\cal W}^1$ fit and the ${\cal B}$ fit. Higgs direct searches data are available up to $M_H=200\units{GeV}$. Above that value the effect of neutral-boson mixing flattens the curve. {\em Right:\/} From darker to lighter, confidence regions with $\Delta \chi^2\leq$ $2$ (blue), $4$ (orange) and $6$ ($95\%$ C.L.) (green), in the plane parametrized by the Higgs mass and the ${\cal W}^1$ coupling to the Higgs.}
\label{fig:Chi2vsmH}
\end{figure}

In Fig.~\ref{fig:Chi2vsmH} left, we plot the minimum of $\chi^2$ as a function of the Higgs mass in three cases: SM, one extra $\mathcal{B}$ and one extra $\mathcal{W}^1$. In all cases, we have used the information from direct searches. The couplings are family universal.
The effect of the ${\cal B}$ and the ${\cal W}^1$ vector bosons is apparent: they flatten the distribution when we go beyond the region disfavoured by the Tevatron searches. This allows to reach large values of $M_H$ with a low cost in $\chi^2$, as compared to the SM case. We also observe that, in the case of ${\cal B}$, the larger number of free parameters is used to lower the $\chi^2$ with respect to the ${\cal W}^1$ case. This effect persists in the flat region, thanks to an improvement in the prediction for LEP 2 hadronic cross sections and for $A_{FB}^{b}$ at the $Z$ pole, which we have already discussed in Section~\ref{Limitsonevector}.

The coefficient $\alpha_\phi^{(3)}$ is proportional to the square of the coupling of the vector bosons to the scalar doublet. Therefore, this coupling must increase when the Higgs mass gets larger. This correlation is shown, for the $\mathcal{W}^1$ boson, in Fig.~\ref{fig:Chi2vsmH} right, where we display several confidence regions in the $M_H$ - $G_{\mathcal{W}^1}^\phi$ plane.


\section{Conclusions}
\label{Conclusions_NewVectors}

In this paper, we have studied general extra particles of spin 1, concentrating on their effects in EWPD. Our results are relevant for model-independent searches and also for explicit models. We have classified all the possibilities that may produce observable effects, and have written the most general couplings consistent with the SM gauge symmetry that are linear in the new fields. We have then derived, to dimension six, the effective Lagrangian that describes the effect of the new vector bosons at energies smaller than their masses. The result is displayed in Tables \ref{Table: B0Table} to \ref{Table: Y5Table}. Our analysis includes the cases of $Z^\prime$ and $W^\prime$ particles, vector leptoquarks and a few vector particles that, to the best of our knowledge, had not been considered previously in the literature. Some of the vector bosons we have studied couple to quarks only, so they are not constrained by EWPD. However, they may in principle be singly produced and seen as resonances at Tevatron and LHC~\cite{delAguila:2007ua,Godfrey:2008vf}.

We have performed model-independent electroweak fits of the different types of new vectors, keeping the new couplings and masses as free parameters.  We have studied scenarios with both family universal and nonuniversal couplings. The main results are collected in Table \ref{table_OneVLimits}. In the fits, low-energy and LEP~2 data are crucial to constrain the different four-fermion interactions that appear upon the integration of the new particles. This translates into limits on the couplings to fermions that, unlike the ones from the $Z$-pole observables, cannot be avoided by making the couplings to the Higgs very small.

In a rough way, we observe that for all vector multiplets the purely leptonic couplings of the extra vectors are constrained to be pretty small, while the limits on quark and leptoquark couplings are weaker. Moreover, in some cases the data show a preference for pretty large quark couplings, driven by the SM discrepancies with the bottom forward-backward asymmetry at the $Z$-pole and with the hadronic cross sections at LEP~2. Finally, small Higgs couplings are also preferred, at least in the fits with just one type of extra vector.

We have also examined the implications of including several of these extra vector bosons at once, and looked for possible cancellations that may relax the electroweak limits. In particular, we have shown that a vector-boson solution to the bottom forward-backward anomaly is possible in a nonuniversal scenario with extra neutral and charged singlets. In this case, the charged vectors are needed to keep the couplings to the RH bottom in the perturbative regime.

An important variable in all the electroweak fits is the mass of the Higgs boson, which enters logarithmically through radiative corrections. We have kept it as a free parameter, and imposed the constraints from direct Higgs searches at LEP and Tevatron. As in the SM, the fits with new vector bosons favor a light Higgs, close to the lower bound of 114~GeV. Nevertheless, there are two types of vector bosons (the neutral singlet and the fermiophobic triplet) that can make the electroweak data consistent with a heavy Higgs, as shown in  Fig.~\ref{fig:Chi2vsmH}. 

The limits we have obtained are somewhat complementary to the ones from Tevatron. Both restrict the discovery potential of LHC. To simplify the following discussion, all limits on the masses of the heavy vector bosons are given with the assumption that the nonvanishing couplings to SM fields have the same strength as the massive gauge bosons in the SM ($\sim 0.2$ for leptonic neutral currents). At hadron colliders, the new vector bosons can be seen as resonances, if kinematically allowed, when the cross section is high enough to distinguish a bump above the SM backgrounds. The most efficient process is Drell-Yan, which requires trilinear couplings of an extra neutral vector ($Z^\prime$) with quarks and with leptons. These two kinds of couplings exist only for the singlet $\mathcal{B}$ and for the neutral component of the triplet $\mathcal{W}$. The lower limits from Tevatron on the mass of neutral vector bosons, coupled to leptons and quarks, are around 1~TeV~\cite{Aaltonen:2008ah}, and the LHC discovery reach is near 5~TeV, assuming 14 TeV operation and an integrated luminosity of 100 fb$^{-1}$~\cite{Diener:2009vq}. For $\sqrt{s}=7\units{TeV}$ and an integrated luminosity of 100 pb$^{-1}$ it should be possible to put bounds above 1 TeV. The limits from precision tests are in general around this value  (see Table~\ref{table_OneVLimits} and \ref{table: ZpLimits}). On the other hand, charged vectors ($W^\prime$) are best seen as resonances produced by quark interactions and decaying into a charged lepton and a neutrino. This is only possible for the representation $\mathcal{W}$ and, if there were sufficiently light RH neutrinos, $\mathcal{B}^1$~\cite{Gninenko:2006br}. Tevatron puts limits around 1 TeV~\cite{Aaltonen:2009qu}, while LHC could discover a $W^\prime$ with mass up to 4 TeV for $\sqrt{s}=$14 TeV and a few fb$^{-1}$ of integrated luminosity~\cite{Aad:2009wy}. The EWPD (for ${\cal W}$, with leptonic coupling $g\approx 0.66$) give a bound around 2.5 TeV.

It is also possible that the new vectors be leptophobic. This is automatic for many of the representations considered here. In this case, the most relevant decay mode is $V \to jj$. The Tevatron limits~\cite{Aaltonen:2008dn} and the LHC reach  \cite{Weiglein:2004hn} are somewhat smaller than when the vectors couple to leptons. Since the quark couplings and the masses of the extra vectors are unconstrained by EWPD, the available parameter space for LHC discovery is pretty large in this case.

Some of the representations we have studied couple leptons to quarks.  If there is enough available phase space, these vector leptoquarks can be pair produced at hadron colliders via renormalizable coupling to gluons in the covariant kinetic term, \refeq{VLag}. This interaction does not contribute to the dimension-six effective Lagrangian, so it is not seen, in our approximation, by EWPD. Single production through trilinear couplings, which are constrained by EWPD, is also possible~\cite{Hewett:1987yg}. The lower limits on leptoquark masses from Tevatron are around 250~GeV \cite{Acosta:2005ge}. On the other hand, leptoquarks coupled to the first family could be singly produced at HERA via trilinear couplings, and their nonobservation puts a lower bound of 290 GeV on their mass \cite{Chekanov:2003af}. The limits on trilinear couplings that we have derived here, making use of low-energy and LEP~2 data, depend a lot on the particular representation of the vector leptoquark, and on the flavor structure of the couplings. They range from $\sim 70$ GeV to $830$ GeV at $95\%$ C.L. assuming a generic coupling $g_V^{\psi_1 \psi_2}= 0.1$ (see Table \ref{table_OneVLimits}) .

The vectors bosons $\mathcal{L}$ only have leptonic couplings and cannot be seen at hadron colliders. At the ILC or muon colliders, they would only appear in the $t$ channel, since they carry two lepton-number units. On the other hand, the vectors $\mathcal{W}^1$ only interact with fermions through their mixing with the SM gauge bosons. Even if this mixing can be relatively large for a heavy Higgs, this vector boson is basically invisible to hadron colliders. We should finally mention that, in principle, other exotic vector bosons in representations not considered here could exist. They cannot be singly produced at colliders, nor contribute significantly to EWPD. However, if they were light enough, they could be pair produced (as in some composite vector models~\cite{Barbieri:2009tx,Hernandez:2010iu}). These exotic vector bosons, if coupled to gluons, would be seen as jets generated by their radiation.

\section*{Acknowledgements}
It is a pleasure to thank Juan Antonio Aguilar Saavedra, Nuno Castro and Jos\'e Santiago for useful conversations, and Paul Langacker for discussions and a critical reading of the manuscript.
This work has been partially supported by 
MICINN 
(FPA2006-05294), and by Junta de Andaluc\'{\i}a (FQM 101 and FQM 437),
and by the European Community's Marie-Curie Research Training
Network under contract MRTN-CT-2006-035505 ``Tools and Precision
Calculations for Physics Discoveries at Colliders''.

\newpage

\appendix

\section{Basis of dimension-six operators}
\label{app: Operators}

As indicated in Section 3, we basically use the basis of dimension-six operators presented in~\cite{Buchmuller:1985jz}. In Table~\ref{d6OpSM} we collect those operators that arise from the integration of any of the vector bosons in Table~\ref{table:newvectors}. The table contains some operators that were not included in~\cite{Buchmuller:1985jz} because they violate $B$ and $L$ (but preserve $B-L$).

All the interactions that can arise in the extensions of the SM considered here can be classified as:
\begin{enumerate}
{\item Four-fermion interactions, with different combinations of chiralities (LLLL, RRRR and LRRL)}.
{\item Operators built exclusively with scalars ({\it S operators}). For notational purposes we have also introduced the dimension-four operator $(\phi^\dagger \phi)^2$, which receives contributions from some extra vectors.}
{\item Operators made of scalars, gauge bosons (or derivatives) and fermion fields ({\it SVF operators}).}
{\item Operator with only scalars and fermions ({\it SF operators}).}
{\item Operators with only scalars and gauge bosons (or derivatives) ({\it Oblique operators}). }
\end{enumerate}
Even though it does not appear in the integration of the extra vectors at tree-level when we write the results in our basis, we have also introduced the oblique operator ${\cal O}_{\WB}$, since it is mentioned in the discussion.

\begin{table}[p]
\begin{center}
\begin{tabular}{c c l c c l }
\ctoprule
&Operator& Notation& &Operator&Notation\\
\cmidrule{2-3}\cmidrule{5-6}
\multirow{4}{*}{\vtext{LLLL}}&$\frac 1 2 \left(\overline{ l_L} \gamma_\mu l_L\right)\left(\overline{l_L} \gamma^\mu l_L\right)$&${\cal O}_{ll}^{(1)}$& &$\frac 1 2 \left(\overline{ l_L} \gamma_\mu\sigma_a l_L\right)\left(\overline{ l_L} \gamma^\mu\sigma_a l_L\right)$&${\cal O}_{ll}^{(3)}$\\
&$\frac 1 2 \left(\overline{q_L} \gamma_\mu q_L\right)\left(\overline{q_L} \gamma^\mu q_L\right)$&${\cal O}_{qq}^{(1,1)}$&&$\frac 1 2 \left(\overline{q_L} \gamma_\mu\sigma_a q_L\right)\left(\overline{q_L} \gamma^\mu\sigma_a q_L\right)$&${\cal O}_{qq}^{(1,3)}$\\
&$\left(\overline{l_L}\gamma_\mu l_L\right)\left(\overline{q_L} \gamma^\mu q_L\right)$&${\cal O}_{lq}^{(1)}$&&$\left(\overline{l_L} \gamma_\mu\sigma_a l_L\right)\left(\overline{q_L} \gamma^\mu\sigma_a q_L\right)$&${\cal O}_{lq}^{(3)}$\\
&$\frac 12\left(\overline{q_L}\gamma_\mu \lambda_A q_L\right)\left(\overline{q_L} \gamma^\mu\lambda_A q_L\right)$&${\cal O}_{qq}^{(8,1)}$&&$\frac 12 \left(\overline{q_L} \gamma_\mu\sigma_a\lambda_A q_L\right)\left(\overline{q_L} \gamma^\mu\sigma_a \lambda_A q_L\right)$&${\cal O}_{qq}^{(8,3)}$\\[0.1cm]
\cmrule
\multirow{6}{*}{\vtext{RRRR}}&$$&$$&&$\frac 1 2 \left(\overline{e_R} \gamma_\mu e_R\right)\left(\overline{e_R} \gamma^\mu e_R\right)$&${\cal O}_{ee}$\\
&$\frac 1 2 \left(\overline{u_R} \gamma_\mu u_R\right)\left(\overline{u_R} \gamma^\mu u_R\right)$&${\cal O}_{uu}^{(1)}$&&$\frac 1 2 \left(\overline{ d_R} \gamma_\mu d_R\right)\left(\overline{ d_R} \gamma^\mu d_R\right)$&${\cal O}_{dd}^{(1)}$\\
&$\left(\overline{e_R} \gamma_\mu e_R\right)\left(\overline{ u_R} \gamma^\mu u_R\right)$&${\cal O}_{eu}$&&$\left(\overline{e_R} \gamma_\mu e_R\right)\left(\overline{ d_R} \gamma^\mu d_R\right)$&${\cal O}_{ed}$\\
&$\left(\overline{u_R} \gamma_\mu u_R\right)\left(\overline{d_R} \gamma^\mu d_R\right)$&${\cal O}_{ud}^{(1)}$&&$ $& $$\\
&$\frac 12\left(\overline{u_R} \gamma_\mu\lambda_A u_R\right)\left(\overline{ u_R} \gamma^\mu \lambda_Au_R\right)$&${\cal O}_{uu}^{(8)}$&&$\frac 12 \left(\overline{d_R} \gamma_\mu \lambda_A d_R\right)\left(\overline{ d_R} \gamma^\mu \lambda_A d_R\right)$&${\cal O}_{dd}^{(8)}$\\
&$\left(\overline{u_R} \gamma_\mu \lambda_A u_R\right)\left(\overline{d_R} \gamma^\mu\lambda_A d_R\right)$&${\cal O}_{ud}^{(8)}$&&$ $& $$\\[0.1cm]
\cmrule
\multirow{5}{*}{\vtext{LRRL}}&$\left(\overline{l_L} e_R\right)\left(\overline{e_R} l_L\right)$&${\cal O}_{le}$&&$\left(\overline{q_L} e_R\right)\left(\overline{e_R} q_L\right)$&${\cal O}_{qe}$\\
&$\left(\overline{l_L} u_R\right)\left(\overline{u_R} l_L\right)$&${\cal O}_{lu}$&&$\left(\overline{l_L} d_R\right)\left(\overline{d_R} l_L\right)$&${\cal O}_{ld}$\\
&$\left(\overline{q_L} u_R\right)\left(\overline{u_R} q_L\right)$&${\cal O}_{qu}^{(1)}$&&$\left(\overline{q_L} d_R\right)\left(\overline{d_R} q_L\right)$&${\cal O}_{qd}^{(1)}$\\
&$\left(\overline{l_L} e_R\right)\left(\overline{d_R} q_L\right)$&${\cal O}_{qde}$&& &\\
&$\left(\overline{q_L}\lambda_A u_R\right)\left(\overline{u_R}\lambda_A q_L\right)$&${\cal O}_{qu}^{(8)}$&&$\left(\overline{q_L}\lambda_A d_R\right)\left(\overline{d_R}\lambda_A q_L\right)$&${\cal O}_{qd}^{(8)}$\\
\multirow{1}{*}{\vtext{B-L}}&$\epsilon_{ABC}\left(\overline{l_L}i\sigma_2q_L^{c\ A}\right)\left(\overline{d_R^{B}}u_R^{c\ C}\right)$&${\cal O}_{lqdu}$&&$\epsilon_{ABC}\left(\overline{q_L^{B}}i\sigma_2 q_L^{c\ C}\right)\left(\overline{e_R}u_R^{c\ A}\right)$&
${\cal O}_{qqeu}$\\[0.1cm]
\cmrule
\vtext{ }&$\left(\phi^\dagger \phi\right)^2$&${\cal O}_{\phi 4}$&&$\frac 1 3 \left(\phi^{\dagger} \phi\right)^3$&${\cal O}_{\phi 6}$\\[0.1cm]
\cmrule
\multirow{5}{*}{\vtext{SVF}}&$\left(\phi^{\dagger}iD_\mu \phi\right)\left(\overline{l_L} \gamma^\mu l_L\right)$&${\cal O}_{\phi l}^{(1)}$&&$\left(\phi^{\dagger}\sigma_a iD_\mu \phi\right)\left(\overline{l_L} \gamma^\mu\sigma_a l_L\right)$&${\cal O}_{\phi l}^{(3)}$\\
&$\left(\phi^{\dagger}iD_\mu \phi\right)\left(\overline{e_R} \gamma^\mu e_R\right)$&${\cal O}_{\phi e}^{(1)}$&&$$&$$\\
&$\left(\phi^{\dagger}iD_\mu \phi\right)\left(\overline{q_L} \gamma^\mu q_L\right)$&${\cal O}_{\phi q}^{(1)}$&&$\left(\phi^{\dagger}\sigma_a iD_\mu \phi\right)\left(\overline{q_L} \gamma^\mu\sigma_a q_L\right)$&${\cal O}_{\phi q}^{(3)}$\\
&$\left(\phi^{\dagger}iD_\mu \phi\right)\left(\overline{u_R} \gamma^\mu u_R\right)$&${\cal O}_{\phi u}^{(1)}$&&$\left(\phi^{\dagger}iD_\mu \phi\right)\left(\overline{d_R} \gamma^\mu d_R\right)$&${\cal O}_{\phi d}^{(1)}$\\
&$\left(\phi^{T}i\sigma_2 iD_\mu \phi\right)\left(\overline{u_R} \gamma^\mu d_R\right)$&${\cal O}_{\phi ud}$&&$ $&$ $\\
\cmrule
\multirow{2}{*}{\vtext{SF}}&$\left(\phi^{\dagger}\phi\right)\left(\overline{l_L}~\!\phi~\!e_R\right)$&${\cal O}_{e \phi }$&&$$&$$\\
&$\left(\phi^{\dagger}\phi\right)\left(\overline{q_L}~\!{\tilde \phi }~\!u_R\right)$&${\cal O}_{u \phi }$&&$\left(\phi^{\dagger}\phi\right)\left(\overline{q_L}~\!\phi~\!d_R\right)$&${\cal O}_{d \phi }$\\[0.1cm]
\cmrule
\multirow{2}{*}{\vtext{Oblique\!\!}}&$\phi^{\dagger}\phi\left(D^\mu \phi\right)^{\dagger}D_\mu \phi$&${\cal O}_{\phi}^{(1)}$&&$\left(\phi^{\dagger}D_\mu \phi\right)(\left(D^\mu \phi\right)^{\dagger}\phi)$&${\cal O}_{\phi}^{(3)}$\\
&$ \phi^\dagger\sigma_a \phi~W^a_{\mu\nu}B^{\mu\nu} $&${\cal O}_{\WB}$&&&\\[0.1cm]
\cbottomrule
\end{tabular}
\caption{Dimension-six operators arising from the integration of the heavy vector bosons considered in the text. We also include the dimension-four operator ${\cal O}_{\phi 4}$ and the dimension-six operator ${\cal O}_{\WB}$ for notational purposes.\label{d6OpSM}}
\end{center}
\end{table}
%

~\newpage

\section{Operator coefficients in the effective Lagrangian}
\label{app: NV_OpCoeff}
Here, we collect the results from the integration of each of the extra vector fields in Table~\ref{table:newvectors}. We give in Tables~\ref{Table: B0Table} to \ref{Table: Y5Table} the corresponding contributions to the dimension-six operator coefficients in the effective Lagrangian. The explicit expressions for the currents coupled to the different new vectors are also written. We generically write LH and RH fermions as $F=l_L,q_L$ and $f=e_R,u_R,d_R$, respectively, and use $\psi$ to denote any SM fermion. Unless otherwise stated in the tables, these generic symbols run over all the possibilities. As stressed in Section~\ref{GEVB}, in order to apply our results the heavy vectors must be in the basis with diagonal mass and kinetic terms. Then, the contributions from several extra vectors are summed independently in the coefficients.

For only one of the SM replicas ${\cal B}$, ${\cal W}$ and ${\cal G}$, we can easily recover the effect of kinetic mixing with the SM fields. Since the rescaling of the heavy vector field ${\cal A}$ in \refeq{redefinition} affects in the same way the current $J^{\cal A}$ and the heavy mass $M_{\cal A}$,  we only need to perform the following replacements in the formulas in Tables~\ref{Table: B0Table}, \ref{Table: W0Table} and \ref{Table: G0Table}:

\bea
(g_{\cal B}^{\psi,\phi})_{ij}&\rightarrow&(g_{\cal B}^{\psi,\phi})_{ij} + g^\prime g_{\cal B}^B Y_{\psi,\phi}\delta_{ij}\nn
(g_{\cal W}^{F,\phi})_{ij}&\rightarrow&(g_{\cal W}^{F,\phi})_{ij} + g g_{\cal W}^W \delta_{ij}\label{App_replacement}\\
(g_{\cal G}^{\psi})_{ij}&\rightarrow&(g_{\cal G}^{\psi})_{ij} + g_s g_{\cal G}^G \delta_{ij}\nonumber
\eea
%

\begin{table}[p]
\begin{center}
{\small
\begin{tabular*}{\textwidth}{c l c c l}
\ctoprule
\multicolumn{5}{c}{\Bfmath{{\cal B}_\mu \sim \left(1,1\right)_0}}\\
\multicolumn{5}{c}{}\\[-0.2cm]
\multicolumn{5}{c}{$\!J^{\cal B}_\mu=(g_{\cal B}^l)_{ij}\overline{l_L^i}\gamma_\mu l_L^j+(g_{\cal B}^q)_{ij}\overline{q_L^i}\gamma_\mu q_L^i+(g_{\cal B}^e)_{ij}\overline{e_R^i}\gamma_\mu e_R^j+(g_{\cal B}^u)_{ij}\overline{u_R^i}\gamma_\mu u_R^j+(g_{\cal B}^d)_{ij}\overline{d_R^i}\gamma_\mu d_R^j+$}\\
\multicolumn{5}{l}{$\phantom{J^{\cal B}_\mu=~}+(g_{\cal B}^\phi~\phi^{\dagger}iD_\mu \phi+\!\hc\!)$}\\
\midrule
\multicolumn{5}{c}{ }\\[-0.3cm]
\multicolumn{5}{c}{{\bf Four-Fermion Operators}}\\
$ $&$ $&$ $&$ $&\\[-0.3cm]
\multicolumn{1}{l}{$\bullet~${\bf LLLL}}& &$\phantom{~~Space~~}$&\multicolumn{1}{l}{$\bullet~${\bf RRRR}}& \\
$ $&$ $&$ $&$ $&\\[-0.3cm]
$\frac{\left(\alpha_{FF^\prime}^{(1(,1))}\right)_{ijkl}}{\Lambda^2}=$&$\!\!\!\!-\frac{\left(g_{\cal B}^F\right)_{ij}\left(g_{\cal B}^{F^\prime}\right)_{kl}}{M_{\cal B}^2}$& &
$\frac{\left(\alpha_{ff^\prime}^{(1)}\right)_{ijkl}}{\Lambda^2}=$&$\!\!\!\!-\frac{\left(g_{\cal B}^{f}\right)_{ij}\left(g_{\cal B}^{f^\prime}\right)_{kl}+\left(g_{\cal B}^{f}\right)_{il}\left(g_{\cal B}^{f^\prime}\right)_{kj}\delta_{ff',ee}}{\left(1+\delta_{ff',ee}\right)M_{\cal B}^2}$\\
$ $&$ $&$ $&$ $&\\[-0.3cm]
$ $&$ $&$ $&$ $&\\[-0.3cm]
\multicolumn{1}{l}{$\bullet~${\bf LRRL}}& & & & \\
$ $&$ $&$ $&$ $&\\[-0.3cm]
$\frac{\left(\alpha_{Ff}\right)_{ijkl}}{\Lambda^2}=$&$\!\!\!\!\frac{2\left(g_{\cal B}^F\right)_{il}\left(g_{\cal B}^f\right)_{kj}}{M_{\cal B}^2}$&$\!\!\!\!\!\scriptmath{\left(Ff=le,lu,ld,qe\right)}$&$ $&\\
$ $&$ $&$ $&$ $&\\[-0.3cm]
$\frac{\left(\alpha_{qf}^{(1)}\right)_{ijkl}}{\Lambda^2}=$&$\!\!\!\!\frac{2\left(g_{\cal B}^q\right)_{il}\left(g_{\cal B}^{f}\right)_{kj}}{3M_{\cal B}^2}$& &$ $&$ $\\
$ $&$ $&$\!\!\!\!\!\scriptmath{\left(f=u,d\right)}$&$ $&\\[-0.3cm]
$\frac{\left(\alpha_{qf}^{(8)}\right)_{ijkl}}{\Lambda^2}=$&$\!\!\!\!\frac{\left(g_{\cal B}^q\right)_{il}\left(g_{\cal B}^{f}\right)_{kj}}{M_{\cal B}^2}$&$ $&$ $&\\
$ $&$ $&$ $&$ $&\\[-0.3cm]
\cmrule
\multicolumn{2}{c}{{\bf SVF and SF Operators}}& &\multicolumn{2}{c}{{\bf Oblique Operators}}\\
\multicolumn{5}{c}{ }\\[-0.3cm]
$\frac{\left(\alpha_{\phi \psi}^{(1)}\right)_{ij}}{\Lambda^2}=$&$\!\!\!\!\!\!\!\!\!\!-\frac{\left(g_{\cal B}^\psi\right)_{ij}g_{\cal B}^\phi}{M_{\cal B}^2}$& &
$\frac{\alpha_{\phi}^{(1)}}{\Lambda^2}=$&$\!\!\!\!\!\!\!\!\!\!\!\!-\frac{\Re{\left(g_{\cal B}^\phi\right)^2}}{M_{\cal B}^2}$\\
$ $&$ $&$ $&$ $&\\[-0.3cm]
$\frac{\left(\alpha_{u\phi}\right)_{ij}}{\Lambda^2}=$&$\!\!\!\!\!\!\!\!\!\!\frac{\left(g_{\cal B}^\phi\right)^2}{2M_{\cal B}^2}V_{ij}^\dagger y_{jj}^u$& &
$\frac{\alpha_{\phi}^{(3)}}{\Lambda^2}=$&$\!\!\!\!\!\!\!\!\!\!\!\!-\frac{2\Re{g_{\cal B}^\phi}^2}{M_{\cal B}^2}$\\
$ $&$ $&$ $&$ $&\\[-0.3cm]
$\frac{\left(\alpha_{f\phi}\right)_{ij}}{\Lambda^2}=$&$\!\!\!\!\!\!\!\!\!\!\frac{\left(\alpha_{u\phi}^\dagger\right)_{ij}}{\Lambda^2}\frac{y^{f}_{ii}\delta_{ij}}{V_{ij}y^u_{jj}}$& &
$\frac{\alpha_{\phi 6}}{\Lambda^2}=$&$\!\!\!\!\!\!\!\!\!\!\!\!\phantom{+}\frac{6\lambda_\phi\Re{\left(g_{\cal B}^\phi\right)^2}}{M_{\cal B}^2}$\\[-0.2cm]
$ $&$\!\!\!\!\!\!\!\!\!\!\!\!\!\scriptmath{\left(f=e,d\right)}$&$ $&$ $&\\[-0.3cm]
$ $&$ $& &$\frac{\alpha_{\phi 4}}{\Lambda^2}=$&$\!\!\!\!\!\!\!\!\!\!\!\!-\frac{\mu_\phi^2}{6\lambda_\phi}\frac{\alpha_{\phi 6}}{\Lambda^2}$\\
$ $&$ $&$ $&$ $&\\[-0.3cm]
\cbottomrule
\end{tabular*}
\caption{Operators arising from the integration of a ${\cal B}$ vector field.}\label{Table: B0Table}
}
\end{center}
\end{table}
%

\begin{table}[p]
\begin{center}
{\small
\begin{tabular}{c l c c l}
\ctoprule
\multicolumn{5}{c}{\Bfmath{{\cal W}_\mu \sim \left(1,\Adj \right)_0}}\\
\multicolumn{5}{c}{ }\\[-0.2cm]
\multicolumn{5}{c}{$J^{\cal W}_{a\ \mu}=(g_{\cal W}^l)_{ij}\overline{l_L^i}\gamma_\mu \frac {\sigma_a}{2}l_L^j+(g_{\cal W}^q)_{ij}\overline{q_L^i}\gamma_\mu \frac {\sigma_a}{2}q_L^j+(g_{\cal W}^\phi~\phi^{\dagger}\frac {\sigma_a}{2}iD_\mu \phi+\hc)$}\\
\midrule
\multicolumn{5}{c}{ }\\[-0.3cm]
\multicolumn{5}{c}{{\bf Four-Fermion Operators}}\\
$ $&$ $&$ $&$ $&\\[-0.3cm]
\multicolumn{1}{l}{$\bullet~${\bf LLLL}}& &$\phantom{~~Space~~}$& & \\
$ $&$ $&$ $&$ $&\\[-0.3cm]
$\frac{\left(\alpha_{FF^\prime}^{((1),3)}\right)_{ijkl}}{\Lambda^2}=$&$\!\!\!\!-\frac{\left(g_{\cal W}^F\right)_{ij}\left(g_{\cal W}^{F^\prime}\right)_{kl}}{4M_{\cal W}^2}$& & & \\
$ $&$ $&$ $&$ $&\\[-0.3cm]
\cmrule
\multicolumn{2}{c}{{\bf SVF and SF Operators}}& &\multicolumn{2}{c}{{\bf Oblique Operators}}\\
\multicolumn{5}{c}{ }\\[-0.3cm]
$\frac{\left(\alpha_{\phi F}^{(3)}\right)_{ij}}{\Lambda^2}=$&$\!\!\!\!\!\!\!\!\!-\frac{\left(g_{\cal W}^F\right)_{ij}g_{\cal W}^\phi}{4M_{\cal W}^2}$& &
$\frac{\alpha_{\phi}^{(1)}}{\Lambda^2}=$&$\!\!\!\!-\frac{\Re{\left(g_{\cal W}^\phi\right)^2}+2\left|g_{\cal W}^\phi\right|^2}{4M_{\cal W}^2}$\\
$ $&$ $&$ $&$ $&\\[-0.3cm]
$\frac{\left(\alpha_{u\phi}\right)_{ij}}{\Lambda^2}=$&$\!\!\!\!\!\!\!\!\!\frac{\left(g_{\cal W}^\phi\right)^2}{8M_{\cal W}^2}V_{ij}^\dagger y_{jj}^u$& &
$\frac{\alpha_{\phi}^{(3)}}{\Lambda^2}=$&$\!\!\!\!\frac{\Im{g_{\cal W}^\phi}^2}{2M_{\cal W}^2}$\\
$ $&$ $&$ $&$ $&\\[-0.3cm]
$\frac{\left(\alpha_{f\phi}\right)_{ij}}{\Lambda^2}=$&$\!\!\!\!\!\!\!\!\!\frac{\left(\alpha_{u\phi}^\dagger\right)_{ij}}{\Lambda^2}\frac{y^{f}_{ii}\delta_{ij}}{V_{ij}y^u_{jj}}$& &
$\frac{\alpha_{\phi 6}}{\Lambda^2}=$&$\!\!\!\!6\lambda_\phi \frac{\Re{\left(g_{\cal W}^\phi\right)^2}}{4M_{\cal W}^2}$\\
$ $&$\!\!\!\!\!\!\!\!\!\!\!\scriptmath{\left(f=e,d\right)}$&$ $&$ $&\\[-0.3cm]
&&&$\frac{\alpha_{\phi 4}}{\Lambda^2}=$&$\!\!\!\!\!\!-\frac{\mu_\phi^2}{6\lambda_\phi}\frac{\alpha_{\phi 6}}{\Lambda^2}$\\[-0.2cm]
$ $&$ $&$ $&$ $&\\
\cbottomrule
\end{tabular}
\caption{Operators arising from the integration of a ${\cal W}$ vector field.}\label{Table: W0Table}
}
\end{center}
\end{table}
%

\begin{table}[p]
\begin{center}
{\small
\begin{tabular}{c l c c l}
\ctoprule
\multicolumn{5}{c}{\Bfmath{{\cal G}_\mu \sim \left(\Adj,1 \right)_0}}\\
\multicolumn{5}{c}{ }\\[-0.2cm]
\multicolumn{5}{c}{$J^{\cal G}_{A\ \mu}=(g_{\cal G}^q)_{ij}\overline{q_L^i}\gamma_\mu \frac {\lambda_A}{2}q_L^j+(g_{\cal G}^u)_{ij}\overline{u_R^i}\gamma_\mu \frac {\lambda_A}{2} u_R^j+(g_{\cal G}^d)_{ij}\overline{d_R^i}\gamma_\mu \frac {\lambda_A}{2} d_R^j$}\\
\midrule
\multicolumn{5}{c}{ }\\[-0.3cm]
\multicolumn{5}{c}{{\bf Four-Fermion Operators}}\\
$ $&$ $&$ $&$ $&\\[-0.3cm]
\multicolumn{1}{l}{$\bullet~${\bf LLLL}}&&$\phantom{~~Space~~}$&\multicolumn{1}{l}{$\bullet~${\bf RRRR}}& \\
$ $&$ $&$ $&$ $&\\[-0.3cm]
$\frac{\left(\alpha_{qq}^{(8,1)}\right)_{ijkl}}{\Lambda^2}=$&$\!\!\!\!-\frac{\left(g_{\cal G}^q\right)_{ij}\left(g_{\cal G}^q\right)_{kl}}{4M_{\cal G}^2} $& &
$\frac{\left(\alpha_{f f^\prime}^{(8)}\right)_{ijkl}}{\Lambda^2}=$&$\!\!\!\!-\frac{\left(g_{\cal G}^{f}\right)_{ij}\left(g_{\cal G}^{f^\prime}\right)_{kl}}{4M_{\cal G}^2}$\\
$ $&$ $&$ $&$ $&$\!\!\!\!\!\!\!\!\!\!\!\!\!\!\!\!\!\!\scriptmath{\left(ff'=uu,dd,ud\right)}$\\[-0.3cm]
$ $&$ $&$ $&$ $&\\[-0.3cm]
\multicolumn{1}{l}{$\bullet~${\bf LRRL}}& & & &\\
$ $&$ $&$ $&$ $&\\[-0.3cm]
$\frac{\left(~\!\alpha_{qf}^{(1)}~\!\right)_{ijkl}}{\Lambda^2}=$&$\!\!\!\!\phantom{+}\frac{8\left(g_{\cal G}^q\right)_{il}\left(g_{\cal G}^{f}\right)_{kj}}{9M_{\cal G}^2}$&$ $&$ $&\\
$ $&$ $&$\!\!\!\!\!\scriptmath{\left(f=u,d\right)}$&$ $&\\[-0.1cm]
$\frac{\left(~\!\alpha_{qf}^{(8)}~\!\right)_{ijkl}}{\Lambda^2}=$&$\!\!\!\!-\frac{\left(g_{\cal G}^q\right)_{il}\left(g_{\cal G}^{f}\right)_{kj}}{3M_{\cal G}^2}$&$ $&$ $&\\[-0.3cm]
$ $&$ $&$ $&$ $&\\
\cbottomrule
\end{tabular}
\caption{Operators arising from the integration of a ${\cal G}$ vector field.\label{Table: G0Table}}
}
\end{center}
\end{table}
%

\begin{table}[p]
\begin{center}
{\small
\begin{tabular}{c c}
\ctoprule
\multicolumn{2}{c}{\Bfmath{{\cal H}_\mu \sim \left(\Adj,\Adj \right)_0}}\\
\multicolumn{2}{c}{ }\\[-0.2cm]
\multicolumn{2}{c}{$J^{\cal H}_{a,A\ \mu}=(g_{\cal G}^q)_{ij}\overline{q_L^i}\gamma_\mu \frac {\sigma_a}{2}\frac {\lambda_A}{2}q_L^j$}\\
\midrule
\multicolumn{2}{c}{ }\\[-0.3cm]
\multicolumn{2}{c}{{\bf Four-Fermion Operators}}\\
$ $&$ $\\[-0.3cm]
\multicolumn{1}{l}{$\bullet~${\bf LLLL}}& \\
$ $&$ $\\[-0.3cm]
$\frac{\left(\alpha_{qq}^{(8,3)}\right)_{ijkl}}{\Lambda^2}=$&$\!\!\!\!-\frac{\left(g_{\cal H}^q\right)_{ij}\left(g_{\cal H}^q\right)_{kl}}{16 M_{\cal H}^2} $\\[-0.2cm]
$ $&$ $\\
\cbottomrule
\end{tabular}
\caption{Operators arising from the integration of a ${\cal H}$ vector field.\label{Table: H0Table}}
}
\end{center}
\end{table}
%

\begin{table}[p]
\begin{center}
{\small
\begin{tabular}{c l c c l}
\ctoprule
\multicolumn{5}{c}{\Bfmath{{\cal B}^1_\mu \sim \left(1,1 \right)_1}}\\
\multicolumn{5}{c}{ }\\[-0.2cm]
\multicolumn{5}{c}{$J^{{\cal B}^1}_\mu=\left(g^{du}_{{\cal B}^1}\right)_{ij}\overline{d_R^i}\gamma_\mu u^j_R+g_{{\cal B}^1}^\phi i D_\mu \phi^Ti\sigma_2\phi$}\\
\midrule
\multicolumn{5}{c}{ }\\[-0.3cm]
\multicolumn{5}{c}{{\bf Four-Fermion Operators}}\\
$ $&$ $&$ $&$ $&\\[-0.3cm]
\multicolumn{1}{l}{$\bullet~${\bf RRRR}}& &$\phantom{~~Space~~}$&&\\
$ $&$ $&$ $&$ $&\\[-0.3cm]
$\frac{\left(\alpha_{ud}^{(1)}\right)_{ijkl}}{\Lambda^2}=$&$\!\!\!\!-\frac{\left(g_{{\cal B}^1}^{du}\right)^\dagger_{ij}\left(g_{{\cal B}^1}^{du}\right)_{kl}}{3M_{{\cal B}^1}^2}$& &
$\frac{\left(\alpha_{ud}^{(8)}\right)_{ijkl}}{\Lambda^2}=$&$\!\!\!\!-\frac{\left(g_{{\cal B}^1}^{du}\right)^\dagger_{ij}\left(g_{{\cal B}^1}^{du}\right)_{kl}}{2M_{{\cal B}^1}^2}$\\
$ $&$ $&$ $&$ $&\\[-0.3cm]
\cmrule
\multicolumn{2}{c}{{\bf SVF Operators}}& &\multicolumn{2}{c}{{\bf Oblique Operators}}\\
\multicolumn{5}{c}{ }\\
$\frac{\alpha_{\phi ud}}{\Lambda^2}=$&$\!\!\!\!\!\!\!\!\!\!\frac{g_{{\cal B}^1}^\phi\left(g_{{\cal B}^1}^{du}\right)^\dagger_{ij}}{M_{{\cal B}^1}^2}$& &
$\frac{\alpha_\phi^{(1)}}{\Lambda^2}=$&$\!\!\!\!\!\!\!\!\!\!\!\!-\frac{3\left|g_{{\cal B}^1}^\phi\right|^2}{2M_{{\cal B}^1}^2}$\\
$ $&$ $&$ $&$ $&\\[-0.3cm]
&&&$\frac{\alpha_\phi^{(3)}}{\Lambda^2}=$&$\!\!\!\!\!\!\!\!\!\!\!\!\phantom{+}\frac{\left|g_{{\cal B}^1}^\phi\right|^2}{M_{{\cal B}^1}^2}$\\[-0.2cm]
$ $&$ $&$ $&$ $&\\
\cbottomrule
\end{tabular}
\caption{Operators arising from the integration of a ${\cal B}^1$ vector field.\label{Table: B1Table}}
}
\end{center}
\end{table}
%

\begin{table}[p]
\begin{center}
{\small
\begin{tabular}{c l c c l}
\ctoprule
\multicolumn{5}{c}{\Bfmath{{\cal W}^1_\mu \sim \left(1,\Adj \right)_1}}\\
\multicolumn{5}{c}{ }\\[-0.2cm]
\multicolumn{5}{c}{$J^{{\cal W}^1}_\mu=g_{{\cal W}^1}^\phi i D_\mu \phi^Ti\sigma_2\frac{\sigma_a}{2}\phi$}\\
\midrule
\multicolumn{5}{c}{ }\\[-0.3cm]
\multicolumn{5}{c}{{\bf Oblique Operators}}\\
\multicolumn{5}{c}{ }\\[-0.3cm]
$\frac{\alpha_\phi^{(1)}}{\Lambda^2}=$&$\!\!\!\!-\frac{\left|g_{{\cal W}^1}^\phi\right|^2}{4M_{{\cal W}^1}^2}$&$\phantom{~~Space~~}$&
$\frac{\alpha_\phi^{(3)}}{\Lambda^2}=$&$\!\!\!\!-\frac{\left|g_{{\cal W}^1}^\phi\right|^2}{4M_{{\cal W}^1}^2}$\\[-0.2cm]
$ $&$ $&$ $&$ $&\\
\cbottomrule
\end{tabular}
\caption{Operators arising from the integration of a ${\cal W}^1$ vector field.\label{Table: W1Table}}
}
\end{center}
\end{table}
%

\begin{table}[p]
\begin{center}
{\small
\begin{tabular}{c l c c l}
\ctoprule
\multicolumn{5}{c}{\Bfmath{{\cal G}^1_\mu \sim \left(\Adj,1 \right)_1}}\\
\multicolumn{5}{c}{ }\\[-0.2cm]
\multicolumn{5}{c}{$J^{{\cal G}^1}_\mu=\left(g^{du}_{{\cal G}^1}\right)_{ij}\overline{d_R^i}\frac{\lambda_A}{2}\gamma_\mu u^j_R$}\\
\midrule
\multicolumn{5}{c}{ }\\[-0.3cm]
\multicolumn{5}{c}{{\bf Four-Fermion Operators}}\\
$ $&$ $&$ $&$ $&\\[-0.3cm]
\multicolumn{1}{l}{$\bullet~${\bf RRRR}}& &$\phantom{~~Space~~}$& &\\
$ $&$ $&$ $&$ $&\\[-0.3cm]
$\frac{\left(\alpha_{ud}^{(1)}\right)_{ijkl}}{\Lambda^2}=$&$\!\!\!\!-\frac{4\left(g_{{\cal G}^1}^{du}\right)^\dagger_{il}\left(g_{{\cal G}^1}^{du}\right)_{kj}}{9M_{{\cal G}^1}^2} $& &
$\frac{\left(\alpha_{ud}^{(8)}\right)_{ijkl}}{\Lambda^2}=$&$\!\!\!\!\frac{\left(g_{{\cal G}^1}^{du}\right)^\dagger_{il}\left(g_{{\cal G}^1}^{du}\right)_{kj}}{6 M_{{\cal G}^1}^2} $\\[-0.2cm]
$ $&$ $&$ $&$ $&\\
\cbottomrule
\end{tabular}
\caption{Operators arising from the integration of a ${\cal G}^1$ vector field.\label{Table: G1Table}}
}
\end{center}
\end{table}
%

\begin{table}[p]
\begin{center}
{\small
\begin{tabular}{c c}
\ctoprule
\multicolumn{2}{c}{\Bfmath{{\cal L}_\mu \sim \left(1,2 \right)_{-\frac 32}}}\\
\multicolumn{2}{c}{ }\\[-0.2cm]
\multicolumn{2}{c}{$J^{{\cal L}}_\mu=\left(g^{el}_{{\cal L}}\right)_{ij}\overline{e_R^{c\ i}}\gamma_\mu l_L^j$}\\
\midrule
\multicolumn{2}{c}{ }\\[-0.3cm]
\multicolumn{2}{c}{{\bf Four-Fermion Operators}}\\
$ $&$ $\\[-0.3cm]
\multicolumn{1}{l}{$\bullet~${\bf LRRL}}& \\
$ $&$ $\\[-0.3cm]
$\frac{\left(\alpha_{le}\right)_{ijkl}}{\Lambda^2}=$&$\!\!\!\!-\frac{2\left(g^{el}_{{\cal L}}\right)^\dagger_{ik}\left(g^{el}_{{\cal L}}\right)_{jl}}{M_{{\cal L}}^2} $\\[-0.2cm]
$ $&$ $\\
\cbottomrule
\end{tabular}
\caption{Operators arising from the integration of a ${\cal L}$ vector field.\label{Table: L3Table}}
}
\end{center}
\end{table}
%

\begin{table}[p]
\begin{center}
{\small
\begin{tabular}{c l c c l}
\ctoprule
\multicolumn{5}{c}{\Bfmath{{\cal U}^2_\mu \sim \left(3,1 \right)_{\frac 23}}}\\
\multicolumn{5}{c}{ }\\[-0.2cm]
\multicolumn{5}{c}{$J^{{\cal U}^2}_\mu=\left(g^{ed}_{{\cal U}^2}\right)_{ij}\overline{e^i_R}\gamma_\mu d_R^j+(g^{lq}_{{\cal U}^2})_{ij}\overline{l^i_L}\gamma_\mu q_L^j$}\\
\midrule
\multicolumn{5}{c}{ }\\[-0.3cm]
\multicolumn{5}{c}{{\bf Four-Fermion Operators}}\\
$ $&$ $&$ $&$ $&$ $\\[-0.3cm]
\multicolumn{2}{l}{$\bullet~${\bf LLLL}}\\
$ $&$ $&$ $&$ $&$ $\\[-0.3cm]
$ $&$ $&$ $&$ $&$ $\\[-0.3cm]
$\frac{\left(\alpha_{lq}^{(1)}\right)_{ijkl}}{\Lambda^2}$=&$\!\!\!\!-\frac{\left(g^{lq}_{{\cal U}^2}\right)^\dagger_{kj}\left(g^{lq}_{{\cal U}^2}\right)_{il}}{2M_{{\cal U}^2}^2} $& &
$\frac{\left(\alpha_{lq}^{(3)}\right)_{ijkl}}{\Lambda^2}$=&$\frac{\left(\alpha_{lq}^{(1)}\right)_{ijkl}}{\Lambda^2}$\\
$ $&$ $&$ $&$ $&$ $\\[-0.3cm]
$ $&$ $&$ $&$ $&$ $\\[-0.3cm]
\multicolumn{2}{l}{$\bullet~${\bf RRRR}}& &\multicolumn{2}{l}{$\bullet~${\bf LRRL}}\\
$ $&$ $&$ $&$ $&$ $\\[-0.3cm]
$\frac{\left(\alpha_{ed}\right)_{ijkl}}{\Lambda^2}$=&$\!\!\!\!-\frac{\left(g^{ed}_{{\cal U}^2}\right)^\dagger_{kj}\left(g^{ed}_{{\cal U}^2}\right)_{il}}{M_{{\cal U}^2}^2} $& &
$\frac{\left(\alpha_{qde}\right)_{ijkl}}{\Lambda^2}$=&$\!\!\!\!\frac{2\left(g^{ed}_{{\cal U}^2}\right)^\dagger_{kj}\left(g^{lq}_{{\cal U}^2}\right)_{il}}{M_{{\cal U}^2}^2} $\\[-0.2cm]
$ $&$ $&$ $&$ $&$ $\\
\cbottomrule
\end{tabular}
\caption{Operators arising from the integration of a ${\cal U}^2$ vector field.\label{Table: U2Table}}
}
\end{center}
\end{table}
%

\begin{table}[p]
\begin{center}
{\small
\begin{tabular}{c c}
\ctoprule
\multicolumn{2}{c}{\Bfmath{{\cal U}^5_\mu \sim \left(3,1 \right)_{\frac 53}}}\\
\multicolumn{2}{c}{ }\\[-0.2cm]
\multicolumn{2}{c}{$J^{{\cal U}^5}_\mu=\left(g^{eu}_{{\cal U}^5}\right)_{ij}\overline{e^i_R}\gamma_\mu u_R^j$}\\
\midrule
\multicolumn{2}{c}{ }\\[-0.3cm]
\multicolumn{2}{c}{{\bf Four-Fermion Operators}}\\
$ $&$ $\\[-0.3cm]
\multicolumn{1}{l}{$\bullet~${\bf RRRR}}& \\
$ $&$ $\\[-0.3cm]
$\frac{\left(\alpha_{eu}\right)_{ijkl}}{\Lambda^2}=$&$\!\!\!\!-\frac{\left(g^{eu}_{{\cal U}^5}\right)^\dagger_{kj}\left(g^{eu}_{{\cal U}^5}\right)_{il}}{M_{{\cal U}^5}^2} $\\[-0.2cm]
$ $&$ $\\
\cbottomrule
\end{tabular}
\caption{Operators arising from the integration of a ${\cal U}^5$ vector field.\label{Table: U5Table}}
}
\end{center}
\end{table}
%

\begin{table}[p]
\begin{center}
{\small
\begin{tabular}{c l c c l}
\ctoprule
\multicolumn{5}{c}{\Bfmath{{\cal Q}^1_\mu \sim \left(3,2 \right)_{\frac 16}}}\\
\multicolumn{5}{c}{ }\\[-0.2cm]
\multicolumn{5}{c}{$J^{{\cal Q}^1}_\mu=\left(g^{ul}_{{\cal Q}^1}\right)_{ij}\overline{u^{c\ i}_R}\gamma_\mu l_L^{j}+\left(g^{dq}_{{\cal Q}^1}\right)_{ij}\epsilon_{ABC}\overline{d^{i\ B}_R}\gamma_\mu i\sigma_2 q_L^{c\ j\ C}$}\\
\midrule
\multicolumn{5}{c}{ }\\[-0.3cm]
\multicolumn{5}{c}{{\bf Four-Fermion Operators}}\\
$ $&$ $&$ $&$ $&\\[-0.3cm]
\multicolumn{1}{l}{$\bullet~${\bf LRRL}}& &$\phantom{~~Space~~}$& &\\
$ $&$ $&$ $&$ $&\\[-0.3cm]
$\frac{\left(\alpha_{lu}\right)_{ijkl}}{\Lambda^2}=$&$\!\!\!\!\!\!\!\!\!\!\!-\frac{2\left(g^{ul}_{{\cal Q}^1}\right)^\dagger_{ik}\left(g^{ul}_{{\cal Q}^1}\right)_{jl}}{M_{{\cal Q}^1}^2} $& & &\\
$ $&$ $&$ $&$ $&\\[-0.3cm]
$\frac{\left(\alpha_{qd}^{(1)}\right)_{ijkl}}{\Lambda^2}=$&$\!\!\!\!\!\!\!\!\!\!\!\phantom{+}\frac{4\left(g^{dq}_{{\cal Q}^1}\right)^\dagger_{lj}\left(g^{dq}_{{\cal Q}^1}\right)_{ki}}{3M_{{\cal Q}^1}^2}$& &
$\frac{\left(\alpha_{qd}^{(8)}\right)_{ijkl}}{\Lambda^2}=$&$\!\!\!\!-\frac{\left(g^{dq}_{{\cal Q}^1}\right)^\dagger_{lj}\left(g^{dq}_{{\cal Q}^1}\right)_{ki}}{M_{{\cal Q}^1}^2}$\\
$ $&$ $&$ $&$ $&\\[-0.3cm]
$ $&$ $&$ $&$ $&\\[-0.3cm]
\multicolumn{1}{l}{$\bullet~${\bf LRRL: B$\bfmath{-}$L}}& & & &\\
$ $&$ $&$ $&$ $&\\[-0.3cm]
$\frac{\left(\alpha_{lqdu}\right)_{ijkl}}{\Lambda^2}=$&$\!\!\!\!\!\!\!\!\!\!\!\phantom{+}\frac{2\left(g_{{\cal Q}^1}^{ul}\right)^\dagger_{il}\left(g_{{\cal Q}^1}^{dq}\right)_{kj}}{M_{{\cal Q}^1}^2}$& & &\\[-0.2cm]
$ $&$ $&$ $&$ $&\\
\cbottomrule
\end{tabular}
\caption{Operators arising from the integration of a ${\cal Q}^1$ vector field.\label{Table: Q1Table}}
}
\end{center}
\end{table}
%

\begin{table}[p]
\begin{center}
{\small
\begin{tabular}{c l c c l}
\ctoprule
\multicolumn{5}{c}{\Bfmath{{\cal Q}^5_\mu \sim \left(3,2 \right)_{-\frac 56}}}\\
\multicolumn{5}{c}{ }\\[-0.2cm]
\multicolumn{5}{c}{$J^{{\cal Q}^5}_\mu=\left(g^{dl}_{{\cal Q}^5}\right)_{ij}\overline{d^{c\ i}_R}\gamma_\mu l_L^{j}+\left(g^{eq}_{{\cal Q}^5}\right)_{ij}\overline{e^{c\ i}_R}\gamma_\mu q_L^{j}+\left(g^{uq}_{{\cal Q}^5}\right)_{ij}\epsilon_{ABC}\overline{u^{i\ B}_R}\gamma_\mu i\sigma_2q_L^{c\ j\ C}$}\\
\midrule
\multicolumn{5}{c}{ }\\[-0.3cm]
\multicolumn{5}{c}{{\bf Four-Fermion Operators}}\\
$ $&$ $&$ $&$ $&\\[-0.3cm]
\multicolumn{1}{l}{$\bullet~${\bf LRRL}}& &$\phantom{~~Space~~}$& &\\
$ $&$ $&$ $&$ $&\\[-0.3cm]
$\frac{\left(\alpha_{ld}\right)_{ijkl}}{\Lambda^2}=$&$\!\!\!\!\!\!\!\!\!\!\!-\frac{2\left(g^{dl}_{{\cal Q}^5}\right)^\dagger_{ik}\left(g^{dl}_{{\cal Q}^5}\right)_{jl}}{M_{{\cal Q}^5}^2} $& &
$\frac{\left(\alpha_{qe}\right)_{ijkl}}{\Lambda^2}=$&$\!\!\!\!\!\!-\frac{2\left(g^{eq}_{{\cal Q}^5}\right)^\dagger_{ik}\left(g^{eq}_{{\cal Q}^5}\right)_{jl}}{M_{{\cal Q}^5}^2} $\\
$ $&$ $&$ $&$ $&\\[-0.3cm]
$ $&$ $&$ $&$ $&\\[-0.3cm]
$\frac{\left(\alpha_{qde}\right)_{ijkl}}{\Lambda^2}=$&$\!\!\!\!\!\!\!\!\!\!\!-\frac{2\left(g^{dl}_{{\cal Q}^5}\right)^\dagger_{ik}\left(g^{eq}_{{\cal Q}^5}\right)_{jl}}{M_{{\cal Q}^5}^2} $& & & \\
$ $&$ $&$ $&$ $&\\[-0.3cm]
$ $&$ $&$ $&$ $&\\[-0.3cm]
$\frac{\left(\alpha_{qu}^{(1)}\right)_{ijkl}}{\Lambda^2}=$&$\!\!\!\!\!\!\!\!\!\!\!\phantom{+}\frac{4\left(g^{uq}_{{\cal Q}^5}\right)^\dagger_{lj}\left(g^{uq}_{{\cal Q}^5}\right)_{ki}}{3M_{{\cal Q}^5}^2}$& &
$\frac{\left(\alpha_{qu}^{(8)}\right)_{ijkl}}{\Lambda^2}=$&$\!\!\!\!\!\!-\frac{\left(g^{uq}_{{\cal Q}^5}\right)^\dagger_{lj}\left(g^{uq}_{{\cal Q}^5}\right)_{ki}}{M_{{\cal Q}^5}^2}$\\
$ $&$ $&$ $&$ $&\\[-0.3cm]
$ $&$ $&$ $&$ $&\\[-0.3cm]
\multicolumn{1}{l}{$\bullet~${\bf LRRL: B$\bfmath{-}$L}}& & & &\\
$ $&$ $&$ $&$ $&\\[-0.3cm]
$\frac{\left(\alpha_{lqdu}\right)_{ijkl}}{\Lambda^2}=$&$\!\!\!\!\!\!\!\!\!\!\!\phantom{+}\frac{2\left(g_{{\cal Q}^5}^{dl}\right)^\dagger_{ik}\left(g_{{\cal Q}^5}^{uq}\right)_{lj}}{M_{{\cal Q}^5}^2}$& &
$\frac{\left(\alpha_{qqeu}\right)_{ijkl}}{\Lambda^2}=$&$\!\!\!\!\!\!-\frac{2\left(g_{{\cal Q}^5}^{eq}\right)^\dagger_{ik}\left(g_{{\cal Q}^5}^{uq}\right)_{lj}}{M_{{\cal Q}^5}^2}$\\[-0.2cm]
$ $&$ $&$ $&$ $&\\
\cbottomrule
\end{tabular}
\caption{Operators arising from the integration of a ${\cal Q}^5$ vector field.\label{Table: Q5Table}}
}
\end{center}
\end{table}
%

\begin{table}[p]
\begin{center}
{\small
\begin{tabular}{c l c c l}
\ctoprule
\multicolumn{5}{c}{\Bfmath{{\cal X}_\mu \sim \left(3,\Adj \right)_{\frac 23}}}\\
\multicolumn{5}{c}{ }\\[-0.2cm]
\multicolumn{5}{c}{$J^{{\cal X}}_\mu=\left(g^{lq}_{{\cal X}}\right)_{ij}\overline{l^i_L}\gamma_\mu \frac{\sigma_a}{2} q_L^j$}\\
\midrule
\multicolumn{5}{c}{ }\\[-0.3cm]
\multicolumn{5}{c}{{\bf Four-Fermion Operators}}\\
$ $&$ $&$ $&$ $&\\[-0.3cm]
\multicolumn{1}{l}{$\bullet~${\bf LLLL}}& &$\phantom{~~Space~~}$& &\\
$ $&$ $&$ $&$ $&\\[-0.3cm]
$\frac{\left(\alpha_{lq}^{(1)}\right)_{ijkl}}{\Lambda^2}=$&$\!\!\!\!-\frac{3\left(g^{lq}_{{\cal X}}\right)^\dagger_{kj}\left(g^{lq}_{{\cal X}}\right)_{il}}{8M_{{\cal X}}^2} $& &
$\frac{\left(\alpha_{lq}^{(3)}\right)_{ijkl}}{\Lambda^2}=$&$\!\!\!\!\frac{\left(g^{lq}_{{\cal X}}\right)^\dagger_{kj}\left(g^{lq}_{{\cal X}}\right)_{il}}{8M_{{\cal X}}^2} $\\[-0.2cm]
$ $&$ $&$ $&$ $&\\
\cbottomrule
\end{tabular}
\caption{Operators arising from the integration of a ${\cal X}$ vector field.\label{Table: XTable}}
}
\end{center}
\end{table}
%

\begin{table}[p]
\begin{center}
{\small
\begin{tabular}{c l c c l}
\ctoprule
\multicolumn{5}{c}{\Bfmath{{\cal Y}^1_\mu \sim \left(\overline{6},2 \right)_{\frac 16}}}\\
\multicolumn{5}{c}{ }\\[-0.2cm]
\multicolumn{5}{c}{$J^{{\cal Y}^1}_\mu=\left(g^{dq}_{{\cal Y}^1}\right)_{ij}\overline{d^{i\left(A\right|}_R}\gamma_\mu i\sigma_2 q_L^{c\ j\left|B\right)}$}\\
\midrule
\multicolumn{5}{c}{ }\\[-0.3cm]
\multicolumn{5}{c}{{\bf Four-Fermion Operators}}\\
$ $&$ $&$ $&$ $&\\[-0.3cm]
\multicolumn{1}{l}{$\bullet~${\bf LRRL}}& &$\phantom{~~Space~~}$& &\\
$ $&$ $&$ $&$ $&\\[-0.3cm]
$\frac{\left(\alpha_{qd}^{(1)}\right)_{ijkl}}{\Lambda^2}=$&$\!\!\!\!-\frac{4\left(g^{dq}_{{\cal Y}^1}\right)^\dagger_{lj}\left(g^{dq}_{{\cal Y}^1}\right)_{ki}}{3M_{{\cal Y}^1}^2} $& &
$\frac{\left(\alpha_{qd}^{(8)}\right)_{ijkl}}{\Lambda^2}=$&$\!\!\!\!-\frac{\left(g^{dq}_{{\cal Y}^1}\right)^\dagger_{lj}\left(g^{dq}_{{\cal Y}^1}\right)_{ki}}{2M_{{\cal Y}^1}^2} $\\[-0.2cm]
$ $&$ $&$ $&$ $&\\
\cbottomrule
\end{tabular}
\caption{Operators arising from the integration of a ${\cal Y}^1$ vector field. $(A|\cdots|B)=\frac 12 (AB+BA)$ stands for the symmetric combination of color indices. \label{Table: Y1Table}}
}
\end{center}
\end{table}
%

\begin{table}[p]
\begin{center}
{\small
\begin{tabular}{c l c c l}
\ctoprule
\multicolumn{5}{c}{\Bfmath{{\cal Y}^5_\mu \sim \left(\overline{6},2 \right)_{-\frac 56}}}\\
\multicolumn{5}{c}{ }\\[-0.3cm]
\multicolumn{5}{c}{$J^{{\cal Y}^5}_\mu=\left(g^{uq}_{{\cal Y}^5}\right)_{ij}\overline{u^{i\left(A\right|}_R}\gamma_\mu i\sigma_2 q_L^{c\ j\left|B\right)}$}\\
\midrule
\multicolumn{5}{c}{ }\\[-0.2cm]
\multicolumn{5}{c}{{\bf Four-Fermion Operators}}\\
$ $&$ $&$ $&$ $&\\[-0.3cm]
\multicolumn{1}{l}{$\bullet~${\bf LRRL}}& &$\phantom{~~Space~~}$& &\\
$ $&$ $&$ $&$ $&\\[-0.3cm]
$\frac{\left(\alpha_{qu}^{(1)}\right)_{ijkl}}{\Lambda^2}=$&$\!\!\!\!-\frac{4\left(g^{uq}_{{\cal Y}^5}\right)^\dagger_{lj}\left(g^{uq}_{{\cal Y}^5}\right)_{ki}}{3M_{{\cal Y}^5}^2} $& &
$\frac{\left(\alpha_{qu}^{(8)}\right)_{ijkl}}{\Lambda^2}=$&$\!\!\!\!-\frac{\left(g^{uq}_{{\cal Y}^5}\right)^\dagger_{lj}\left(g^{uq}_{{\cal Y}^5}\right)_{ki}}{2M_{{\cal Y}^5}^2} $\\[-0.2cm]
$ $&$ $&$ $&$ $&\\
\cbottomrule
\end{tabular}
\caption{Operators arising from the integration of a ${\cal Y}^5$ vector field.\label{Table: Y5Table}}
}
\end{center}
\end{table}
%

~\newpage

\section{Experimental data and fit method}
\label{app: fits}

In this appendix we briefly describe the data included in our fits. Table~\ref{Table: Exp-SM} gathers, and updates in some cases, all the experimental measurements used in references~\cite{delAguila:2008pw,delAguila:2009vv}. This includes:
\begin{itemize}
{\item $Z$-pole observables: $Z$ decay widths, LR and forward-backward asymmetries, etc.}
{\item $W$ data: $W$ mass and width, constraints on the unitarity of the first row of the CKM matrix.} {\item Low-energy effective couplings from neutrino scattering with nucleons and electrons and from parity violation in atoms and in M{\o}ller scattering}
\end{itemize}
The first class of observables provides the strongest constraints on extra corrections to trilinear couplings. The low-energy data, in turn, puts restrictions on possible four-fermion interactions, whose effects are only noticeable away from the $Z$ resonance. This table also shows the values for the experimental determinations of the SM parameters entering in the fit, such as the top mass or the strong coupling constant at the $Z$ peak. 

We also include in our fits results from the measurements at LEP~2 of $e^+e^-\rightarrow \overline{f}f$ at energies above $M_Z$~\cite{Alcaraz:2006mx,Abbiendi:2003dh}. Although the precision for each of these measurements is in general smaller than for other observables, this is compensated in the $\chi^2$ by the large amount of available data. Therefore, they are a valuable source of constraints to four-fermion interactions, complementary to those coming from the low-energy data. They cannot compete with the $Z$-pole data, however, in constraining the trilinear couplings.\\

We have used {\bf ZFITTER 6.43} \cite{Arbuzov:2005ma} for the computation of the SM predictions for observables. Some low-energy observables not available (or not well described) in that code, such as M{\o}ller scattering, have been computed with our own dedicated routines. The new physics effects induced by  the dimension six operators in the effective Lagrangian are summed to the SM value. As discussed in the main text, in most cases we only incorporate tree-level contributions from new physics and keep their interference with the SM amplitudes. When we find large couplings we also include quadratic corrections in the new physics for LEP 2 observables. In general, we keep $M_H$, $m_t$ and $\alpha_s\left(M_Z\right)$ as floating parameters in all our fits, as far as the $Z$-pole and low-energy observables are concerned, while $M_Z$ and  $\Delta \alpha_\mt{had}^{(5)}\left(M_Z\right)$ are kept fixed at their SM best-fit values. On the other hand, these SM parameters are fixed in the predictions for the LEP 2 data since, as we have checked, the effect of their variation is small. We have included, however, the leading oblique correction from the variation of the Higgs mass, as this is the parameter with the largest uncertainty. 

Although the Higgs boson has not been discovered yet, the results from direct searches at LEP 2 \cite{Barate:2003sz} and Tevatron~\cite{Collaboration:2009je} constraints the possible values for $M_H$. These results have been also included in our fits.\\

In order to estimate the values of the free parameters of each class of models, which are generically denoted by $\theta$ in the following, we test the predictions for the observables against the experimental data using a $\chi^2$ analysis. Specifically, we compute the quantity
\be
\chi^2\left(\theta\right)=\left[\mathrm{O}_\mathrm{exp}-\mathrm{O}_\mathrm{th}\left(\theta\right)\right]^T U_\mathrm{exp}^{-1}\left[\mathrm{O}_\mathrm{exp}-\mathrm{O}_\mathrm{th}\left(\theta\right)\right],
\ee
where $\left(U_\mathrm{exp}\right)_{ij}=\sigma_i \rho_{ij} \sigma_j$ is the covariance matrix, with $\sigma$ and $\rho$ the experimental errors and correlation matrix. $\mathrm{O}_\mathrm{exp}$ denotes the experimental values for the observables and $\mathrm{O}_\mathrm{exp}\left(\theta\right)$ the theoretical prediction depending on the free parameters $\theta$. Experimental correlations can be obtained from the references in the tables. The minimization and computation of limits and contours have been performed with the aid of the program {\bf MINUIT} \cite{James:1975dr}. 

%

%
\begin{table}[p]
\begin{center}
{\scriptsize
\begin{tabular}{l r | c | c r}
\ctoprule
\crowcolor {\bf Observable}&&{\bf Experimental Value}&{\bf Standard Model}&\!{\bf Pull}\!\\
\cmrule
$m_t\left[\mbox{GeV}\right]$&\cite{Tevatron:2009ec}&$173.1\pm 1.3$&$173.4$&$-0.3 $\\
$\Delta \alpha^{(5)}_\mathrm{had}\left(M_Z^2\right)$&\!\!\!\!\!\!\!\!\!\!\!\!\!\!\!\cite{Burkhardt:2005se,Teubner:2010ah}&$0.02760\pm 0.00014 $&$0.02760$&$\phantom{+}0\phantom{.0} $\\[0.05cm]
$\alpha_s\left(M_Z^2\right)$&\!\!\!\!\!\!\!\!\!\!\!\!\!\!\!\cite{Bethke:2009jm}&$0.1184\pm 0.0007$&$0.1184 $&$\phantom{+}0\phantom{.0}  $\\[0.05cm]\cmrule
$M_W\left[\mbox{GeV}\right]$&\cite{Alcaraz:2009jr}&$80.399\pm 0.023$&$80.367$&$+1.4 $\\
$\Gamma_W\left[\mbox{GeV}\right]$& &$2.098\pm0.048 $&$2.091$&$+0.1$\\
$\mbox{Br}\left(W\rightarrow e\nu\right)$\!&\cite{Amsler:2008zzb}&$0.1075\pm0.0013$&$0.1083 $&$-0.6 $\\
$\mbox{Br}\left(W\rightarrow \mu\nu\right)$\!&&$0.1057\pm0.0015 $&$ $&$-1.7$\\
$\mbox{Br}\left(W\rightarrow \tau\nu\right)$\!&&$0.1125\pm0.0020 $&$ $&$+2.1$\\[0.05cm]\cmrule
$M_Z\left[\mbox{GeV}\right]$&\cite{LEPEWWG:2005ema}&$91.1876\pm 0.0021$&$91.1876 $&$\phantom{+}0\phantom{.0} $\\
$\Gamma_Z \left[\mbox{GeV}\right]$&&$2.4952\pm 0.0023$&$2.4955$&$-0.1$\\
$\sigma_\mt{had}\left[\mbox{nb}\right]$&&$41.541\pm 0.037$&$41.479$&$+1.7$\\
$R_e$&&$20.804\pm 0.050$&$20.740$&$+1.3$\\
$R_\mu$&&$20.785\pm 0.033$&$20.740$&$+1.4 $\\
$R_\tau$&&$20.764\pm 0.045$&$20.787$&$-0.5$\\
$A^{e}_\mathrm{FB}$&&$0.0145\pm 0.0025$&$0.0163$&$-0.7$\\
$A^{\mu}_\mathrm{FB}$&&$0.0169\pm 0.0013$&$ $&$+0.5$\\
$A^{\tau}_\mathrm{FB}$&&$0.0188\pm 0.0017$&$ $&$+1.5$\\[0.05cm]\cmrule
$A_e\left(\mbox{SLD}\right)$&\cite{LEPEWWG:2005ema}&$0.1516\pm 0.0021$&$0.1474$&$+2.0$\\
$A_\mu\left(\mbox{SLD}\right)$&&$0.142\pm 0.015$&$ $&$-0.4$\\
$A_\tau\left(\mbox{SLD}\right)$&&$0.136\pm 0.015$&$ $&$-0.8$\\[0.05cm]\cmrule
$A_e\left(P_\tau\right)$&\cite{LEPEWWG:2005ema}&$0.1498\pm 0.0049$&$ $&$+0.5$\\
$A_\tau\left(P_\tau\right)$&&$0.1439\pm 0.0043$&$ $&$-0.8$\\[0.05cm]\cmrule
$R_b$&\cite{LEPEWWG:2005ema}&$0.21629\pm 0.00066$&$0.21580$&$+0.7$\\
$R_c$&&$0.1721\pm 0.0030$&$0.1722$&$-0.1$\\
$A^{b}_\mathrm{FB}$&&$0.0992\pm 0.0016$&$0.1033$&$-2.6$\\
$A^{c}_\mathrm{FB}$&&$0.0707\pm 0.0035$&$0.738$&$-0.9$\\ 
$A_b$&&$0.923\pm 0.020$&$0.935$&$-0.6 $\\
$A_c$&&$0.670\pm 0.027$&$0.668$&$+0.1$\\[0.05cm]\cmrule
$A^{s}_\mathrm{FB}$&\cite{LEPEWWG:2005ema}&$0.098\pm 0.011$&$0.1034$&$-0.5 $\\
$A_s$&&$0.895\pm 0.091$&$0.936 $&$-0.5 $\\ 
$R_u/R_{u+d+s}$&&$0.258\pm 0.045$&$0.282$&$-0.5$\\[0.05cm]\cmrule
$Q_\mathrm{FB}^\mathrm{had}$&\cite{LEPEWWG:2005ema}&$0.0403\pm 0.0026$&$0.0423$&$-0.8$\\
$\sin^2{\theta_\mathrm{eff}^\mathrm{lept}}$&\cite{Acosta:2004wq}&$0.2315 \pm 0.0018 $&$0.2315$&$\phantom{+}0\phantom{.0} $\\[0.05cm]\cmrule
$g^2_L$&\cite{ErlerLanginPDG}&$0.3012\pm 0.0013$&$0.3039$&$-2.0 $\\
$g^2_R$&&$0.0310\pm 0.0010$&$0.03013$&$+0.9$\\
$\theta_L$&&$2.500\pm 0.033$&$2.46 $&$+1.1 $\\
$\theta_R$&&$4.58\pm 0.41$&$5.18$&$-1.5 $\\[0.05cm]\cmrule
$g_V^{\nu e}$&\cite{ErlerLanginPDG}&$-0.040\pm 0.015$&$-0.0398 $&$\phantom{+}0\phantom{.0}  $\\
$g_A^{\nu e}$&&$-0.507\pm 0.014$&$0.507 $&$\phantom{+}0\phantom{.0}  $\\[0.05cm]\cmrule
$Q_W\left(^{133}_{55}\mbox{Cs}\right)$&\cite{Porsev:2009pr}&$-73.16 \pm 0.35$&$-73.14$&$-0.1$\\
$Q_W\left(^{205}_{81}\mbox{Tl}\right)$&\cite{Vetter:1995vf} &$-116.4 \pm 3.6 $&$-116.7$&$+0.1 $\\
$\cos{\gamma} C_{1d}\!-\!\sin{\gamma}C_{1u}\!\!\!\!\!\!\!$&\cite{Erler:2009jh}&$0.342 \pm 0.063 $&$0.388$&$-0.7$\\
$\sin{\gamma} C_{1d}\!+\!\cos{\gamma}C_{1u}\!\!\!\!\!\!\!$& &$-0.0285 \pm 0.0043$&$-0.0335 $&$+1.2 $\\
$Q_W\left(e\right)\left(\mbox{M{\o}ller}\right)$\!\!\!\!\!\!\!&\cite{Anthony:2005pm}&$-0.0403\pm 0.0053$&$-0.0471$&$+1.3$\\[0.05cm]\cmrule
$\sum_i\left|V_{ui}\right|^2$&\cite{BluchMarciinPDG}&$0.9999 \pm 0.0006$&$1 $&$-0.2 $\\[0.05cm]\cmrule
$\sigma^{\nu e\rightarrow \nu \mu}/\sigma^{\nu e\rightarrow \nu \mu}_\SM $&\cite{Mishra:1990yf}&$0.981\pm 0.057$&$1 $&$-0.3 $\\
\cbottomrule  
\end{tabular}}
\caption{Measurements of the observables included
  in our fits, compared with the best-fit values in the SM.
  }
\label{Table: Exp-SM}
\end{center}
\end{table}

~\newpage


\end{document}